\documentclass[a4paper]{amsart}
\usepackage[active]{srcltx}

\usepackage{latexsym}
\usepackage{amsmath}
\usepackage{amssymb}
\usepackage{amsfonts}
\usepackage{paralist}
\usepackage{mathrsfs}

\usepackage{esint}
\usepackage{graphicx}
\usepackage{accents}

\let\oldtocsection=\tocsection
\let\oldtocsubsection=\tocsubsection
\let\oldtocsubsubsection=\tocsubsubsection
\renewcommand{\tocsection}[2]{\hspace{0em}\oldtocsection{#1}{#2}}
\renewcommand{\tocsubsection}[2]{\hspace{2em}\oldtocsubsection{#1}{#2}}
\renewcommand{\tocsubsubsection}[2]{\hspace{3em}\oldtocsubsubsection{#1}{#2}}

\usepackage{color}
\definecolor{Blue}{rgb}{0.,0.,1.}
\definecolor{Red}{rgb}{1.,0.,0.}
\newcounter{smallarabics}
\newenvironment{arabicenumerate}
{\begin{list}{{\normalfont\textrm{(\arabic{smallarabics})}}}
  {\usecounter{smallarabics}\setlength{\itemindent}{0cm}
   \setlength{\leftmargin}{5ex}\setlength{\labelwidth}{4ex}
   \setlength{\topsep}{0.75\parsep}\setlength{\partopsep}{0ex}
   \setlength{\itemsep}{0ex}}}
{\end{list}}

\newcounter{smallroman}

\newcommand{\ben}{\begin{arabicenumerate}}  
\newcommand{\een}{\end{arabicenumerate}}

\def\init{\setcounter{equation}{0}}

\newtheorem{theoreme}{Theorem}[section]
\newtheorem{proposition}[theoreme]{Proposition}
\newtheorem{lemma}[theoreme]{Lemma}
\newtheorem{definition}[theoreme]{Definition}
\newtheorem{corollary}[theoreme]{Corollary}
\newtheorem{remark}[theoreme]{Remark}
\newtheorem{example}[theoreme]{Example}

\newcommand{\beq}{\begin{equation}}
\newcommand{\eeq}{\end{equation}}

\newcommand{\bex}{\begin{example}}
\newcommand{\eex}{\end{example}}
\def\bel{\begin{lemma}}
\def\eel{\end{lemma}}
\def\bet{\begin{theoreme}}
\def\eet{\end{theoreme}}
\def\bed{\begin{definition}}
\def\eed{\end{definition}}
\def\ber{\begin{remark}}
\def\eer{\end{remark}}

\def\Im{{\mathrm{Im}}}
\def\Re{{\mathrm{Re}}}

\def\rr{{\mathbb R}}

\def\cc{{\mathbb C}}
\def\nn{{\mathbb N}}

\def\Re{{\rm Re}}

\def\bar{\overline}

\def\coinf{C_0^\infty}

\def\qed{$\Box$\medskip}
\def \p{ \partial}

\def\12{\frac{1}{2}}

\def\e{{\rm e}}

\def\bbbone{{\mathchoice {\rm 1\mskip-4mu l} {\rm 1\mskip-4mu l}
{\rm 1\mskip-4.5mu l} {\rm 1\mskip-5mu l}}}
\def\one{\bbbone}
\def\cH{{\mathcal H}}

\def\C{{\cc}}

\def\cG{{\mathcal G}}

\def\N{\nn}

\def\cK{{\mathcal K}}

\def\bep{\begin{proposition}}
\def\eep{\end{proposition}}

\newcommand{\mbc}{\mathbb{C}}
\newcommand{\mbr}{\mathbb{R}}
\newcommand{\R}{\mathbb{R}}

\newcommand{\ca}{\mathcal{A}}
\newcommand{\cb}{\mathcal{B}}
\newcommand{\ch}{\mathcal{H}}

\newcommand{\cj}{\mathcal{J}}
\newcommand{\ck}{\mathcal{K}}

\newcommand{\cm}{\mathcal{M}}
\newcommand{\cn}{\mathcal{N}}

\newcommand{\cR}{\mathcal{R}}

\def\rond{\mathcal}

\newcommand{\rb}{\rond{B}}

\newcommand{\rd}{\rond{D}}
\newcommand{\re}{\mathcal{E}}

\newcommand{\rk}{{\mathcal E}}
\newcommand{\rh}{{\mathcal K}}

\newcommand{\rl}{\rond{L}}

\def\rmc{\mathrm{c}}

\def\rme{\mathrm{e}}

\def\rms{\mathrm{s}}
\def\rmu{\mathrm{u}}

\def\veps{\varepsilon}
\def\vphi{\varphi}

\def\rag{\rangle}
\def\lag{\langle}

\def\braket#1#2{\langle{#1}|{#2}\rangle}

\def\jap#1{\langle {#1} \rangle}

\def\what{\widehat}
\def\what#1{\widehat{ #1\,}}
\def\wtilde{\widetilde}

\def\nin{\notin}

\def\Ker{\mbox{\rm Ker\,}}

\def\sign{\text{ sign\,}}

\def\nin{\notin}

\def\qed{\hfill \raisebox{0.5ex}{\framebox[1.6ex]{
                                       \rule[0ex]{0ex}{0.3ex} }}}

\def\build#1_#2^#3{\mathrel{\mathop{\kern 0pt#1}\limits_{#2}^{#3}}}
\newcommand{\mat}[4]{\left(\begin{array}{cc}#1 &#2 \\ #3
      &#4 \end{array}\right)}  
\newcommand{\lin}[2]{\left(\begin{array}{c}#1 \\#2\end{array}\right)}

\begin{document}
\def\4{\frac{1}{4}}
\def\what{\widehat}
\def\cE{\mathcal{E}}
\def\cG{\mathcal{G}}
\def\Dom{{\rm Dom}}
\def\i{{\rm i}}
\def\h{\langle h \rangle}
\def\epsi{\langle \epsilon\rangle}
\def\cN{{\mathcal N}}
\def\jh{\langle h\rangle}
\def\sh{|h|}
\def\jr{\langle x\rangle}
\def\t{{\scriptscriptstyle\#}}

\title[Boundary values of resolvents ]{Boundary values of resolvents of \\ 
self-adjoint operators in Krein spaces \\[4mm] }   
\author{V. Georgescu}
\address{D\'epartement de Math\'ematiques,
Universit\'e de Cergy-Pontoise,
95302 Cergy-Pontoise Cedex, France }
\email{Vladimir.Georgescu@math.cnrs.fr}
\author{C. G\'erard}
\address{D\'epartement de Math\'ematiques, Universit\'e de Paris XI,
  91405 Orsay Cedex France} 
\email{christian.gerard@math.u-psud.fr}
\author{D. H\"{a}fner}
\address{Universit\'e de Grenoble 1, Institut Fourier, UMR 5582
  CNRS, BP 74 38402 Saint-Martin d'H\`eres France} 
\email{Dietrich.Hafner@ujf-grenoble.fr}
\date{\today}
\begin{abstract}
 We prove in this paper resolvent estimates for the boundary values of  resolvents of selfadjoint operators on a Krein space: if $H$ is a selfadjoint operator on a Krein space $\cH$, equipped with the Krein scalar product $\langle \cdot| \cdot \rangle$, $A$ is the generator of a $C_{0}-$group on $\cH$  and  $I\subset \rr$ is an interval such that:
\begin{itemize}
 \item[]1) $H$ admits a Borel functional calculus on $I$, 
 \item[]2)  the  spectral projection $\one_{I}(H)$ is positive in the Krein sense,
 \item[]3)  the following {\em positive commutator estimate} holds:
 \[
\Re \langle u| [H, \i A]u\rangle\geq c \langle u| u\rangle, \ u \in {\rm Ran}\one_{I}(H), \ c>0.
\]
\end{itemize}
  then assuming some smoothness of $H$ with respect to the group $\e^{\i t A}$, the following resolvent estimates hold:
  \[
\sup_{z\in I\pm \i]0, \nu]}\| \langle A\rangle ^{-s}(H-z)^{-1}\langle A\rangle^{-s}\| <\infty, \ s>\12.
\]
As an application we consider  abstract Klein-Gordon equations
\[
\p_{t}^{2}\phi(t)- 2 \i k \phi(t)+ h\phi(t)=0,
\]
and obtain resolvent estimates for their generators in {\em charge spaces} of Cauchy data.
\end{abstract}
\maketitle

\section{Introduction}\label{introd}\init
30 years ago, E. Mourre showed that a local in energy positive commutator estimate for a selfadjoint operator $H$ entails a limiting absorption principle for this operator and thus the absence of singular continuous spectrum, see \cite{M1}. This result had a very deep impact in scattering theory leading in particular to asymptotic completeness results for quantum $N-$ particle systems. Among many other applications we mention applications to Quantum Field Theory or sccattering problems in General Relativity. A lot of efforts had been made to weaken the original hypotheses in the work of Mourre, see
e.g. \cite{ABG}. A central requirement remained however that the hamiltonian $H$ is a selfadjoint operator on a Hilbert space. Whereas this is a very natural requirement for the Schr\"odinger equation, it turns out that it is in general not fulfilled for the Klein-Gordon equation when this equation is coupled to an electric field or associated to a lorentzian metric which is not stationary. The natural setting in this situation seems to be the one of a selfadjoint operator on a so called Krein space (which is a generalization of a Hilbert space). The present paper is devoted to the proof of weighted estimates for boundary values on the real line of selfadjoint operators on Krein spaces. Our result generalizes the result of Mourre to the Krein space setting. Applications to the Klein-Gordon equation are given. Let us now briefly describe the results and methods of this work.

\subsection{Selfadjoint operators on Krein spaces}
A {\em Krein space } is a Hilbertizable Banach space $\cH $ equipped with a non-degenerate hermitian form $\langle u| v\rangle$, $u, v\in \cH$ called a {\em Krein scalar product}. Orthogonals to vector subspaces and adjoint of linear operators on $\cH$ are defined with respect to $\langle \cdot| \cdot \rangle$.   

In contrast to Hilbert spaces, the hermitian form is not assumed to
be positive definite. Note however that the notion of positivity of a subspace $\cK\subset \cH$ resp.  of an operator $A$ on $\cH$ still makes sense, by requiring  that $\langle u| u\rangle\geq 0$ for all $u\in \cK$ resp. $\langle u| Au\rangle\geq 0$ for all $u\in \Dom A$.

Of special interest are {\em selfadjoint operators} on Krein spaces. Typically a selfadjoint operator $H$ on a Krein space arises as the  generator of a $C_{0}-$group $\{\e^{\i tH}\}_{t\in \rr}$ preserving the quadratic quantity $\langle u| u\rangle$.

In general, not much of interest can be said about the spectrum, functional calculus  or the behavior of the resolvent of  selfadjoint operators on a Krein space. Namely the spectrum is invariant under complex conjugation, the functional calculus is limited to the Dunford-Taylor functional calculus, and the behavior of the resolvent, both near the real axis or near infinity, can be arbitrary. 

However, there is a class of selfadjoint operators, called {\em definitizable}, first defined and studied by Langer \cite{La}, which admit a rich (i.e. Borel outside a finite subset of $\rr$) functional calculus. A selfadjoint operator $H$ on $\cH$ is definitizable if its resolvent set $\rho(H)$ is not empty and if there exists a (real) polynomial $p$ such that $p(H)\geq 0$.   Real zeroes of $p$ in the spectrum of $H$ are called {\em critical points}.
\subsection{Positive commutator method}
If $H$ is definitizable  and $I\subset \rr$ is a bounded interval with $\p I$ disjoint from the critical points of $H$, then the spectral projection 
$\one_{I}(H)$  is well defined and bounded on $\cH$. Moreover if $I$ does not contain any critical point, then $\one_{I}(H)$ is definite in the Krein sense, i.e. $\one_{I}(H)\geq 0$ or $-\one_{I}(H)\geq 0$.

This local definiteness of the Krein scalar product opens the way for an extension to the Krein space framework of the  well-known {\em positive commutator method}, which is a standard way to prove weighted resolvent estimates for usual selfadjoint operators on a Hilbert space. 
In the Hilbert space framework, the positive commutator method introduced by Mourre \cite{M1} relies on an estimate
\beq\label{tira}
\one_{I}(H)[H, \i A]\one_{I}(H)\geq c \one_{I}(H), \ c>0,
\eeq
where $H$ is the selfadjoint operator under study, $I\subset \rr$ is an interval, and $A$ is another selfadjoint operator, called a {\em conjugate operator}. From (\ref{tira}), assuming some regularity of $H$ with respect to the unitary group $\e^{\i tA}$, one obtains the resolvent estimates:
\beq\label{732}
\sup_{z\in I \pm \i]0, +\infty[}\| \langle A\rangle ^{-s}(H-z)^{-1}\langle A\rangle^{-s}\| <\infty, \ s>\12,
\eeq
see \cite{M1}, \cite{PSS}, \cite{ABG}.  The original proofs relied on differential inequalities. Some years ago  Gol\'enia and Jecko \cite{GJ} gave a new proof of the limiting absorption principle in an
abstract framework, by a contradiction argument. A direct proof, based on energy estimates  was given in \cite{G1}.  The argument in \cite{G1} is  closer to a method of Putnam \cite{P2}, which was an ancestor of  the positive commutator method. It turns out that the proof of \cite{G1} can be adapted to the Krein space framework.

Several difficulties must be faced before an estimate like (\ref{732}) can be obtained for a selfadjoint operator on a Krein space. First of all $H$ should have a Borel functional calculus in order to be able to define spectral projections. Second the conjugate operator $A$ is in general not unitary for a compatible Hilbert space structure on $\cH$.  In particular the definition of $\langle A\rangle^{-s}= (A^{2}+1)^{-s/2}$ is not obvious.

However on a Krein space, an estimate like (\ref{tira}) has still a meaning, if it is understood formally as
\beq\label{732-c}
\Re \langle u| [H, \i A]u\rangle\geq c \langle u| u\rangle, \ u \in {\rm Ran}\one_{I}(H), \ c>0.
\eeq
The main result of this paper, Thm. \ref{th:bvr}, states  that if $H$ is a selfadjoint operator on a Krein space, which is of class $C^{\alpha}$ with respect to $A$ for some $\alpha>3/2$, and $I\subset \rr$ is an interval such that:
\begin{itemize}
\item[]1) $H$ admits a Borel functional calculus near $I$, $\one_{I}(H)\geq 0$,
 \item[]2) the Mourre estimate (\ref{tira}) holds,
\end{itemize}
  then the resolvent estimates (\ref{732}) hold, possibly replacing $A$ by $\epsilon A$ for $0<\epsilon\ll 1$ and restricting $z$ to $I\pm\i ]0,\nu]$ for some $\nu>0$, due to the possible presence of complex eigenvalues.  We also prove a {\em virial theorem}, which has the same consequences as in the Hilbert space case.
 \subsection{Abstract Klein-Gordon equations}
  
In a subsequent paper \cite{GGH1}, we apply the abstract results of this paper to the generators of abstract Klein-Gordon equations
\begin{equation}\label{eq:kgab}
\p_{t}^{2}\phi(t)- 2 \i k \phi(t)+ h\phi(t)=0,
\end{equation}
where $\phi: \rr\to \cH$, $\cH$ is a Hilbert space and $h$, $k$ are
selfadjoint, resp. symmetric operators on $\cH$.  
The simplest example  is the {\em Klein-Gordon equation on Minkowski space} minimally coupled with an external electric field:
\beq\label{ei.2}
(\p_{t}- \i v(x))^{2}\phi(t,x)- \Delta_{x}\phi(t,x)+ m^{2}\phi(t,x)=0,
\eeq
for which $\cH= L^{2}(\rr^{d}, dx)$, $h=-\Delta_{x}+ m^{2}- v^{2}(x)$, $k= v(x)$ is a (real) electric potential and  $m\geq 0$ is the mass of the Klein-Gordon field.

There is a large literature devoted to the spectral theory of the abstract Klein-Gordon equation \eqref{eq:kgab} or the concrete one \eqref{ei.2} in the Krein space framework. Our basic references are the two papers \cite{LNT1,LNT2}, whose assumptions on the operators $h$ and $k$ are more general and the results more precise than the previous ones. On the historical side, we note that the equation  was first treated in the charge Krein space setting, which is of special interest for us, in \cite{V1} and further studied in \cite{N1,N2,V2}. The relevance of the scale of Krein spaces $\cK_{\theta}$ (see  Subsect. \ref{notato} for the notation) has been pointed out in \cite{N1}.
 
In contrast to Schr\"{o}dinger  equations, there is no preferred topology on the space of Cauchy data $(\phi(t),-\i\p_{t}\phi(t))$.
It turns out, cf.\ \cite{V1,LNT1,LNT2} for example, that two spaces of Cauchy data are natural, the {\em energy space} $\cE= \langle h\rangle^{-\12}\cH\oplus \cH$ and the {\em charge space} $\cK_{1/4}= \langle h\rangle^{-1/4}\cH\oplus \langle h\rangle^{1/4}\cH$. In \cite{GGH1} resolvent estimates are proved on the energy space, and then extended to the charge space by duality and interpolation. 
This extension argument is a consequence of our Theorem \ref{th:bvr}.

In this paper we give another application of Thm. \ref{th:bvr} by directly proving resolvent estimates on the charge space. We also discuss in detail various realizations of the Klein-Gordon generator starting from the dual space $\cE^{*}= \ch\oplus \langle h\rangle^{\12} \cH$, and 
the functional calculus of 'free' Klein-Gordon generators, corresponding to $k=0$. 
  \subsection{Plan of the paper}
We now briefly describe the plan of this paper.  In Sect. \ref{bvfcalc} we describe some basic results on the smooth and Borel functional calculus for linear operators on Banach spaces. The Dunford-Taylor functional calculus for a linear operator $H$ can be extended to smooth functions on an interval $I\subset \rr$ if the resolvent $(H-z)^{-1}$ is of polynomial growth near the real axis. If this functional calculus is continuous for the sup norm, then it uniquely extends to bounded Borel functions on $I$.

In Sect. \ref{sec2} we recall basic results on $K-$spaces, which are natural generalizations of Krein spaces.  Sect. \ref{defidefi} is devoted to the construction of a Borel functional calculus for definitizable selfadjoint operators on Krein spaces. Although various versions of this construction can be found in the literature (see in particular \cite{La}, \cite{J}, or more recently \cite{Wora}), we believe our presentation might have some interest. In particular we precise the optimal class of admissible functions, namely bounded Borel functions on $\rr$ having a precise asymptotic expansion at each critical point of $H$.

In Sect. \ref{sec0} we collect some rather standard facts on the
smoothness of an operator with respect to a $C_{0}-$group. In the
usual Hilbert space framework, the $C_{0}-$groups of practical
interest for the Mourre method are unitary, with selfadjoint
generators. In this case a very comprehensive study can be found in
\cite{ABG}. In our applications to Krein spaces, no natural Hilbert
space structure is present and part of the formalism has to be
generalized.

These results are used in Sect.  \ref{sec1} to prove {\em commutator
  expansions}.  Roughly speaking if $H$ is an operator and $A$ the
generator of a $C_{0}-$group on a Banach space $\cH$, we need to
expand the commutator $[H, \i f(A)]$ for some class of functions $f$
as $f'(A)[H, \ i A]+ R$ with a careful estimate of the error term
$R$. Again in the Hilbert space case, such commutator expansions are
a basic tool of spectral and scattering theory, see among many other
references \cite{GJ}.

In Sect. \ref{s:bvr} we prove the main result of this paper,
Thm. \ref{th:bvr}, by adapting the Hilbert space proof in \cite{G1}
to the Krein space framework.  In the last section of this paper,
Sect. \ref{kgsect}, we discuss abstract Klein-Gordon operators
and give a concrete application of  Thm. \ref{th:bvr}.

\section{Boundary values of resolvents and functional
  calculus}\label{bvfcalc}

\init In this section we present some results on the smooth and
Borel functional calculus for linear operators on Banach spaces,
under some general assumptions on the growth of their resolvents
near the real axis.

\subsection{Notations}\label{ss:not}

If $\cH$ is a Banach space we denote $\cH^*$ its adjoint space,
i.e. the set of continuous anti-linear functionals on $\cH$ equipped
with the natural Banach space structure.  The canonical anti-duality
between $\cH$ and $\cH^{*}$ is denoted $\lag u, w \rag\equiv w(u)$,
where $u\in\cH$ and $w\in\cH^{*}$.  So
$\lag\cdot,\cdot\rag:\cH\times\cH^{*}\to\C$ is anti-linear in the
first argument and linear in the second one. On the other hand, we
denote by $\braket{\cdot}{\cdot}$ hermitian forms on $\cH$, again
anti-linear in the first argument and linear in the second one.

We say that $\cH$ is {\em Hilbertizable} if there is a scalar product on
$\cH$ such that the norm associated to it defines the topology of
$\cH$; such a scalar product and the norm associated to it will be
called {\em admissible}.  Scalar products are denoted by 
$(\cdot|\cdot)$.

If $\cH$ is a reflexive Banach space then the canonical
identification $\cH^{**}=\cH$ is obtained by setting
$u(w)=\overline{w(u)}$ for $u\in\cH$ and $w\in\cH^{*}$. In other
terms, the relation $\cH^{**}=\cH$ is determined by the rule $\lag
w,u \rag=\overline{\lag u,w \rag}$.

Let $\cG,\cH$ be reflexive Banach spaces and $\re=\cG\oplus\cH$.
The usual realization $(\cG\oplus\cH)^{*}=\cG^{*}\oplus\cH^{*}$ of
the adjoint space will not be convenient later, we shall
rather identify $\re^{*}=\cH^{*}\oplus\cG^{*}$ in the obvious
way. For example, if $\cH=\cG^{*}$, so $\cH^*=\cG$, the adjoint
space of $\re=\cG\oplus\cG^{*}$ is identified with itself
$\re^{*}=\re$.

If $S$ is a closed densely defined operator on a Banach space $\cH$, we denote by $\rho(S)$, $\sigma(S)$ its resolvent set and spectrum.

We use the notation $\jap{a}=(1+a^{2})^\12$ if $a$ is real number or an
operator for which this expression has a meaning.
\subsection{Polynomial growth condition}\label{ss:sfc}

Let $H$ be a closed densely defined operator on a Banach space
$\cH$. We first give a meaning to the boundary values
$R(\lambda\pm\i0)$ of the resolvent of $H$ as $B(\cH)$-valued
distributions on a certain real open  set defined by a growth
condition on $\|R(\lambda\pm\i\mu)\|$ as $\mu\downarrow0$.
We recall that if $\cb$ is a Banach space then a $\cb$-valued
distribution on a  real open set $I$ is a continuous linear map
$T:C^\infty_{0}(I)\to\cb$. We often use the formal notation
$T(\chi)=\int T(\lambda)\chi(\lambda)d\lambda$ for
$\chi\in C^\infty_{0}(I)$. The topology on this space of 
distributions is defined as in the scalar case.

\begin{lemma}\label{lm:bv}
  Assume that $I\subset \rr$ is open with $I\pm \i ]0, \nu]\subset \rho(H)$ for some $\nu>0$ and that there exists $n \in \nn$ and $C>0$ such that
\begin{equation}\label{eq:div}
\|R(z)\|\leq C |{\rm Im}z|^{1-n}, \ z\in I\pm \i ]0, \nu].
\end{equation}
Then the boundary values
$R(\lambda\pm\i0):=\lim_{\mu\downarrow0}R(\lambda\pm\i\mu)$ exist as
$B(\cH)$-valued distributions of order $n$ on $I$. More explicitly,
if $\chi\in C_0^n(I)$ and we set
\[
\chi_{(n)}(\lambda+\i\mu)=\sum_{k=0}^n\chi^{(k)}(\lambda)(\i\mu)^k/k!, \lambda, \mu\in\rr,
\]
 then
\begin{equation}\label{eq:bvf}
\begin{array}{rl}
&\int_{\R}R(\lambda+\i0) \chi(\lambda) d\lambda\\[2mm]
=&\int_{\R} \Big( R(\lambda+\i\nu)\chi_{(n)}(\lambda+\i\nu)
+\int_0^\nu R(\lambda+\i\mu)\frac{d(\i\mu)^n}{n!} 
\chi^{(n)}(\lambda) \Big) d\lambda.
\end{array}
\end{equation}
\end{lemma}
\proof We use a well-known elementary argument, valid for any
holomorphic function, cf \cite[ Thm. 3.1.11]{Ho} and the
comment after its proof: make a Taylor expansion up to order $n$ of
the function $\mu\mapsto R(\lambda+\i\mu)$ on the interval
$[\varepsilon,\nu]$ with $0<\varepsilon<\nu$ and note that
$\frac{d}{d\mu}R(\lambda+\i\mu)=\i\frac{d}{d\lambda}R(\lambda+\i\mu)$
by holomorphy.  The remainder is the derivative of order $n$ of a
bounded function hence we may let $\varepsilon\to0$ and get
\begin{equation}\label{eq:bv}
R(\lambda+\i0)
=\sum_{k=0}^{n-1}\frac{\nu^k}{k!}(-\i\partial_\lambda)^kR(\lambda+\i\nu)
+(-\i\partial_\lambda)^n \int_0^\nu R(\lambda+\i\mu)\frac{d\mu^n}{n!}
\end{equation}
as $B(\cH)$-valued distributions on $I$. This relation
is equivalent to \eqref{eq:bvf}.  \qed

In the next definition we define the maximal open real set on
which the distributions $R(\cdot\pm\i0)$ make sense.

\begin{definition}\label{df:bv}
  Let $\beta(H)$ be the set of $\lambda\in\rr$ such that there is a
  real open neighborhood $I$ of $\lambda$ and there are numbers
  $\nu>0, n\in\nn, C>0$ such that 
  \[
  \|R(z)\|\leq C |{\rm Im}z|^{1-n}, \ z\in I\pm \i]0, \nu].
\] 
The boundary values
  $R(\lambda\pm\i0)=\lim_{\mu\downarrow0}R(\lambda\pm\i\mu)$ of the
  resolvent of $H$ are well defined $B(\cH)$-valued distributions on
  $\beta(H)$.
\end{definition}

\begin{remark}\label{re:lap}{\rm
If $\mathscr{X}$ is a Banach space such that $B(\cH)$ is continuously embedded
in $\mathscr{X}$ then $R(\cdot\pm\i 0)$ may be viewed as $\mathscr{X}$-valued
distributions on $\beta(H)$. It may happen that on some open set 
$I\subset\beta(H)$ these $\mathscr{X}$-valued distributions are defined by
locally bounded $\mathscr{X}$-valued functions: this is the case if  the 
\emph{limiting absorption principle} holds on $I$ relatively to $\mathscr{X}$, i.e. if $\| R(z)\|_{\mathscr{X}}\leq  C$ for $z\in I\pm \i ]0, \nu]$ for some $\nu>0$.

The usual strategy (adopted here) is to construct Banach spaces $\ck$ 
with $\ck\subset\cH$ continuously and densely, which allows one to 
take $\mathscr{X}=B(\ck,\ck^{*})$, such that $R(\lambda\pm\i 0)$, when viewed 
as a $B(\ck,\ck^*)$-valued distributions, is well defined and  a continuous 
function of $\lambda$.
}\end{remark}

\subsection{Smooth functional calculus}\label{ss:fcs}

We now describe an elementary functional calculus which makes
sense under very general conditions. In the selfadjoint case these
techniques were introduced in \cite{HS}. A detailed presentation may
be found in \cite{Da1} and an extension to non selfadjoint
operators in \cite{Da2}.

Under the conditions of Lemma \ref{lm:bv}, for any
$\chi\in C_0^n(I)$ we define a bounded operator on $\cH$ by
\begin{equation}\label{eq:sfc}
\chi(H)=\frac{1}{2\pi\i}\int
\big(R(\lambda+\i0)-R(\lambda-\i0)\big) \chi(\lambda) d\lambda.
\end{equation}
The right hand side above can be made quite explicit by using
\eqref{eq:bv} and a similar relation for $R(\lambda-\i0)$.

Note that the map $\chi\mapsto\chi(H)$ is an {\em algebra morphism}. 
Indeed, linearity is obvious and in order to prove that it is multiplicative
it suffices to show that $R(z)\chi(H)=(r_{z}\chi)(H)$ for $\Im z\neq0$,
where $r_{z}(\lambda)=(\lambda-z)^{-1}$. For this it suffices to note
that $R(z)R(\lambda\pm\i 0)=(R(z)-R(\lambda\pm\i 0))r_{z}(\lambda)$.

The Helffer-Sj\"ostrand version of the formula for $\chi(H)$ may be
obtained with the help of an almost analytic extension of $\chi$ as
in \cite{HS} (or see \cite[p. 24]{Da1}). For example, choose
$\theta\in C_c^\infty(\R)$ with $\theta(\lambda)=1$ if $|\lambda|<\nu/2$
and $\theta(\lambda)=0$ if $|\lambda|>\nu$. If for $z=\lambda+\i\mu$
we define $\tilde\chi(z)= \theta(\mu/\jap{\lambda})\chi_{(n)}(z)$  
and we set $\bar\partial=(\partial_\lambda+\i\partial_\mu)/2$ then
$\bar\partial\tilde\chi(z)=O(|\mu|^n)$ and
\begin{equation}\label{eq:hs}
\chi(H)=-\frac{1}{2\pi\i} \int_\C R(z) \bar\partial\tilde\chi(z) d z
\wedge d \bar{z}.
\end{equation}

\subsection{Borel functional calculus}\label{ss:borel}

The functional calculus \eqref{eq:sfc} introduced under the
conditions of Lemma \ref{lm:bv} is a priori well defined only for
$\chi\in C^n_{0}(I)$ but often it extends to larger classes of
functions by continuity. 

We shall say that \emph{$H$ admits a
  $C^0$-functional calculus on $I$} if $I\subset\beta(H)$ and
$\|\chi(H)\|\leq C \sup_{\lambda\in I}|\chi(\lambda)|$ for some
finite number $C$ and all $\chi\in\coinf(I)$. Then clearly the
smooth functional calculus has a unique continuous extension to an
algebra morphism $C_0(I)\to B(\ch)$.  If $\cH$ is reflexive one can extend the functional calculus to Borel functions, as shown in Thm. \ref{th:Borel} below. 

Let $\cb(I)$ be the set of bounded Borel functions on $I$. 
A sequence of functions $\vphi_{n}$ on $I$ is \emph{boundedly convergent} 
if $\sup_{n,\lambda}|\vphi_{n}(\lambda)|<\infty$ and
$\lim_{n}\vphi_{n}(\lambda)=\vphi(\lambda)$ exists for all $\lambda\in I$. 
Note that $\vphi\in\cb(I)$ if $\vphi_{n}\in\cb(I)\ \forall \ n$.
The following result is a straightforward application of the Riesz theorem,
see \cite[Cor.\ 9.1.10]{Wora} for example.

\begin{theoreme}\label{th:Borel}
  Assume that $\ch$ is a reflexive Banach space  and let
  $F_0:C_0(I)\to B(\ch)$ be a norm continuous algebra morphism. Then
  $F_{0}$ extends uniquely to an algebra morphism $F:\cb(I)\to B(\ch)$ 
  such that: $\vphi_{n}\to\varphi \text{ boundedly } 
  \Rightarrow F(\varphi_n)\to F(\varphi)$ weakly.
\end{theoreme}

\begin{remark}\label{re:fck}{\rm
If $H$ is a selfadjoint operator on a Krein space (see Def. \ref{df:K}) 
and if $H$ admits a $C^0$-functional calculus on $I$ then it is clear that
$\chi(H)^*=\bar\chi(H)$ for all bounded Borel functions $\chi$ on $I$. 
}\end{remark}

\section{$K-$spaces}\label{s:krein}\init\label{sec2}

In this section, we discuss $K-${\em spaces}, a
  generalization of Krein spaces which is natural in the context of
  the phase spaces considered in Subsect.\ \ref{ss:phspace}.  We
  refer to \cite{Bo} for the general theory of Krein spaces;
  topological vector spaces equipped with hermitian forms are
  considered in \cite[Ch.\ 3]{Bo}.

\subsection{Definition of  $K-$spaces}\label{sec2.2}

\begin{definition}\label{df:K}
A \emph{$K-$space } is a Banach space $\ch$ equipped with a 
continuous hermitian  form $\braket{\cdot}{\cdot}$
such that for any continuous linear form $\vphi$ on $\ch$ there
is a unique $u\in\ch$ such that $\vphi=\braket{u}{\cdot}$. 
The form $\braket{\cdot}{\cdot}$ is called the \emph{Krein structure}. 
If $\ch$ is Hilbertizable then $\ch$ is called a \emph{Krein space}.
\end{definition}

Let $J:\ch\to\ch^*$ be the linear continuous map defined by
$Ju=\braket{\cdot}{u}$, so that $\braket{u}{v}=\lag u, Jv \rag$. 
Since $\braket{\cdot}{\cdot}$ 
is hermitian, we have  $\lag u, Jv \rag=\overline{\lag v,Ju \rag}$.
The \emph{topological non-degeneracy} condition imposed on 
$\braket{\cdot}{\cdot}$ above means that $J$ is bijective. 
Thus the Krein structure $\braket{\cdot}{\cdot}$ allows us to identify 
$\ch^{*}$ and $\ch$  with the help of $J$.

\begin{proposition}
A $K-$space is  reflexive.
\end{proposition}
\proof
Let $I: \cH\to \cH^{**}$ the canonical injection. Since $J: \cH\to \cH^{*}$ is an isomorphism, so are $J^{*}: \cH^{**}\to \cH^{*}$ and $(J^{*})^{-1}\circ J: \cH\to \cH^{**}$. We note then that $(J^{*})^{-1}\circ J= I$. \qed

\begin{remark}{\rm
One may also say that a $K-$space structure on a reflexive Banach space $\ch$ 
is a hermitian isomorphism $J:\ch\to\ch^{*}$. A Hilbert structure is 
a positive Krein structure, i.e. a positive isomorphism $J:\ch\to\ch^{*}$.  
}\end{remark}

\begin{remark}{\rm
Assume that $\braket{\cdot}{\cdot}$ is a hermitian form on
a complex vector space $\ch$ which is algebraically non-degenerate, i.e.
$u=0$ if $\braket{u}{v}=0$ for all $v\in\ch$. Then \emph{there is at most 
one normed space topology on $\ch$ such that the conditions of Definition
\ref{df:K} be satisfied}. Indeed, any such norm on $\ch$ is complete because
$\ch^{*}$ is always a Banach space. And if the adjoint spaces associated to
two complete norms on $\ch$  are equal then the corresponding classes of
bounded sets are identical by the uniform boundedness principle, hence the
norms are equivalent. See \cite[p. 60-67]{Bo} for better results of this nature.
}\end{remark}

\subsection{Adjoints on $K$-spaces}\label{sec2.3}

If $T\in B(\ch)$ then the adjoint $T^*\in B(\ch^{*})$ of $T$ in the 
Banach space sense is defined on $\ch^*$ as usual and then we may 
transport it on $\ch$ with the help of $J$. In other terms, the Krein
structure $\braket{\cdot}{\cdot}$ allows us to define an involution 
$T\mapsto T^*$ on $B(\ch)$ such that $\braket{T^* u}{v}=\braket{u}{Tv}$.
This definition extends as usual to closed densely defined  operators. 

Clearly $B(\ch)$ becomes a $*$-algebra with a continuous involution. The 
selfadjoint operators are defined as usual by the relation $S^{*}=S$, 
where $S$ may be unbounded. We say that $S$ is {\em positive} and we write 
$S\geq 0$ if $\braket{u}{Su}\geq0$ for all $u\in \Dom S$. If $S$ is bounded 
and $S\geq 0$   then $T^{*}ST\geq0$ for all $T\in B(\ch)$, but the identity
operator is not positive unless $\ch$ is a Hilbert space. So $T^{*}T\geq0$
holds only in exceptional cases. 
To each positive bounded operator $S$ we associate a semi-norm on $\ch$,
namely $\|u\|_{S}=\sqrt{\braket{u}{Su}}$, which satisfies
$|\braket{u}{Sv}|\leq \|u\|_{S}\|v\|_{S}$.   

We say that a linear subspace $\ck$ is a \emph{Hilbert subspace of $\ch$} 
if $\big(\ck,\braket{\cdot}{\cdot}|_{\ck\times\ck}\big)$ is a Hilbert space.  
Equivalently, this means that $\ck$ is a closed subspace of $\ch$ such that
$\braket{u}{u}\geq c\|u\|^{2}$ for some number $c>0$ and all $u\in\ck$. We
equip such a subspace with the natural Hilbert norm 
$\|u\|_{\ck}=\sqrt{\braket{u}{u}}$ which is equivalent to 
$\|\cdot\||_{\ck}$.

\subsection{Projections on $K$-spaces}\label{sec2.4}

A \emph{projection} on $\ch$ is an element $\Pi\in B(\ch)$ such that
$\Pi^{2}=\Pi$. A selfadjoint projection is also called an {\em orthogonal
projection}. A \emph{positive projection} is a projection $\Pi$ such
that $\Pi\geq0$. In particular, $\Pi$ will be orthogonal. For the proof of
the following fact, see \cite{Bo}. 

\begin{proposition}\label{pr:positive}
The range of a positive projection is a Hilbert subspace of $\ch$.
Reciprocally, if $\ck$ is a Hilbert subspace of $\ch$ then there is a
unique selfadjoint projection $\Pi$ such that $\Pi\ch=\ck$ and this
projection is positive.
\end{proposition}

If $\Pi$ is a positive projection then $\|u\|_{\Pi}=\|u\|_{\Pi\ch}$
for all $u\in\Pi\ch$. If $S\in B(\ch)$ we denote $\|S\|_{\Pi}$ the
norm of the operator $\Pi S \Pi$ on the Hilbert space $\Pi\ch$. If
$S=S^{*}$ then $\|S\|_{\Pi}=\sup\{|\braket{u}{Su}| \mid u\in\Pi\ch,
\braket{u}{u}=1\}$. It follows that if $S\in B(\ch)$ and $ S= S^{*}$
then
\begin{equation}
\label{e2.1}
\pm \braket{\Pi u}{S\Pi u}\leq \|S\|_{\Pi}\braket{\Pi u}{\Pi u}, \ u\in \cH. 
\end{equation}

\subsection{Phase spaces}\label{ss:phspace}

A typical construction of $K-$spaces starts with a reflexive Banach
space $\cG$ thought as configuration space of a system. Then the 
\emph{phase space of $\cG$}  is $\ch=\cG\oplus\cG^{*}$ and
its $K-$space structure is

\begin{equation}\label{eq:phasesp}
\braket{u}{v}= v_1(u_0)+\overline{u_1(v_0)}=\langle u_{0}, v_{1}\rangle + \langle u_{1}, v_{0}\rangle,\ 
u=(\begin{smallmatrix}u_0\\u_1\end{smallmatrix}), \ v=(\begin{smallmatrix}v_0\\v_1\end{smallmatrix})\in \cH.
\end{equation}
 Recall that according to the convention adopted in Subsect.
 \ref{ss:not} we identify $\ch^{*}=\cG\oplus\cG^*=\ch$. Thus $J$
 is the identity operator and \eqref{eq:phasesp} satisfies the
 required topological non-degeneracy condition.

 Note that we think of elements of $\ch$ as column matrices hence we
 may represent bounded operators on $\ch$ as matrices
\[
S=\mat{a}{b}{c}{d}
\]  
where $a:\cG\to\cG$, $b:\cG^*\to\cG$, $c:\cG\to\cG^*$,
$d:\cG^*\to\cG^*$. A computation gives
\begin{equation}\label{eq:adj}
S^*=\mat{d^*}{b^*}{c^*}{a^*}
\end{equation}
hence
\begin{equation}\label{eq:ksym}
S=S^{*} \Longleftrightarrow S=\mat{a}{b}{c}{a^{*}} \quad\text{with}\quad
a\in B(\cG), \  b=b^{*}:\cG^*\to\cG,\ 
c=c^{*}:\cG\to\cG^*.
\end{equation}

\begin{lemma}\label{lm:kpositive}
An operator $S:\ch\to\ch$ is positive if and only if it is as in
\eqref{eq:ksym} with $b\geq0$, $c\geq0$, and 
\begin{equation}\label{eq:kpos}
|\braket{au_{0}}{u_{1}}|^{2}\leq \braket{u_1}{bu_1}\braket{u_0}{cu_0}
\quad \text{for all } u_{0}\in\cG, u_{1}\in\cG^{*}.
\end{equation}
If $\cG$ is a Hilbert space and $\cG^{*}=\cG$ then this means 
$a,b,c\in\ B(\cG)$ with $b,c\geq0$ and 
$$
\|c^{-1/2}ab^{-1/2}\|\equiv
{\textstyle\sup_{\veps>0}}\|(c+\veps)^{-1/2}a(b+\veps)^{-1/2}\|\leq1.
$$
\end{lemma}
\proof
The symmetric operator $S$ as given in \eqref{eq:ksym} is positive if and
only $\braket{u}{Su}\geq 0$ for all $u\in\ch$ with
\[
\braket{u}{Su}= 2\Re\braket{au_{0}}{u_{1}}
+\braket{u_1}{bu_1} +\braket{u_0}{cu_0}.
\]
Taking successively $u_0=0$ and $u_1=0$ we see that $b\geq0$ and
$c\geq0$ are necessary conditions. Then by changing $u_{1}$ in
$-\omega u_{1}$ with
$\omega=\overline{\braket{au_{0}}{u_{1}}}|\braket{au_{0}}{u_{1}}|^{-1}$
if the denominator is not zero and $\omega=1$ otherwise, we see that
positivity of $S$ is equivalent to $2|\braket{au_{0}}{u_{1}}|\leq
\braket{u_1}{bu_1} +\braket{u_0}{cu_0}$ for all $u_{0}\in\cG$ and
$u_{1}\in\cG^{*}$. Replace $u_{0},u_{1}$ by $\veps^{1/2} u_{0}$ and
$\veps^{-1/2} u_{1}$ with $\veps>0$. If one of the terms in the
right hand side is zero then we get $\braket{au_{0}}{u_{1}}=0$ by
making $\veps\to0$ or $\veps\to\infty$. If not then $\veps=
\braket{u_0}{cu_0}^{1/2}\braket{u_1}{bu_1}^{-1/2}$ gives
\eqref{eq:kpos}.  \qed

\begin{remark}\label{re:gg}{\rm If $\cG$ is a Hilbert space
    identified with its adjoint space $\cG^*$ with the help of the
    Riesz isomorphism then on the phase space $\ch=\cG\oplus\cG$ we
    have the direct sum Hilbert structure
    $(u|v)_H=(u_{0}| v_{0})+ (u_{1}| v_{1})$ and the
    Krein structure $\braket{u}{v}_K$ defined by
    \eqref{eq:phasesp}. Clearly $\braket{u}{v}_K=\braket{u}{Jv}_H$
    with $J=(\begin{smallmatrix}0&1\\1&0\end{smallmatrix})$. Observe
    that now we have two natural ways of identifying $\ch$ with its
    adjoint space, namely by using $(\cdot | \cdot)_H$
    (i.e. the Riesz isomorphism) or $\braket{\cdot}{\cdot}_K$. In
    our framework it is more convenient to use the second one which
    could be called \emph{Krein isomorphism}.  This is coherent with
    the convention $(X\oplus Y)^*=Y^*\oplus X^*$ adopted in
    Subsect. \ref{ss:not}. }\end{remark}

\section{ Definitizable operators on Krein spaces}\label{defidefi}
\init

The definitizable operators on a Krein space,
  introduced by H.\ Langer in 1965, are remarkable because they
admit a functional calculus almost as rich as that of selfadjoint
operators on a Hilbert space. In fact the functions $\vphi$ for
which $\vphi(H)$ may be given a natural meaning can be arbitrary
bounded Borel functions outside a finite set of ``critical points'',
cf. \cite{La}. In this section we shall
consider only continuous functions because, thanks to
Thm. \ref{th:Borel}, this is sufficient to our needs. The main point
in the approach we present below is the estimate in
Prop. \ref{pr:jestime} due to P.\ Jonas \cite[Thm. 1]{J}.  Another
presentation of the Langer-Jonas functional calculus may be found in
\cite[Ch.\ 9]{Wora}.

\subsection{Definitizable operators }

In this section we fix a Krein space $\ch\equiv(\ch, \braket{\cdot}{\cdot})$.

\begin{definition}\label{de:def}
A selfadjoint operator $H$ on $\ch$ is \emph{definitizable} if
$\rho(H)\neq\emptyset$ and there is a real polynomial $p\neq0$ such that 
$p(H)\geq 0$, i.e. $\braket{u}{p(H)u}\geq0$ 
for all $u\in \Dom H^n$ where $n$ is the degree of $p$. 
Such a $p$ is called a
\emph{definitizing polynomial} for $H$. 
\end{definition}

\begin{remark}\label{re:pos}{\rm
The assumption $\rho(H)\neq\emptyset$ is important, some natural
selfadjoint operators on a Krein space have empty resolvent 
set, see \cite[p.\ 148]{Bo}. For example, let $\ch$ be the phase space of a Hilbert space $\cG$ (cf.\ Remark \ref{re:gg}) and let $b$ be a positive
injective operator on $\cG$. If $b$ or $c:=b^{-1}$ is unbounded, then
$(\begin{smallmatrix}0&{b}\\c&0\end{smallmatrix})$
is strictly positive, i.e. $\braket{u}{Hu}>0$ for all $u\neq0$ in the domain
of $H$, and $\rho(H)=\emptyset$.  
}\end{remark}

The next result gives informations on the non-real spectrum of a
definitizable operator. The proof is easy, see \cite[Lemma 1]{J}.

\begin{proposition}\label{frogu}
Let $H$ be definitizable. Then: 
 \ben
\item If $z\in \sigma(H)\backslash \rr$  then 
$p(z)=0$ for each definitizing polynomial $p$.

\item There is a definitizing polynomial $p$ such that
  $\sigma(H)\setminus\mbr$ is exactly the set of non-real zeroes of
  $p$.

\item Moreover, this $p$ may be chosen such that if
  $\lambda\not\in\mbr$ is a zero of multiplicity $k$ of $p$ then
  $\lambda$ is an eigenvalue of $H$ of Riesz index $k$.

\item The non-real spectrum of $H$ consists of a finite number of 
eigenvalues of finite Riesz index distributed symmetrically with respect to 
the real axis.
\een
\end{proposition}

The following consequence is easily proved with the help of the Riesz
projection associated to the finite set $\sigma(H)\setminus\R$. 
A \emph{Krein subspace of $\ch$} is a closed subspace which is a Krein 
space when equipped with the hermitian form induced by 
$\braket{\cdot}{\cdot}$.

\begin{corollary}\label{co:def}
There are Krein subspaces $\ch_{1},\ch_{2}$ of $\ch$ such that 
$\ch=\ch_{1}\oplus\ch_{2}$, where the sum is direct and orthogonal 
with respect to the Krein structure of $\ch$, such that $H=H_1\oplus H_2$ with 
$H_1$ a bounded selfadjoint operator in $\cH_{1}$ with finite spectrum and
$H_{2}$ a definitizable operator in $\ch_2$ with  $\sigma(H_2)\subset\mbr$.
\end{corollary}

The above decomposition is canonical in a sense easy to make precise. 
Thus for any ``reasonable'' function $\vphi$ we should have 
$\vphi(H)=\vphi(H_{1})\oplus\vphi(H_{2})$. Since the definition of 
$\vphi(H_{2})$ is rather obvious, when we discuss the functional calculus 
of a definitizable operator it suffices to consider the case when it has only real spectrum.

\subsection{Rational functional calculus}

Before going on we make a general remark concerning the rational functional
calculus associated to an arbitrary closed operator $H$ with non-empty
resolvent set on a Banach space $\ch$. This makes things completely elementary
and avoids the use of the (analytic) Dunford calculus. In the sequel we denote by $\hat{\cc}$, $\hat{\rr}$ the one-point compactifications of $\cc$, $\rr$.

Denote $\cR_H$ the set of rational functions whose poles belong to  $\rho(H)$ and which are bounded near infinity.  This space is an
unital algebra. If $\rho(H)= \overline{\rho(H)}$, as in the case of a selfadjoint operator on a Krein space, then $\cR_H$ becomes a $*$-algebra 
if we define the adjoint $\bar\varphi$ of $\varphi$ by 
$\bar\varphi(\lambda)=\overline{\varphi(\bar \lambda)}$.

\begin{lemma}\label{lm:rfc}
There is a unique unital algebra morphism 
$\cR_{H}\ni\varphi\mapsto\varphi(H)\in B(\ch)$ with 
$\varphi(H)=(H-z)^{-1}$ if $\varphi(\lambda)=(\lambda-z)^{-1}$ 
for some $z\in \rho(H)$. If $\ch$ is a Krein space and $H$ is selfadjoint
then $\varphi\mapsto\varphi(H)$ is a $*$-morphism.
\end{lemma}

\proof  Let $\Omega\subset\hat{\cc}\times \nn$ be the set of couples $\omega=(z,s)$ with $z\in \rho(H)$ and $s\in\N^{*}$ or $\omega=(\infty,0)\equiv\infty$. 
For $\omega\in  \Omega$ we set:
\[
\rho_{\omega}(\lambda):= (\lambda-z)^{-s} \hbox{ if }\omega\in \rho(H)\times \nn^{*} , \ \rho_{\omega}(\lambda):= 1 \hbox{ if }\omega= (\infty, 0).
\]
Then $\{\rho_{\omega}\}_{\omega\in\Omega}$ is a vector space basis in
$\cR_{H}$. Hence there is a unique linear map
$\varphi\mapsto\varphi(H)$ from $\cR_H$ into $B(\ch)$ which sends
$\rho_{\omega}$ into $\varphi(H)=(H-z)^{-s}$ if $\omega\neq\infty$
and $1$ into the identity operator.  From the first resolvent
identity it follows that this map is an algebra morphism. In the Krein
space case note that $\varphi(H)^*=\bar\varphi(H)$ for any 
$\vphi$ follows from the fact that the adjoint of $(H-z)^{-1}$ 
is $(H-\bar{z})^{-1}$.
\qed

\subsection{${C^{\alpha}}$ functional calculus}

The set $\cR$ of \emph{bounded rational functions 
$\varphi:\mbr\to\mbc$} is a unital $*$-algebra for the 
usual algebraic operations. By Lemma \ref{lm:rfc} if $H$ is 
a definitizable operator  with only real spectrum then there is a unique 
unital $*$-morphism 
$\varphi\mapsto\varphi(H)$ of $\cR$ into $\in B(\ch)$ such that 
$\varphi(H)=(H-z)^{-1}$ 
if $\varphi(\lambda)=(\lambda-z)^{-1}$ with $z\in\C\setminus\R$. We now
extend this calculus to a class of continuous functions 
$\varphi:\R\to\C$ which have a certain degree of regularity at a finite set
of real points and/or at infinity.

\begin{definition}
Let $\omega=(\xi,s)\in\hat\R\times\N$ and   $\vphi:\R\to\C$.
\ben
\item If $\xi\in\R$, then $\vphi$ \emph{is of class $C^{s}$ at $\xi$} 
if there is a polynomial $P$ such that $\vphi(x)=P(x-\xi)+o(|x-\xi|^{s})$.
\item $\vphi$ \emph{is of class $C^{s}$ at infinity} if there
  is a polynomial $P$ such that $\vphi(x)=P(1/x)+o(|x|^{-s})$.
   \een
Denote $C^{\omega}(\R)=\{\vphi\in C(\hat\R) \mid \varphi 
\text{ is of class } C^{s} \text{ at } \xi\}$, for $\omega= (\xi, s)$
\end{definition}

Under the conditions of the definition, the terms of degree $\leq s$ of $P$ 
are uniquely determined hence 
if $\xi\in\R$ there is a unique polynomial $T^{+}_{\omega}\vphi$ of
degree $\leq s$ such that
$\vphi(x)=T^{+}_{\omega}\vphi(x)+o(|x-\xi|^{s})$ and if $\xi=\infty$
there is a unique rational function of the form
$T^{+}_{\omega}\vphi(x)=\sum_{k\leq s} a_{k}x^{-k}$ such that
$\vphi(x)=T^{+}_{\omega}\vphi(x)+o(|x|^{-s})$. Some new notations
will allow us to write this in a more convenient form. 

Equip $\hat\R\times\N$ with the following order relation: $\mu\leq\nu$ means
$\mu=(\xi,s)$ and $\nu=(\eta,t)$ with $\xi=\eta$ and $s\leq t$.  If
$\omega=(\xi,s)\in\hat\R\times\N$ let $\chi_{\omega}$ be the
rational function defined by $\chi_{\omega}(x)=(x-\xi)^{s}$ if
$\xi\in\R$ and $\chi_{\omega}(x)=x^{-s}$ if $\xi=\infty$. Set
$\rho_{\omega}=\chi_{\omega}^{-1}$

Now it is clear that there is a unique sequence of complex numbers
$\{\delta_\mu(\varphi)\}_{\mu\leq\omega}$ such that
$T^{+}_{\omega}\vphi=\sum_{\mu\leq\omega}\delta_{\mu}(\vphi)\chi_{\mu}$. Set
$T_{\omega}\vphi=\sum_{\mu<\omega}\delta_{\mu}(\vphi)\chi_{\mu}$  and
\begin{equation}\label{eq:R}
R_{\omega}\vphi=\rho_{\omega}(\varphi-T_{\omega}\vphi)
\quad\text{hence}\quad 
\varphi = T_{\omega}\vphi + \chi_{\omega}R_{\omega}\vphi.
\end{equation}
Since $C^\omega\subset C^\mu$ if $\mu\leq\omega$ the quantity 
$\|\varphi\|_\omega= \sum_{\mu\leq\omega}\sup|R_\mu\vphi|$  
is a well defined real number if $\varphi\in C^\omega$. 

An element $\omega\in\hat\R\times\N$ may be thought of as a
function $\hat\R\to\N$ with support containing at most one point.
More generally, consider functions with finite support $\alpha:
\hat{\rr}\to \nn$, which we also call \emph{order functions}. We write
$\omega\preceq\alpha$ if $\omega=(\xi,s)\in\hat\R\times\N$ and
$s\leq\alpha(\xi)$. Then $\omega\prec\alpha$ means
$\omega\preceq\alpha$ and $s< \sum_\eta \alpha(\eta)$.

\begin{lemma}\label{lm:dens}
  If $\alpha$ is an order function then $C^{\alpha}(\rr)=
  \cap_{\omega\preceq\alpha}C^{\omega}(\rr)$ is an involutive Banach algebra
  with unit for the usual algebraic operations and the norm
  $\|\varphi\|_\alpha=\sup_{\omega\preceq\alpha}\|\varphi\|_\omega$.
  The space $\cR$ is a dense $*-$sub-algebra of $C^{\omega}$.
\end{lemma}

The proof is elementary and will not be given. Next we show that the
functional calculus for definitizable operators extends to an
algebra of the form $C^{\alpha}(\rr)$. We start by associating an
order function $\alpha$ to each definitizable operator.
 
\begin{definition}\label{isap}
  Let $H$ be a definitizable operator on $\cH$ with
  $\sigma(H)\subset \rr$.  \ben
\item To each definitizing polynomial $p$ for $H$ we associate an
  order function $\beta$ as follows: if $\xi\in\R$ then $\beta(\xi)$
  is the multiplicity of $\xi$ as zero of $p$ and $\beta(\infty)=0$
  if $p$ is of even degree and $\beta(\infty)=1$ if $p$ is of odd
  degree.
\item The order function $\alpha_{H}$ of $H$ is the infimum over all
  definitizing polynomials for $H$ of the above functions $\beta$.
  
  \een
\end{definition}

\begin{theoreme}\label{th:main}
  Let $H$ be a selfadjoint definitizable operator on the Krein
  space $\ch$ with $\sigma(H)\subset\mbr$.Then there is a unique
  linear continuous map $\vphi\mapsto\vphi(H)$ from
  $C^{\alpha_{H}}(\R)$ into $B(\ch)$ such that if
  $\vphi(\lambda)=(\lambda-z)^{-1}$ for $z\in \cc\backslash \rr$ then
  $\vphi(H)=(H-z)^{-1}$. This map is a morphism of unital
  $*$-algebras.
\end{theoreme}

The theorem follows from the next proposition and Lemma
\ref{lm:dens}.

\begin{proposition}\label{pr:jestime}
  There is a constant $C$ such that 
  $\|\varphi(H)\|\leq C\|\varphi\|_\alpha \ \forall \ \varphi\in\cR$.
\end{proposition}

The rest of this section is devoted to the proof of this
proposition.  We begin with three simple observations concerning the
$*$-algebra $\cR$.

\begin{lemma}\label{lm:pos}
If  $\varphi\in\cR$ then $\varphi\geq0$ as function on $\R$ if and
  only if there is $\psi\in\cR$ such that
  $\varphi=\overline\psi\psi$. 
\end{lemma}
\proof We have $\varphi=P/Q$ where $P,Q$ are polynomials, $Q$ has no
real zeroes, and the degree of $P$ is less or equal to that of
$Q$. Since $\varphi=P\overline Q/Q\overline Q$, we may assume
$Q\geq0$. Then the degree of $Q$ is $2n$ and one may write
$Q=\overline Q_0 Q_0$ where $Q_0$ is a polynomial of degree $n$
whose zeroes are exactly the zeroes of $Q$ in the upper half-plane.
If $\varphi\geq0$ then $P$ is a positive polynomial hence its degree 
is $2m$ with $m\leq n$ and one may similarly factorize 
$P=\overline P_0 P_0$ (the real zeroes of $P$ being of even multiplicity). 
Then we take $\psi=P_0/Q_0$. \qed

As a consequence, if $\theta$ is a positive linear form on $\cR$
then $|\theta(\varphi)|\leq\theta(1)\sup|\varphi|$. The following
version of this assertion is more convenient for our purposes.

\begin{lemma}\label{lm:positive}
  Let $\ch$ be a complex vector space equipped with a positive
  sesquilinear form $(\cdot,\cdot)$ and the associated semi-norm
  $|u|=(u,u)^{1/2}$. Let $M:\cR\to L(\ch)$ be a unital algebra
  morphism such that
  $(u,M(\varphi)v)=(M(\overline\varphi)u,v)$. Then $|M(\varphi)u|\leq
  \sup|\varphi| |u|$ for all $\varphi\in\cR$ and $u\in \ch$. 
\end{lemma}
\proof It suffices to show that 
$|M(\varphi)u|^2=(u,M(|\varphi|^2)u)\leq (u,u)$ if
$\sup|\varphi|=1$. We have $1-|\varphi|^2\geq0$ as function on $\R$
and $1-|\varphi|^2\in\cR$ hence by Lemma \ref{lm:pos} there is
$\psi\in\cR$ such that $1-|\varphi|^2=\overline\psi \psi$. Since
$M(1)=1$ we obtain $(u,(1-M(|\varphi|^2)u)=(u,M(\overline\psi
\psi)u)=|M(\psi)u|^2\geq0$ which proves the assertion. \qed

The third observation is an analogue of the division algorithm in
the algebra $\cR$.  To each $\psi\in\cR$ we associate an order
function $\alpha_\psi$ by defining
$\alpha_\psi(\xi)$ as the multiplicity of $\xi$ as zero of
$\psi$. In other terms, $\alpha_\psi(\xi)=k$ means that the limit
$\lim_{\lambda\to\xi}\psi(\lambda)\rho_\omega(\lambda)$ exists in
$\C$ and is not zero for $\omega= (\xi, k)$.  The proof of the next lemma is quite
elementary and we skip the details.

\begin{lemma}\label{lm:division}
  Let $\psi\in\cR$ with only real zeros and set
  $\alpha=\alpha_\psi$.  Then there are numbers $a_\omega\in\C$ and
  functions $b_\omega\in\cR$ such that for each $\varphi\in\cR$ we
  have:
\begin{equation}\label{eq:division}
\varphi=\psi\sum_{\omega\preceq\alpha}
a_\omega R_\omega\vphi
+\sum_{\omega\prec\alpha}\delta_\omega(\varphi)b_\omega.
\end{equation}
\end{lemma}

{\it Proof of Prop. \ref{pr:jestime}. }  By a simple argument
its suffices to show that $\|\varphi(H)\|\leq C\|\varphi\|_\alpha$ where
$\alpha$ is the order function of a definitizing polynomial $p$ with
only real zeros.  Let $n$ be the degree of $p$, define
$m=\left[\frac{n+1}{2}\right]$, let $\lambda\in\C\setminus\R$, and
let $\psi(x)=p(x)(x-\lambda)^{-m}(x-\bar\lambda)^{-m}$. Then
$\psi\in\cR$ and $(u|u):=\braket{u}{\psi(H)u}\geq0$ for all
$u\in\ch$. Set $|u|=(u|u)^{1/2}$. Since the rational functional
calculus is an algebra morphism we get from \eqref{eq:division}:
\begin{equation}\label{eq:repres}
\varphi(H)=
\psi(H)\sum_{\omega\preceq\alpha} a_{\omega} (R_{\omega}\vphi)(H)
+ \sum_{\omega\prec\alpha} \delta_{\omega}(\varphi)b_{\omega}(H), \ \varphi\in\cR.
\end{equation}
Since the $b_{\omega}(H)$ are bounded operators, there is a constant 
$C$ such that for any $u,v\in\ch$:
\begin{equation}\label{eq:tmp}
|\braket{u}{\varphi(H)v}| \leq
\Big|\Big(u \big|\sum_{\omega\preceq\alpha}
a_{\omega}(R_{\omega}\vphi)(H)v\Big)\Big| 
+ C\sum_{\omega\prec\alpha} |\delta_{\omega}(\varphi)|\, \|u\|\|v\|, \ \varphi\in\cR,
\end{equation}
where we used the positive scalar product
$(f|g):=\braket{f}{\psi(H)g}$ introduced above. It is easy to prove
that $\sum_{\omega\prec\alpha}
|\delta_{\omega}(\varphi)|\leq\|\varphi\|_\alpha$.  On the other hand,
by Cauchy-Schwarz inequality and Lemma \ref{lm:positive} we get:
\[
\Big|\Big(u \big|\sum_{\omega\preceq\alpha}
a_{\omega}(R_{\omega}\vphi)(H)v\Big)\Big| \leq
\sup\Big|\sum_{\omega\preceq\alpha} a_{\omega} R_{\omega}\vphi \Big|
|u| |v| \leq C \|\varphi\|_\alpha \|u\|\|v\|.  
\]
Thus \( |\braket{u}{\varphi(H)v}| \leq C
\|\varphi\|_\alpha \|u\|\|v\|, \) which finishes the proof of
Prop. \ref{pr:jestime}.  \qed

From Thm. \ref{th:main} we can deduce  an optimal estimate of the resolvent of
a definitizable operator. We first recall some terminology, cf. \cite{La, J}.
\begin{definition}
Denote $\sigma_{\cc}(H):= \sigma(H)\backslash \rr$ and
$c(H):= \{\omega\in \hat{\rr}\ : \ \alpha_{H}(\xi)\neq 0\}$. Then the set 
$c(H)$ is called {\em the set of critical points} of $H$.
\end{definition}

Let $H$ be a definitizable operator.  Recall that   $\alpha_{H}$ is defined in Def. \ref{isap}. 

\begin{proposition}\label{pr:reso}
  With the preceding notations, there exists $c>0$ such that
\begin{equation}\label{eq:reso}
c\|(H-z)^{-1}\|\leq  \sum_{\xi\in\sigma_{\cc}(H)} |z-\xi|^{-\alpha_H(\xi)}
+|\mathrm{Im} z|^{-1}
\Big(1+\sum_{\xi\in c(H)} |z-\xi|^{-\alpha_H(\xi)} +
|z|^{\alpha_H(\infty)} \Big)
\end{equation}
for all $z\nin\sigma_c\cup\R$.  Note that $\alpha_H(\infty)$ is
either $0$ or $1$.
\end{proposition}

\proof It is clearly sufficient to assume that the spectrum of $H$
is real. If $z\nin\R$ and $\varphi(x)=(z-x)^{-1}$ then
$\varphi\in\cR$ and thus $\|(z-H)^{-1}\|\leq
C\|\varphi\|_{\alpha_H}$ by Thm. \ref{th:main}. To simplify
notations we set $\alpha_H(\xi)=k_\xi$ and $T_{(\xi,k)}=T_\xi^k$. 
Since $\varphi(\infty)=0$ we have
$$
\|\varphi\|_{\alpha_H} \leq \sup |\varphi| +
\sup_{\xi\in\sigma_r}\sum_{k\leq k_\xi} 
\sup_{x\in\R}\big|\varphi(x)-T_\xi^k\varphi(x)\big|\,|x-\xi|^{-k}
+\alpha_{H}(\infty)\sup_{x\in\R} |x\varphi(x)|. 
$$ 
We have $\sup|\varphi|=|\Im z|^{-1}$ and
$\sup|x\varphi(x)|=|z||\Im z|^{-1}$ hence it remains to estimate
$(\varphi(x)-T_\xi^k\varphi(x))(x-\xi)^{-k}$.  We shall prove the
following extension of the first order resolvent identity:
\begin{equation}\label{eq:mice}
\varphi(x)-T_\xi^k\varphi(x)=(x-\xi)^{k}\vphi^{k}(\xi)\varphi(x)
\quad \text{ if } x,\xi\neq z.
\end{equation}
This implies the next estimate, which proves the proposition:
$$
\sup_x \big|\varphi(x)-T_\xi^k\varphi(x) \big||x-\xi|^{-k}=
\sup_x|z-\xi|^{-k}|z-x|^{-1}= |z-\xi|^{-k}|\Im z|^{-1} .
$$
Observe that \eqref{eq:mice} is trivial if $k=0$ because
$T_\xi^0\varphi=0$ and is just the first order resolvent identity if
$k=1$. Now assume \eqref{eq:mice} holds for $k$. Since
$\varphi^{(k)}=k!\varphi^{k+1}$ we have  
$$
T_\xi^{k+1}\varphi(x)=T_\xi^{k}\varphi(x)+
\frac{1}{k!}\varphi^{(k)}(\xi)(x-\xi)^k
=T_\xi^{k}\varphi(x)+\varphi^{k+1}(\xi)(x-\xi)^k,
$$ 
which when used in \eqref{eq:mice} gives the same identity with $k$
replaced by $k+1$. 
\qed

\begin{remark}\label{re:takecare}{\rm The interpretation of the
    points $\xi\in \rr$ with $\alpha_H(\xi)>0$ as ``critical points'' of
    $H$ is misleading from the point of view of the functional
    calculus. For example, the operator of $q$ of multiplication by
    $x$ in the Krein space $L^2(\R,\sign{x}\, dx)$ is positive and
    $\alpha_q$ has value $1$ at $0$ and $\infty$ but the functional
    calculus extends continuously from the algebra $C^{\alpha_q}$ to
    $C(\hat\R)$ defined by the order function $\alpha=0$.
  }\end{remark}

\section{ $C_0$-groups and regular operators }\label{sec0}
\init 

In this section we collect some standard facts on smoothness
of operators with respect to  $C_{0}-$groups (see also \cite{ABG,GGM}).   

\subsection{$C^{\alpha}(A)$ classes of bounded operators}\label{sec1.2}

Let $W=\{W_t\}$ be a $C_0$-group on a Banach space $\ch$ with
generator $A$ defined such that $W_t=\rme^{\i tA}$. Then there are
numbers $M\geq1$ and $\gamma\geq0$ such that
\begin{equation}\label{e1.2toto}
\|W_{t}\|\leq M\rme^{\gamma|t|} \quad \text{for all\ \ } t\in \rr.
\end{equation}
The spectrum of the operator $A$ is included in the strip $\{z\in
\cc \mid \ |\Im z|\leq \gamma\}$ and it could be equal to this strip.

One may naturally associate to $A$ three operators acting on the
Banach space $B(\ch)$, namely {\em left multiplication} by $A$, denoted
$\ca_{\ell}$, {\em right multiplication} by $A$, denoted $\ca_{r}$, and
 {\em commutation} by $A$, denoted $\ca$ and defined by
$\ca(T)=[T,A]$, so that $\ca=\ca_{r}-\ca_{\ell}$. Since $A$ is
unbounded, it is convenient to define these operators as generators
of one parameter groups of bounded operators on $B(\ch)$. More
precisely, if $t\in \rr$ and $T\in B(\ch)$ we have:
\begin{equation}\label{e1.03}
\rme^{\i t\ca_{\ell}}(T)=\rme^{\i tA}T, \quad
\rme^{\i t\ca_{r}}(T)=T\rme^{\i tA}, \quad
\rme^{\i t\ca}(T)=\rme^{-\i tA}T\rme^{\i tA}\equiv T(t).
\end{equation}
These operators commute in the sense that the elements of the groups
they generate are pairwise commuting, and $\ca=\ca_{r}-\ca_{\ell}$
i.e.  $\rme^{\i t\ca}=\rme^{-\i t\ca_{\ell}}\rme^{\i t\ca_{r}}$.

These are $C_0$-groups if we equip $B(\ch)$ with the strong operator
topology.  If we assume (\ref{e1.2toto}) then
\begin{equation}\label{e1.3}
  \|\e^{\i t \ca_{\ell}}\|\leq M \e^{\gamma|t|}, \ \|\e^{\i t
    \ca_{r}}\|\leq M \e^{\gamma|t|}, \ 
  \|\e^{\i t \ca}\|\leq M^{2} \e^{2\gamma|t|} \quad\text{for all\ \ } t\in \rr.
\end{equation}
Let $0<\alpha<1$. We say that $ S\in B(\ch)$ is of class
$C^{\alpha}(A)$, and we write $S\in C^\alpha(A)$, if the map $
\rr\ni t\mapsto S(t)= \e^{\i t \ca}S\in B(\ch)$ is of class $
C^{\alpha}$ (i.e.\ is H\"older continuous of order $\alpha$) for the
strong operator topology of $B(\ch)$.  By the uniform boundedness
principle, this is equivalent to $\|S(t)- S\|\leq C |t|^{\alpha}$
for $ |t|\leq 1$ and from this estimate we easily get that
 \begin{equation}
\label{e1.4}
\| S(t)- S(s)\|\leq C\e^{2\gamma|t|}|t-s|^{\alpha}, \hbox{ for }|t-s|\leq 1.
\end{equation} 
We say that $ S\in B(\ch)$ is of class $C^{1}(A)$ if $ t\mapsto
S(t)$ is of class $C^{1}$ for the strong operator topology. If $\ch$
is reflexive then $ S\in C^{1}(A)$ if and only if $t\mapsto S(t)$ is
locally Lipschitz (this property holds in the strong topology if and
only if it holds in the norm topology).  Then we may define
\begin{equation}
\label{e1.5}
S':= \frac{ d}{ dt}S(t)_{\mid t=0}
\end{equation}
so that $ S(b)- S(a)= \int_{a}^{b}S'(t)dt$ in the strong sense.
Note that $S\in C^{1}(A)$ if and only if $S\Dom A\subset \Dom A$ and the
operator $[S,\i A]$ with domain $\Dom A$ extends to a bounded operator
on $\ch$ which is exactly $S'$.  For this reason we often abuse
notation and denote $S'=\i\ca S= [S,\i A]$.

If $1<\alpha\leq 2$, we say that $S$ is of class $C^{\alpha}(A)$ if
$S\in C^{1}(A)$ and $S'\in C^{\alpha-1}(A)$. The class $C^\alpha(A)$
is similarly defined for $\alpha>2$. Note however that for integer
$\alpha$ it would be more natural to define this class in terms of
Zygmund type conditions.  The next lemma follows easily from the
fact that $\e^{\i t \ca}$ are automorphisms of $B(\cH)$.

\begin{lemma}\label{1.3}
The following properties hold for any number $\alpha>0$:
 \ben
\item the classes $C^{\alpha}(A)$ are sub-algebras of $ B(\cH)$,
\item  $\ca$ is a derivation of $B(\cH)$, i.e.
\(
(S_{1}S_{2})'= S_{1}'S_{2}+ S_{1}S_{2}' \quad\text{if \ } 
S_{1}, S_{2}\in C^{1}(A),
\)
\item if $S\in B(\cH)$ is boundedly invertible and $S\in
  C^{\alpha}(A)$ then $S^{-1}\in C^{\alpha}(A)$.  Moreover if
  $S\in C^{1}(A)$ then $ (S^{-1})'= - S^{-1}S'S^{-1}$.  
\een
\end{lemma}

\subsection{$C^{\alpha}(A)$ classes of unbounded operators}\label{sec1.3c}

In this subsection we fix $0<\alpha\leq 2$ and $S$ a closed,
densely defined operator on $\cH$ with $\rho(S)\neq \emptyset$.  We
set $ R(z)= (S-z)^{-1}$ for $ z\in \rho(S)$.

We say that $S$ is {\em regular} if there is a sequence $ (z_{n})\in
\rho(S)$ with $\lim |z_{n}|=+\infty$ and
\[
\| (S-z_{n})^{-1}\|\leq C |z_{n}|^{-1} \quad\text{for some constant
  \ } C\geq 0.
\]
Note that this is not an innocent condition, some natural realizations of
the free Klein-Gordon operator considered later do not have this property: if $S=H_{0}$ 
as in Remark \ref{re:bad} we may have $\sigma(S)=\R$ and
$\|(S-z)^{-1}\|\geq1$ for all $z\nin\R$. 

\begin{definition}
  We say that $ S\in C^{\alpha}(A)$ for $0<\alpha\leq 2$ if $R(z_{0})\in C^{\alpha}(A)$
  for some $z_{0}\in\rho(S)$.
 \end{definition}
 
 \begin{lemma}\label{1.4}
 \ben
 \item if $R(z_{0})\in C^{\alpha}(A)$ for some $ z_{0}\in \rho(S)$
   then $R(z)\in C^{\alpha}(A)$ for all $ z\in \rho(S)$, 
 \item If $ S\in C^{1}(A)$ then
 \[
[A, R(z)]= (S- z_{0})R(z)[A, R(z_{0})]R(z)(S-z_{0}), \  z_{0}, z\in \rho(S).
\]
\item If $S\in C^{1}(A)$ then the space $D:= R(z)\Dom A$ is
  independent on $z\in \rho(S)$, included in $ \Dom A\cap \Dom S$ and is
  a core for $S$.
\item If moreover $S$ is regular, then $D$ is dense in $ \Dom A\cap \Dom S$. 
\een
\end{lemma}
\proof (1) follows from (3) of Lemma \ref{1.3} and the first
resolvent formula. Then (2) follows from (2) of Lemma \ref{1.3} and
again the first resolvent formula.  Let us prove (3). Since $\Dom A$
is dense in $ \cH$, the set $D_{z}:=R(z)\Dom A$ is a core for $S$. By
Subsect. \ref{sec1.2} we know that $D_{z}\subset
\Dom A$. Using the first resolvent formula, we see that
$D_{z_{1}}\subset D_{z_{2}}$ for all $ z_{1}, z_{2}\in \rho(S)$,
hence $ D_{z}$ is independent on $z$.

If $S$ is regular, then $J_{n}:= -z_{n}R(z_{n})$ tends strongly to
the identity in $\cH$ and in $ \Dom S$. Let $u\in \Dom A\cap \Dom
S$. Then $u_{n}:=J_{n}u\in D$ and $u_{n}\to u$ in $\Dom S$.  From
(2) we obtain that:
\[
[A, J_{n}]= (S- z_{0})R(z_{n})[A, R(z_{0})]J_{n}(S- z_{0}).
\] 
Since $S$ is regular, we see that $ (S-z_{0})R(z_{n})\to0$ strongly
on $\cH$. So $[A, J_{n}]\to0$ strongly on $\cH$ hence $u_{n}\to u$
in $\Dom A$ and $D$ is dense in $ \Dom A\cap \Dom S$.  \qed

We now assume that the Banach space $\cH$ is reflexive. Then 
\begin{equation}\label{e1.12}
\begin{array}{rl}
\|u\|= \sup_{w\in \cH^{*}, \|w\|=1}|\lag w, u\rag | \quad\text{if }  
u \in \cH,\\[2mm]
\| S\|= \sup_{u\in \cH, w\in \cH^{*}, \|u\|=\|w\|=1}|\lag w, Su\rag
| 
\quad\text{if } S\in B(\cH).
\end{array}
\end{equation}
 
From (\ref{e1.12}) we obtain that for $S\in B(\cH)$ we have $S\in
C^{\alpha}(A)\Leftrightarrow S^{*}\in C^{\alpha}(A^{*})$.  This
extends to $S$ closed and densely defined.  Moreover, if $S$ is
closed densely defined and regular, then so is $ S^{*}$.

We consider the sesquilinear form:
\[
[A, S](w, u):= \lag A^{*}w, S u\rag- \lag  S^{*}w, Au\rag,
\ u\in \Dom S\cap \Dom A,\  w\in \Dom S^{*}\cap \Dom A^{*}. 
\]
We equip $\Dom S$ and $\Dom S^{*}$ with their graph norms.

\begin{proposition} Let $S$ be regular. Then the following are equivalent:
\ben

\item $S$ is  of class $C^{1}(A)$,
 
\item  the following three conditions are satisfied:
\begin{compactenum}
\item[{\rm(i)}] $|[A, S](w, u)|\leq C \| w\|_{\Dom S^{*}}\|u\|_{\Dom S}$, $ u\in
 \Dom S\cap \Dom A$, $w\in \Dom S^{*}\cap \Dom A^{*}$,  

\item[{\rm(ii)}] $\{u\in \Dom A\ : \ R(z)u\in \Dom A\}$ is a core for $A$ for some
 $z\in \rho(S)$, 
 
\item[{\rm(iii)}] $\{w\in \Dom A^{*}\ : \ R(z)^{*}w\in \Dom A^{*}\}$ is a core for $
 A^{*}$ for some $ z\in \rho(S)$. 
\end{compactenum}
\een
\end{proposition}

For the proof, see \cite[Props. 2.19, 2.21]{GGM}. 

Assume that $S\in C^{1}(A)$  is regular. Then by Lemma \ref{1.4}
$\Dom A\cap \Dom S$ is dense in $ \Dom S$ and $\Dom A^*\cap \Dom S^*$ in $
\Dom S^*$. As in the proof of \cite[Prop. 2.19]{GGM} we see
that $ [A, S]$ uniquely extends to a bounded sesquilinear form $ [A,
S]^{\circ}$ on $\Dom S^{*}\times \Dom S$ and \( [A, R(z)]= - R(z)[A,
S]^{\circ}R(z).  \) Here, the left $R(z)$ acts on $\ch^*$ as
$R(z)^*$.

\begin{remark}\label{re:adj}{\rm On a Krein space (see Sect. \ref{df:K}), if $S= S^{*}$ and
    if the Krein structure is of class $ C^{1}(A)$, (see Subsect.
    \ref{ss:sec3}), then (iii) follows from (ii), because we can
    consider $ S^{*}$, $ A^{*}$ as operators on $ \cH$ and $ A-
    A^{*}$ is bounded.  }\end{remark}

We now give some regularity properties with respect to $A$ of a
function of $S$.

\begin{lemma}\label{1.4b}
  If $S\in C^{\alpha}(A)$ then $ \chi(S)\in C^{\alpha}(A)$ for any
  $\chi\in \coinf(\beta(S))$.
 \end{lemma}
 \proof We prove more, namely that $\int
 R(\lambda\pm\i0)\chi(\lambda)d\lambda$ are of class $C^\alpha$.  From the definition of $\beta(S)$  (see Def. \ref{df:bv}) and 
 using a partition of unity, we may assume that the assumptions of
 Lemma \ref{lm:bv} are fulfilled. We begin with the case $
 0<\alpha\leq1$. We claim first that
\begin{equation}
\label{e1.15}
\|\e^{\i t \ca}R(z)- R(z)\|\leq C |{\Im z}|^{- 2n}|t|^{\alpha}, \
0\leq |t|\leq 1, \ z\in I \pm \i ]0, \nu].
\end{equation}
This implies the lemma if $0<\alpha\leq1$ using  \eqref{eq:bvf}
with $n$ replaced by $2n$.

We now prove (\ref{e1.15}). If $T\in B(\cH)$ with $ T^{-1}\in
B(\cH)$ then from $ \e^{\ i t \ca}T^{-1}= (\e^{\i t \ca}T)^{-1}$ we get
\begin{equation}
\label{e1.13}
\| \e^{\i t \ca}T^{-1}- T^{-1}\| \leq C \| T^{-1}\|^{2}\| \e^{\i t
  \ca}T- T\|, \ |t|\leq 1. 
\end{equation}
 The same argument gives for $T_1,T_2\in B(\cH)$:
 \begin{equation}
\label{e1.14}
\| \e^{\i t \ca}(T_1T_2)- T_1T_2\| \leq C \| T_1\|\| \e^{\i t \ca}T_2-
T_2\| + C \| T_2\| \| \e^{\i t \ca}T_1- T_1\|, \ |t|\leq 1. 
\end{equation}
For $ z_{0}\in \rho(S)$ and $ z\in I\pm \i ]0, \nu]$ we have:
\[
R(z)= R(z_{0})(1 + (z-z_{0})R(z_{0}))^{-1}.
\]
Applying (\ref{e1.13}), (\ref{e1.14}) and the hypothesis that
$R(z_{0})\in C^{\alpha}(A)$,  we obtain 
\[
\|\e^{\i t \ca}R(z)- R(z)\|\leq C \| R(z)\|^{2}|t|^{\alpha}, \ 0\leq |t|\leq 1,
\]
which proves (\ref{e1.15}).  Note that in the case $\alpha=1$ the
formula \eqref{eq:bvf} gives an explicit expression for the
commutator $[\int R(\lambda+\i0)\chi(\lambda)d\lambda,A]$ involving
expressions of the form $R(z)[S,A]^{0}R(z)$. 

In the case $1<\alpha\leq 2$ we repeat the same arguments applied to
the first derivative, using again (\ref{eq:bvf}).  \qed

\subsection{Some Fourier transforms}\label{sec1.1}

For simplicity of future notation, we normalize the Fourier
transform of tempered distributions in such a way that $f(\tau)=
\int\e^{\i \tau t }\what{f}(t)dt$. We set
 \begin{equation}
\label{e1.0}
f_{s}(\tau):= f(s\tau),  \ f\in {\mathcal S}'(\rr), \ s\in \rr.
\end{equation}
Then $ \what{f_{s}}(t)= s^{-1}\what{f}(s^{-1}t)$. If $ \delta:= \tau
\frac{d}{ d \tau}$, then $f_{s}= \e^{-t \delta}f$ for $s= \e^{-t}$. We
will set
\begin{equation}\label{eq:tilde}
 \tilde{f}(\tau):= \delta f(\tau)= \tau f'(\tau), \ f\in {\mathcal
   S}'(\rr).
\end{equation}
We denote by $ S^{\sigma}(\rr)$ for $ \sigma\in \rr$ the space of
functions $ f\in C^{\infty}(\rr)$ such that
\(
| f^{(n)}(\tau)|\leq C_{n}\langle \tau\rangle^{\sigma-n}, \ n\in \nn.
 \)
 
\begin{lemma}\label{1.1}
The classes $S^\sigma$ have the following properties:
\ben
 \item If $ f\in S^{\sigma}(\rr)$ then $ \what{f}\in
   C^{\infty}(\rr\backslash\{0\})$ and  
 \[
 |\what{f}(t)|\leq C_{n}\langle t\rangle^{-n}\hbox{ in }|t|\geq 1, \
 \forall \ n\in \nn. 
\]
\item If $ f\in S^{\sigma}(\rr)$ for $\sigma<0$ then $ \what{f}\in L^{1}(\rr)$,
\item If $ f\in S^{\sigma}(\rr)$ for $-1<\sigma<0$ then 
\[
|t^{k}\what{f}^{(k)}(t)|\leq C_{k}|t|^{-\sigma-1}, \ \forall \ k\in \nn.
\]
\een
\end{lemma}
These facts are well known.  The typical example of a symbol in
$S^{-\sigma}(\rr)$ is the function $\jap{\cdot}^{-\sigma}$
whose Fourier transform is the  {\em Bessel potential} $
G_{\sigma}$. For all $ t\neq 0$, $G_{\sigma}(t)$ is given by the
following absolutely convergent integral (see
e.g. \cite[Sect. V.3]{St}):
 \begin{equation}
\label{e1.1}
G_{\sigma}(t)= \frac{1}{2^{\sigma}\sqrt\pi\Gamma(\sigma/2)}
\int_{0}^{+\infty} \rme^{-t^{2}/r-r/4} r^{(\sigma-1)/2} \frac{dr}{r}.
\end{equation}
 The following lemma is easy.

\begin{lemma}\label{1.2}
The distributions $G_\sigma$ have the following properties:
 \ben
 \item 
 \(
G_{\sigma}'(t)= C_{\sigma} t G_{\sigma-2}(t), \  t\neq 0, \ \sigma\in \rr,
\)
\smallskip
\item 
 \(
|t^{k} G_{\sigma}^{(k)}(t)|\leq C_{k, \sigma} |t|^{\sigma-1}, \
t\neq 0, \ \sigma\in \rr, \ k\in \nn, 
\)
\smallskip
\item
 \(
|G_{\sigma}^{(k)}(t)|\leq C_{k, \sigma}|t|^{k} \e^{- | t|/2}, \
|t|\geq 1, \sigma\in \rr, \ k \in \nn, 
\)
\smallskip
\item
\(
 \e^{ c|t|}\delta^{k}G_{\sigma}\in L^{1}(\rr), \ \sigma>0, \ 
 c <\12,\ k\in \nn,
\)
\smallskip\item
\(
  \e^{ c|t|}G'_{\sigma}\in L^{1}(\rr), \ \sigma>1, \ c <\12.
\)
 \een
\end{lemma}
\proof We get (1) by differentiating (\ref{e1.1}) under the integral
sign. Relation (2) for $k=0$ follows from
 \[
 \e^{- t^{2}/r- r/4}\leq \e^{- t^{2}/r}.
 \] Using (1) we obtain (2) for arbitrary $k$. Similarly using the inequality
 \[
t^{2}/r + r/4 \geq |t|/2 + 1/2r + r/8, \ |t|\geq 1,
\]
and the fact that the integral $\int_{0}^{+\infty} \e^{- 1/2r
  -r/8}r^{(\sigma-1)/2} \frac{dr}{r}$ is finite for all $ \sigma\in
\rr$ we obtain (3) for $k=0$, and then for arbitrary $k$ using
(1). Finally, (4) and (5) follow from (2) and (3). \qed

\subsection{Functional calculus associated to $A$}\label{sec1.3} 
 Let us fix a $C_{0}-$group $W$ on the Banach space $\cH$ with generator $A$.

Let $\cm_\gamma$ be the set of functions $f:\R\to\C$ whose Fourier
transforms are complex measures such that:
\begin{equation}\label{eq:cm}
\|f\|_{\cm}:=\int\rme^{\gamma|t|}|\what f(t)|dt<\infty.
\end{equation}

$\cm$ is a unital Banach $*$-algebra for the usual operations of
addition and multiplication and  the involution $f^{*}(\tau)=\bar{f}(-\tau)$.
 \label{p:sym} Such functions $f$ admit a holomorphic
extension in the strip $\{\tau\ : \ |\Im{\tau}|<\gamma\}$, in particular do not
have compact support. We define 
\[
f(A):=\int \rme^{\i tA}
\what{f}(t)dt
\] and note that $\cm\ni f\mapsto f(A)\in B(\ch)$ is a
linear multiplicative map. Clearly $f\in\cm_\gamma\Rightarrow
f_s\in\cm_\gamma$ if $0\leq s\leq1$ and
\begin{equation}\label{e1.02}
\|f(sA)\|\leq M\|f\|_{\cm_\gamma} \quad\text{where\ } f(s A)=f_{s}(A). 
\end{equation} 
 By Lemma
\ref{1.2} we see that  if $\sigma>0$ then  $\jap{\cdot}^{-\sigma}\in \cm_\gamma$ if
$\gamma<1/2$ hence $\jap{s A}^{-\sigma}$ is a well defined bounded
operator on $\ch$ if $0\leq 2s\gamma < 1$. 

A similar assertion holds for a large class of analytic symbols of
strictly negative order but the problem of the boundedness of the
operator $f(A)$ for symbols of class $S^0$ which are not Fourier
transforms of measures is much more delicate.

We will be interested in the apparently trivial case when the
derivative of $f$ satisfies $f'(\tau)=\jap{\tau}^{-\sigma}$ with
$\sigma>1$. To understand the nature of the problem note that for
such an $f$ the operator $f(P)$ with $P=-\i\frac{d}{dx}$ is bounded
in $L^p(\R)$ if $1<p<\infty$ but not in $L^1(\R),L^\infty(\R)$, or
$C_0(\R)$.

If $W$ is a bounded $C_{0}$-group and $\ch$ is Hilbertizable then
$\|f(A)\|\leq C\sup|f|$ because such a group is unitary for an
admissible Hilbert norm.  In our applications this is not sufficient
because $W$ is of exponential growth. But we have:

\begin{proposition}\label{pr:bdd}
  If $\ch$ is Hilbertizable and $f$ is holomorphic on the strip
  $\{z\ : \ |\Im z|<\gamma'\}$ for some $\gamma'>\gamma$ then
\begin{equation}\label{eq:bdd}
\|f(A)\|\leq C \sup_{\rr+\i]-\gamma', \gamma'[}|f(z)|
 \end{equation}
\end{proposition}

For the proof, see \cite[Prop. 3.7.1]{ABG}. The
hilbertizability assumption is rather annoying but we expect that
the result remains true in UMD spaces.

One may define $f(A)$ for unbounded functions $f$ by allowing
$\what{f}$ to be a distribution of exponential decay instead of a measure.
In other terms, $\what{f}$ may be a sum of derivatives of exponentially
decaying measures, or $f$ a sum of functions in $\cm_{\gamma}$ multiplied
by  polynomials. We assume $\gamma<1/2$ and explain this in
detail only for the functions $f(\tau)=\jap{\tau}^s$ with $0<s<1$
which are important here. Let us set $\sigma=2-s$, so
that $1<\sigma<2$.  Note that from
$$
\lag \tau\rag^{-s}\lag \tau\rag^{-\sigma}
=\lag \tau\rag^{-2}=(1-\i\tau)^{-1}(1+\i\tau)^{-1},
$$
identity  valid in the algebra $\cm_\gamma$, we get by the already
defined functional calculus
$$
\lag A\rag^{-s}\lag A\rag^{-\sigma}=(1+A^{2})^{-1}
=(1-\i A)^{-1}(1+\i A)^{-1}.
$$
Thus $B=\lag A\rag^{-\sigma}(1+A^{2})$ is a well defined operator on
 $\Dom A^{2}$ and there we have 
$\jap{A}^{-s}B=B\lag A\rag^{-s}=1 $. Hence we must define
$\jap{A}^s$ as the closure of $B$. Then we have on
$\Dom A^{2}$:
\begin{align}\label{eq:s>0}
\lag A\rag^{s} =\int (1+A^{2})\rme^{\i tA} G_\sigma(t) dt=
\int \Big((1-\partial_{t}^{2})\rme^{\i tA} \Big) G_\sigma(t) dt=
\int \rme^{\i tA} \big(G_\sigma(t) - G''_\sigma(t) \big)dt 
\end{align}
where we interpret the derivatives in the sense of distributions.
If we set  $P=-\i\frac{d}{dt}$ as operator acting on $\ch$-valued
distributions then we may write
$$
\jap{A}^s u =\jap{A}^{-\sigma}u-\int \rme^{\i tA}  uG''_\sigma(t)dt
=\int W_{t}u \cdot  (1+P^2)G_\sigma(t) dt, , \ u \in \Dom A^{2}.
$$ 
This representation gives the following useful estimate:

\begin{proposition}\label{pr:holder}
  If $\gamma<1/2$ and $0<s<m<1$ then there exists $C\geq 0$ such that
\begin{equation}\label{eq:holder}
\|\jap{A}^su\|\leq C \|u\| + 
C {\textstyle\sup_{|x|<1}} |x|^{-m}\|(W(x)-1)u\|.
\end{equation} 
\end{proposition} 
\proof Let $\theta$ be a $C^\infty$ function such that $\theta(t)=1$
for $|t|<1$ and $\theta(t)=0$ if $|t|>2$. Set $V(t)=\theta(t)W_{t}u$.
Then \[
\jap{A}^su=\int V(t)\cdot (1+P^2)G_\sigma(t) dt +\int
(1-\theta)W_{t}u (1+P^2) G_\sigma(t) dt.
\] By Lemma \ref{1.2}  (3) the second term is bounded by 
$ C\|u\|$. Since  $V$ is a continuous function with compact support,
 for any $s<\mu<m$ we have:
$$
\int V(t)\cdot (1+P^2)G_\sigma(t) dt
=\int \jap{P}^\mu V(t) \cdot \jap{P}^{2-\mu} G_\sigma(t) dt.
$$
Since $\what{Pf}(t)= t\what{f}(t)$ and $ \sigma= 2-s$ we have
$\jap{P}^{2- \mu}G_{\sigma}= G_{\mu-s}$, hence
$$
\left\|\int V(t)\cdot(1+P^2)G_\sigma(t) dt \right\|
=\left\|\int\jap{P}^\mu V(t)\cdot G_{\mu-s}(t)dt \right\|\leq
\|\jap{P}^\mu V\|_{L^\infty} \|G_{\mu-s}\|_{L^1},
$$
where we used that $\mu-s>0$ and Lemma \ref{1.2} (4). 
Then it remains to note that 
$\|\jap{P}^\mu V\|_{L^\infty}\leq C \|V\|_{C^m}$ if $0<\mu<m<1$,
$V$ has compact support, and 
$$
\|V\|_{C^m}=\sup_t\|V(t)\| +\sup_{t\neq s} |t-s|^{-m} \|V(t)-V(s)\|.
$$
This is easy to prove by a standard Littlewood-Paley type argument.
\qed

\subsection{$C_{0}$-groups on $K-$spaces}\label{ss:sec3}

In this subsection $\ch$ is a $K-$space equipped with the hermitian form $\braket{\cdot}{\cdot}$. Since $\ch$ is
reflexive $W^*=\{W^*_t\}$ is also a $C_0$-group of operators on
$\ch$ whose generator is $-A^*$. In other terms, $W^*_t=\rme^{-\i
  tA^*}$.  Clearly $\|W^{*}_{t}\|\leq M'\rme^{\gamma|t|}$ with the
same $\gamma$ hence the operators $A,A^{*}$ admit an $\cm_\gamma$
functional calculus and we have $f(A)^{*}=\overline{f}(A^{*})$ for
all $f\in\cm_{\gamma}$. For example, 
$(\jap{\varepsilon A}^{-\sigma})^{*}=\jap{\varepsilon A^{*}}^{-\sigma}$ for
$\varepsilon>0$ small enough.

We shall say that \emph{the Krein structure is of class $C^1(A)$} if
the conditions of the next proposition are verified.

\begin{proposition}\label{pr:rega}
The following assertions are equivalent:
\begin{compactenum}
\item the function $t\mapsto\braket{W_t u}{W_t u}$ is derivable at
  zero for each $u\in\ch$,
\item the function $t\mapsto\braket{W_t u}{W_t u}$ is of class $C^1$ for
  each $u\in\ch$,
\item the map $t\mapsto W^{*}_tW_t$ is locally Lipschitz,
\item $A^*=A+B$ where $B$ is a bounded operator.
\end{compactenum}
\end{proposition}

\proof
For $u,v\in \Dom A$ we have 
\begin{equation}
-\i\frac{d}{dt}\braket{W_{t}u}{W_{t}u}=
\braket{W_{t}u}{AW_{t}u} -\braket{AW_{t}u}{W_{t}u} .
\end{equation}
If the derivative in the left hand side exists at zero for each 
$u,v\in\ch$ then the map $t\mapsto W^{*}_tW_t$ is weakly 
differentiable at $t=0$, hence by the uniform boundedness principle 
the derivative is a bounded operator and so there is a number $C$ 
such that $|\braket{u}{Av} -\braket{Au}{v}|\leq C \|u\|\|v\|$
for all $u,v\in \Dom A$. Thus if we fix $v\in \Dom A$ then
$|\braket{Au}{v}|\leq C' \|u\|$ for all $u\in \Dom A$ hence 
$v\in \Dom A^{*}$ and 
$|\braket{u}{(A-A^{*})v}|\leq C \|u\|\|v\|$ for $u,v\in \Dom A$.
Thus $\Dom A\subset \Dom A^{*}$ and $\|(A-A^{*})v\|\leq C''\|v\|$
for $v\in \Dom A$. If we denote $A^{*}_{0}$ the restriction of 
$A^{*}$ to $\Dom A$ then we get $A^{*}_{0}=A+B$ for a bounded 
operator $B$. If $a>0$ is large enough then
$$
A^{*}_{0}+\i a=(A+\i a)+B=[1+B(A+\i a)^{-1}](A+\i a)
$$
and $\|B(A+\i a)^{-1}\|<1$ hence $A^{*}_{0}+\i a:\Dom A\to\ch$ is bijective.
But $A^{*}+\i a:\Dom A^{*}\to\ch$ is also bijective for large $a$, so
$\Dom A=\Dom A^{*}$ and $A^{*}=A+B$. This proves $(1)\Rightarrow(4)$.
Then $(4)\Rightarrow(2)\Rightarrow(1)$ follows from
\begin{equation}\label{eq:int}
\braket{W_{t_{2}}u}{W_{t_{2}}v}-\braket{W_{t_{1}}u}{W_{t_{1}}v}=
\i\int_{t_{1}}^{t_{2}}
\big(\braket{W_{t}u}{AW_{t}v} -\braket{AW_{t}u}{W_{t}v} \big) dt
\end{equation}
which holds for $u,v\in \Dom A$ and extends to all $u,v\in\ch$ under
the assumption $(4)$.
Finally, $(2)\Rightarrow (3)$ follows from uniform boundedness
principle and $(3)\Rightarrow(2)$ follows from (\ref{eq:int}) and a
density argument.
\qed

\begin{remark}\label{re:B}{\rm
  Note that $B=\i\frac{d}{dt}W^{*}_tW_t|_{t=0}$.
}\end{remark}

\begin{remark}\label{re:rega}{\rm If $A$ is selfadjoint for a
    Hilbert norm $(\cdot|\cdot)^{1/2}$ and $\braket{u}{v}=(u|Jv)$ then 
    (4) means $J\in C^1(A)$.  }\end{remark}

\begin{corollary}\label{co:rega}
  If the Krein structure is of class $C^{1}(A)$ then the Besov
  scales $\ch_{s,p}$ associated to the groups $W$ and $W^*$ coincide
  for $-1< s < 1$.
\end{corollary} 
\proof We have $\Dom A=\Dom A^*$ by $(4)$ of Prop.
\ref{pr:rega}. The spaces $\ch^A_{s,p}$ with $0<s<1$ associated to
$W$ are obtained by interpolation between $\Dom A$ and $\ch$ and
similarly for $W^*$, hence $\ch^A_{s,p}=\ch^{A^*}_{s,p}$ if
$0<s<1$. Then $\ch^{A}_{s,p}=\ch^{A^{*}}_{s,p}$ follows by duality
if $-1<s<0$ (supplemented by an obvious density argument if
$p=\infty$). The case $s=0$ is covered by interpolating between
$\ch^A_{1/2,p}$ and $\ch^A_{-1/2,p}$.  \qed

\begin{proposition}\label{pr:est2}
If the Krein structure is of class $C^{1}(A)$ then for $0<\sigma<1$ 
and $\veps>0$ small we have:
\begin{align}
& \|\lag\veps A\rag^{\sigma}-\lag\veps A^{*}\rag^{\sigma}\|
\leq C \varepsilon , \label{eq:sobk1} \\
& \lag\veps A\rag^{-\sigma}-\lag\veps A^{*}\rag^{-\sigma}=
\lag\veps A\rag^{-\sigma}-\big(\lag\veps A\rag^{-\sigma}\big)^{*}=
\lag\veps A\rag^{-\sigma} O(\veps)\lag\veps A\rag^{-\sigma}.
\label{eq:sobk2}
\end{align}
\end{proposition}

\proof Set for simplicity of notation $\cH_{s}= \cH^{A}_{s,2}$. From
\eqref{eq:s>0} we get
$$
\lag \veps A\rag^{\sigma}-\lag \veps A^{*}\rag^{\sigma} =
\veps\int \frac{\rme^{\i t\veps A} - \rme^{\i t\veps A^{*}}}{\veps t} 
t\big(G_{2-\sigma}(t) - G''_{2-\sigma}(t) \big)dt. 
$$
This holds in $B(\ch_{1},\ch_{-1})$ by  Corollary
\ref{co:rega}.  Using that $\|\rme^{\i\veps t A} - \rme^{\i\veps t
  A^{*}}\|\leq C |\veps t| \rme^{a\veps |t|}$ since $A- A^{*}$ is
bounded, and the estimates for $G_{2-\sigma}$ in Lemma \ref{1.2}, we
obtain \eqref{eq:sobk1}. This implies
$\|\lag\varepsilon A\rag^{\sigma}u\|\leq c\|\lag\varepsilon
A^{*}\rag^{\sigma}u\|$ hence by using a similar estimate with $A$
and $A^*$ interchanged and then taking adjoints we obtain:
\begin{equation}\label{eq:soc}
\begin{array}{rl}
&\|\lag\veps A\rag^{\sigma}\lag\veps A^{*}\rag^{-\sigma}\|\leq C, \ 
\|\lag\veps A^*\rag^{\sigma}\lag\veps A\rag^{-\sigma}\|\leq C, \\[2mm]
&\|\lag\veps A\rag^{-\sigma}\lag\veps A^{*}\rag^{\sigma}\|\leq C, \ 
\|\lag\veps A^{*}\rag^{-\sigma}\lag\veps A\rag^{\sigma}\|\leq C,
\end{array}
\end{equation}
where the number $C$ is independent of $\veps$. The left hand side
of \eqref{eq:sobk2} is 
\[
\lag\veps A\rag^{-\sigma}
\big(\lag\veps A^{*}\rag^{\sigma}-\lag\veps A\rag^{\sigma}\big)
\lag\varepsilon A^{*}\rag^{-\sigma},
\]
 and so if we use \eqref{eq:sobk1}
and \eqref{eq:soc} we get \eqref{eq:sobk2}.
\qed

\section{Commutator expansions}\label{sec1} \init
In this section we prove some results on commutator expansions. These results are well-known in the Hilbert space setting. In the Banach space setting considered here they seem to be new.

\subsection{Functional calculus associated to $\ca$}\label{sec1.4} 

We now discuss the functional calculus associated to the operator
$\ca$ acting on $ B(\cH)$ introduced in \eqref{e1.03}.  By \eqref{e1.3} the operator $f(\ca)=\int\rme^{\i
  t\ca}\what{f}(t)dt$ is well defined if $f\in \cm_{2\gamma}$ and
$f\mapsto f(\ca)$ is a linear multiplicative map with values in the
Banach algebra of bounded operators on $B(\ch)$ such that
\begin{equation}
\label{e1.6b}
\|f(\ca)\|\leq M^{2}\|f\|_{\cm_{2\gamma}}.
\end{equation}
Let $ \cn$ be the set of functions whose Fourier transforms are
measures supported in $|t|\leq 1$.  Then $\cn$ is a linear subspace
of $\cm_{2\gamma}$ which contains the constants, is stable under
derivations, and:
\begin{equation}\label{e1.6}
\|f_{s}\|_{\cm_{2\gamma}}\leq \e^{2\gamma|s|} \|\what{f}\|_{L^1(\R)}, \ s\in \rr, \ f\in \cn.
\end{equation}
Below we use the notation $\wtilde{f}$ introduced in \eqref{eq:tilde}.

\begin{lemma}\label{lm:normal}

\ben
\item   If $f\in\cn$ then
\[
\|f(s \ca)\|\leq  C \e^{2\gamma |s|} , \ s\in \rr.
\]
\item If  $f,\wtilde{f}\in\cn$ and $T\in C^{\alpha}(A)$
  for some $0<\alpha<1$ then 
  \[
  \|f(s_{2}\ca)T-f(s_{1}\ca)T\|\leq
  C|s_{2}- s_{1}|^{\alpha}\rme^{2\gamma|s_{1}|} \hbox{ for }|s_{2}- s_{1}|\leq 1.
  \]
  \een
\end{lemma}
\proof (1) follows from (\ref{e1.6b}), (\ref{e1.6}). Let us now
prove (2). We first claim that if $T\in C^{\alpha}(A)$ and $g\in
\cn$ with $ g(0)=0$ then:
\begin{equation}
\label{e1.7}
\|g(s\ca)T\|\leq C |t|^{\alpha}\e^{2\gamma|t|}, \ t \in \rr.
\end{equation} 
In fact since 
$\|\big(\rme^{\i t\ca}-1\big) T\|\leq C |t|^{\alpha}\rme^{2\gamma|t|}$ if
$T\in C^{\alpha}$, we have:
\begin{align*}
\|g(s\ca)T\| &=\|g(s\ca)T-g(0\ca)T\|
=\|\int\big(\rme^{\i st\ca}-1\big) T \what{g}(t)dt\| \\
& \leq \int \|\big(\rme^{\i st\ca}-1\big) T\| |\what{g}(t)|dt
\leq C\int |st|^{\alpha}\rme^{2\gamma|st|} |\what{g}(t)|dt
\leq C' |s|^{\alpha}\rme^{2\gamma|s|}. 
\end{align*}
 We write now
 $$
 f(s_{2}\ca)-f(s_{1}\ca)
 =\int_{s_{1}}^{s_{2}}\frac{d}{ds}f(s\ca)ds 
 =\int_{s_{1}}^{s_{2}}\ca f'(s\ca)ds 
 =\int_{s_{1}}^{s_{2}} \wtilde{f}(s\ca)\frac{ds}{s}.
 $$
 Since $\wtilde{f}\in\cn$ and $\wtilde{f}(0)=0$ we get
 $\|\wtilde{f}(s\ca)T\|\leq C |s|^{\alpha}\rme^{2\gamma|s|}$ by
 (\ref{e1.7}). So if $0\leq s_{1}<s_{2} < s_{1}+1$:
 \begin{align*}
 \|f(s_{2}\ca)T-f(s_{1}\ca)T\| & \leq
 \int_{s_{1}}^{s_{2}} \|\wtilde f(s\ca)T\|\frac{ds}{s} \leq
 \int_{s_{1}}^{s_{2}} C|s|^{\alpha} \rme^{2\gamma s}\frac{ds}{s} \\
& \leq
 \frac{C}{\alpha}\big(s_{2}^{\alpha}-s_{1}^{\alpha}\big) \rme^{2\gamma(s_{1}+1)}
 \leq \frac{C}{\alpha}(s_{2}-s_{1})^{\alpha} \rme^{2\gamma(s_{1}+1)}.
 \end{align*}
 If $s_{1}<s_{2}\leq0$ the argument is similar.  The case
 $s_{1}<0<s_{2}$ follows from the preceding ones.  \qed

The next lemma will be needed later on. 
\begin{lemma}\label{lm:cont}
Let $\rb$ a  normed vector space. Let $\xi=\theta\eta$ where $\theta:\R\to\rb$ with $\theta(0)=0$
and $\eta:\R\setminus\{0\}\to\C$ is a function of class $C^{1}$. 
Assume that for some real numbers $a,b,\beta,m,\mu$ satisfying 
$0<m<\beta<1$ and $\mu\geq 3\gamma$ we have: 
\begin{align}
&\|\theta(s_{1})-\theta(s_{2})\| \leq a |s_{1}-s_{2}|^{\beta}\rme^{2\gamma|s_{1}|} 
\quad\text{if \ } |s_{1}-s_{2}|\leq 1,
\label{eq:cont2}\\
&|\eta(s)|+|\wtilde\eta(s)| \leq b|s|^{-m-1}\rme^{-\mu|s|} . 
\label{eq:cont3}
\end{align}
Then:
\[
\int\rme^{\gamma|s|}\|\xi(s+t)-\xi(s)\|ds\leq C_{\beta,m}ab |t|^{\beta-m}, \ |t|<1.
\] 
\end{lemma}
\proof It suffices to consider the case $0<t<1$.  From
$\|\xi(s)\|\leq ab |s|^{\beta-m-1}$ and since $\beta-m>0$ we get:
$$
\int_{|s|\leq2t } \|\xi(s+t)-\xi(s)\| ds\leq
2\int_{|s|\leq3t } \|\xi(s)\| ds\leq 2ab (\beta-m)^{-1}
(3t )^{\beta-m}.
$$
We estimate next $\int_{2t }^{+\infty}$, the integral 
$\int^{-2t }_{-\infty}$ is treated similarly. 
Clearly
\begin{align*} 
&\int_{2t }^{\infty} \rme^{\gamma s} \|\xi(s+t)-\xi(s)\| ds\\
\leq&
\int_{2t }^{\infty} \rme^{\gamma s}\|\theta(s+t)-\theta(s)\| |\eta(s+t)|ds
+\int_{2t }^{\infty} \rme^{\gamma s}\|\theta(s)\| |\eta(s+t)-\eta(s)|ds \\
\leq&
ab \int_{t }^{\infty} \rme^{\gamma s}
t ^{\beta} \rme^{2\gamma s} (s+t)^{-m-1} \rme^{-\mu(s+t)}ds
+a\int_{2t }^{\infty} \rme^{\gamma s} s^{\beta} \rme^{2\gamma s}
\left|\int_{s}^{s+t}\eta'(y)dy \right|ds.
\end{align*}
The first  integral is less than $ab m^{-1}t^{\beta-m}$ and the last
integral is less than
\[
\begin{array}{rl}
&ab\int_{2t}^{\infty} \int_{s}^{s+t} s^{\beta} \rme^{3\gamma s} y^{-m-2}
\rme^{-\mu y}dy ds \\[2mm]
 \leq &ab \int_{2t}^{\infty} s^{\beta-m-2}t ds
\leq \frac{ab}{1-\beta+m}(2t)^{\beta-m-1}t\\[2mm]
=& C_{\beta,m}abt^{\beta-m}.
\end{array}
\]
This completes the proof of the lemma. 
\qed

In the next lemma we will use Lemma \ref{lm:cont} for $\rb= B(\cH)$.

\begin{lemma}\label{lm:ave}
  Assume that either $K\in\cn$ with $K(0)=0$ and $\wtilde{K}\in\cn$
  or that $K(\tau)=1-\rme^{-\i   \tau}$.  Let $\zeta$ be a complex function in
  $C^1(\R\setminus\{0\})$ such that $|\zeta(s)|+|\wtilde\zeta(s)| \leq
  C|s|^{-m}\rme^{-\mu|x|}$ with $0<m<1$ and $\mu>3\gamma$. Set
  \[
  \cj_{\veps}=\int \rme^{\i \veps s\ca_{r}} K(\veps s\ca)
  \zeta(s) \frac{ds}{s} \quad\text{for \ } 0<\varepsilon<1.
  \]
   Then for $T\in C^{\beta}(A)$ with
  $m<\beta<1$ we have $\cj_\veps(T)\in B(\ch)$ and
  \[
  \|\cj_{\veps}(T)(W(\veps t)-1)\|\leq C\veps^{\beta}|t|^{\beta-m}, \ |t|<1. 
\]
  In particular 
  $\|\cj_{\veps}(T)\langle \veps A\rangle^{s}\|\leq C\veps^{\beta}$ if
  $s<\beta-m$ and $2\varepsilon\gamma<1$.
\end{lemma}
\proof The function $K$ is such that $K(0)=0$ and
\[
\|K(s_{1}\ca)T-K(s_{2}\ca)T\|\leq
C|s_{1}-s_{1}|^\beta\rme^{2\gamma|s_{1}|}
\] if
\mbox{$|s_{1}-s_{1}|<1$}. Indeed, this follows from Lemma
\ref{lm:normal} for the first choice of $K$ and is obvious in the
second case.  Since $K(0)=0$ we obviously get $\|K(s\ca)T\|\leq
C|s|^\beta\rme^{2\gamma|s|}$ for any $s$. Then
$$
\|\rme^{\i \veps sA}K(\veps s\ca)(T)\zeta(s)\|\leq 
C\veps^{\beta}|s|^{\beta-m} \rme^{-|s|(\mu-3\veps\gamma)},
$$  
hence the integral defining $\cj_{\veps}(T)$ is absolutely convergent in
norm and $\|\cj_{\veps}(T)\|\leq C\veps^{\beta}$.  
Then we put $\xi(s)=K(\veps s\ca)(T)\zeta(s)/s$ and we write
$$
\|\cj_{\veps}(T)(\rme^{-\i   \veps tA}-1)\|=\left\|
\int \xi(s)\rme^{\i \veps sA}  ds(\rme^{-\i   \veps tA}-1) \right\|=
\left\| \int \big(\xi(s+t)-\xi(s)\big)\rme^{\i \veps sA} ds \right\|
$$
which is less than $\int \rme^{\varepsilon\gamma|s|} \|\xi(s+t)-\xi(s)\| ds$. 
Now we apply Lemma \ref{lm:cont} with $\theta(s)=K(\veps s\ca)(T)$ and
$\eta(s)=\zeta(s)/s$. The last assertion follows from Prop.
\ref{pr:holder} by using the estimates 
$\|\cj_{\veps}(T)\|\leq C\veps^{\beta}$
and $\|\cj_{\veps}(T)(W(\veps t)-1)\|\leq C\veps^{\beta}|t|^{\beta-m}$ 
for $|t|<1$.
\qed

\subsection{Commutator expansions}\label{ss:hoce}

Our proof of Thm. \ref{th:bvr} is based on the strategy
introduced in \cite{G1} and involves two ingredients: a version of
the Putnam argument, cf.\ Props. \ref{pr:putd} and
\ref{pr:putnam-krein} below, and a commutator expansion estimate,
cf. \cite[Sec.\ 2]{G1}, which we discuss in this and next
subsections.

More precisely, we are interested in developing the commutator
$[S,f(A)]$ in terms of iterated commutators $\ca^j(S)$ with
estimates on the remainder for ``nice'' functions $f:\R\to\C$.  If
$A$ is selfadjoint such results were obtained in \cite{GJ} using
the Helffer-Sj\"ostrand formula \eqref{eq:hs} (with $H$ replaced by
$A$). If $A$ is the generator of a $C_{0}$-group then $f(A)$ cannot
be expressed by a relation of the type \eqref{eq:hs} (the imaginary
part of the spectrum of $A$ may be too large) but a version of the
Dunford functional calculus could certainly be used. On the other
hand, the method we use below is quite classical and elementary (a
detailed presentation in the case of groups of polynomial growth may
be found in \cite[Sect.\ 5.5]{ABG}).

In this section we make some general remarks on commutator
expansions. We first discuss the ``truncated exponentials'' $E_k$
defined for any $k\in\N$ as follows:
\begin{equation}\label{eq:texp}
E_k(\tau)=\frac{1}{(\i\tau)^k}
\left(\rme^{\i\tau} -{\textstyle\sum_{0\leq j <k}}(\i\tau)^j/{j!}\right).
\end{equation}
The following properties are easy to check.

\begin{lemma}\label{zlub}
\[
\begin{array}{rl}
(1)&E_k(0)=\frac{1}{k!},\\[2mm]
(2)&E_k(\tau)=\frac{1}{k!}+\i\tau E_{k+1}(\tau),\\[2mm]
(3)&E_k(\tau)=\int_0^1 \rme^{\i\tau\theta} \frac{(1-\theta)^{k-1}}{(k-1)!} d\theta 
=-\int_0^1 \rme^{\i\tau\theta} d\frac{(1-\theta)^{k}}{k!},  \\[2mm]
(4)&  \i E_k'=kE_{k+1}-E_k,\\[2mm]
(5)&\delta E_k= E_{k-1} -kE_k, \hbox{ for }1\leq k, \hbox{ where }\delta=\tau\partial_\tau, \\[2mm]
(6)&  \tau^m\partial_{\tau}^m E_k =
  \sum_{n=0}^m C_k^m(n) E_{k-j},\hbox{ for each }0\leq m\leq k, \hbox{ and } C_{k}^{m}(n)\in \nn,\\[2mm]
(7)&\tau^m\partial_{\tau}^m E_k \in \cn, \hbox{ for } m\in \nn.
\end{array}
\]
 \end{lemma}
\proof 
For example, (3) is clearly true if $k=0,1$ and the function defined
by the right hand side of (3) satisfies the induction relation (2),
hence (3) holds for any $k$. To prove (6) observe first that
$\tau^m\partial_\tau^m= \sum_{l=1}^m b_\ell^m \delta^\ell$ for some
integers $b_\ell^m$ and then use (5). Since $E_k\in\cn$ because of
(3), we get (7). \qed

We write $[S,f(A)]=\big(f(\ca_{r})-f(\ca_{\ell})\big)(S)$ and
develop the operator $f(\ca_{r})-f(\ca_{\ell})$ acting on $B(\ch)$
in terms of powers of $\ca=\ca_{r}-\ca_{\ell}$ by using a Taylor
expansion. The class of functions $f:\R\to\C$ for which this makes
sense is easy to specify and depends only on the behavior for large
$t$ of the group $\rme^{\i tA}$, for example $f$ could be the
Fourier transform of an exponentially decaying distribution.

\begin{lemma}\label{lm:Taylor}
  For any integer $k\geq1$ we have
\begin{equation}\label{eq:Taylor}
f(\ca_{r})=\sum_{0\leq j<k} \ca^j f^{(j)}(\ca_\ell)/j! + \ca^k\cR_k(f^{(k)}),
\end{equation}
where 
\begin{equation}\label{eq:remains}
\cR_k(g)=\int \rme^{\i t\ca_\ell} E_k(t\ca) \what{g}(t) \, dt. 
\end{equation}
\end{lemma}

\proof We use the
notation:
$$
\ca_{0}:=\ca_{\ell},\quad \ca_{1}:=\ca_{r},\quad 
\ca_{\theta}:=\ca_{0}+\theta\ca=(1-\theta)\ca_{0}+\theta\ca_{1}.
$$
We have the following Taylor formula for $f(\ca_{1})=f(\ca_{0}+\ca)$:
\begin{equation}\label{eq:taylor}
f(\ca_{1})=\sum_{0\leq j<k}\frac{\ca^j}{j!}f^{(j)}(\ca_0)
-\frac{\ca^k}{k!}\int_0^1 f^{(k)}(\ca_\theta) \, d (1-\theta)^{k}.
\end{equation}
This is easy to prove by induction: if $k=1$ then 
\begin{equation}\label{eq:taylor1}
f(\ca_{1})
=f(\ca_{0})+\int_{0}^{1}\frac{d}{d\theta}f(\ca_{0}+\theta\ca) d\theta
=f(\ca_{0})+\int_{0}^{1}f'(\ca_{\theta})\ca \, d\theta
\end{equation}
and to pass from the $k$ to the $k+1$ step of the induction process
it suffices to integrate by parts the last term in
\eqref{eq:taylor}. If we set $g=f^{(k)}$ we get \eqref{eq:Taylor}
with
\begin{equation}\label{eq:remain}
\cR_k(g)=\int_0^1 g(\ca_\theta)  \frac{(1-\theta)^{k-1}}{(k-1)!} \, d\theta
\end{equation}
From $\ca_\theta=\ca_0+\theta\ca$ we
get
$$
g(\ca_\theta)=\int\rme^{\i t\ca_\theta} \what{g}(t) \, dt
=\int\rme^{\i t\ca_0}\rme^{\i\theta t\ca} \what{g}(t) \, dt
$$ 
which inserted in \eqref{eq:remain} gives \eqref{eq:remains}. This
proves the lemma. Another easy proof by induction follows from
$\cR_k(g)=\frac{1}{k!} g(\ca_0) +\ca \cR_{k+1}(g')$ which is an
immediate consequence of the definition \eqref{eq:remains} and of
the relation (2) in Lemma \ref{zlub}.  \qed

We now explain how to estimate an operator like $\cR_{k}(g)T$ when
$T\in B(\ch)$; in our case $T=\ca^{k}S$ for some bounded operator
$S$ of class $C^{k}(A)$. Observe that $\cR_{k}(g)$ looks like the
Fourier transform of the function $t\mapsto E_{k}(t\ca)\what{g}(t)$
evaluated at the point $\ca_{\ell}$. Hence we expect that decay of
$\cR_{k}(g)$ with respect to $\ca_{\ell}$ follows from regularity of
the function $t\mapsto E_{k}(t\ca)\what{g}(t)$. In fact, an
integration by parts argument which can easily be justified under
convenient conditions on $g$ gives:
\begin{align}
(-\i\ca_{\ell})^{j} \cR_{k}(g) 
&=\int \Big((-\partial_{t})^{j}  \rme^{\i t\ca_\ell} \Big) 
E_k(t\ca) \what{g}(t) \, dt
=\int \rme^{\i t\ca_\ell} 
\partial_{t}^{j} \Big( E_k(t\ca) \what{g}(t) \Big) \, dt 
\nonumber \\
&=\sum_{m=0}^{j} C_{m}^{j}\int \rme^{\i t\ca_\ell} 
E_{k}^{m}(t\ca) \frac{\what{g}^{(j-m)}(t)}{t^{m}} \, dt \nonumber 
\end{align}
with $E_{k}^{m}(\tau)=\tau^{m}\partial_{\tau}^{m}E_{k}$. We saw
before that $E_{k}^{m}\in\cn$ if $m\leq k$ and then
$\|E_{k}^{m}(t\ca)\|\leq C \rme^{2\gamma|t|}$ by Lemma
\ref{lm:normal}. The exponential decay of $\what{g}^{(j-m)}(t)$ will
compensate the divergence of this factor hence there are no problems
at infinity if $j\leq m$. Only the singularity at $0$ of
$\what{g}^{(j-m)}(t) t^{-m}$ could make the integral divergent.

Our main purpose in the next subsection is to show that
$\|\jap{A}^{s} X\jap{A}^s\|$ is finite for some $X=\cR_k(g)T \in
B(\ch)$ and $0<s<1$. For this it suffices to prove that $X$ sends
$\ch_{-\mu,1}$ into $\ch_{\mu,\infty}$ for a number $\mu$ with
$s<\mu<1$. If $\ch$ is reflexive then this is a consequence of an
estimate of the form $\|(\rme^{\i xA}-1)T(\rme^{\i yA}-1)\|\leq C
|x|^\mu |y|^\mu$ for small $x,y$. Hence $(\rme^{\i
  x\ca_\ell}-1)(\rme^{\i y\ca_r}-1)\cR_k(g)$ is the object one has
to estimate.

\subsection{First order estimates}\label{sec1.5}

The main results of this subsection concern estimates of the remainders 
in some commutator expansions
of interest later on.  We will denote $O(\veps)$ any bounded operator on
$\ch$ depending on the parameter $\veps>0$, defined at least for
small $\veps$, and such that $\|O(\veps)\|\leq C\veps$.

\begin{proposition}\label{pr:est1}
If $\,0<s<\beta<1$ and $S\in B(\ch)$ is of class $C^\beta(A)$ then
$[\lag\veps A\rag^{-s}, S]=
\lag\veps A\rag^{-s}O(\veps^{\beta})\lag\veps A\rag^{-s}$.
\end{proposition}   

\proof The idea of the proof is very simple at a formal level: we
write
$$
[\lag\veps A\rag^{-s}, S]=
\lag\veps A\rag^{-s}
[S,\lag\veps A\rag^{s}]
\lag\veps A\rag^{-s}
$$ 
and show that $[S,f(\veps A)]=O(\veps^{\beta})$ for $f(\tau)=\lag
\tau\rag^{s}$. In order to justify this formal computation we first
take $\varepsilon=1$ (we assume, without loss of generality,
$\gamma<1/2$) and assume $S\in C^{2}(A)$, so that $S$ leaves
invariant $\Dom A^{2}$.  If $B=\jap{A}^{-\sigma}(1+A^{2})$ with
$\sigma=2-s$ (see Subsect. \ref{sec1.3}) then on $\Dom A^{2}$ we have:
$$
[\lag A\rag^{-s},S]= 
\lag A\rag^{-s}SB\lag A\rag^{-s}-
\lag A\rag^{-s}BS\lag A\rag^{-s}=
\lag A\rag^{-s}[S,B]\lag A\rag^{-s}.
$$ 
Then by \eqref{eq:s>0} we have
$[S,B]=\int [S,\rme^{\i tA}]\big(G_\sigma(t)-G''_\sigma(t)\big)dt$ 
on $\Dom A^{2}$ hence 
\begin{align}\label{eq:ident}
[\lag A\rag^{-s},S] = 
\lag A\rag^{-s} \left(\int [S,\rme^{\i tA}]
G_{\sigma}(t)dt
-\int [S,t^{-1}\rme^{\i tA}] tG''_{\sigma}(t) dt
\right)\lag A\rag^{-s},
\end{align}
where $G_{\sigma}(t)$ is the Bessel potential considered in
Subsect. \ref{sec1.1}. By Lemma \ref{1.2}, $|tG''_{\sigma}(t)|\leq
C|t|^{-s}$ for $ t\neq 0$ and $|tG''_{\sigma}(t)|\leq
C|t|^3\rme^{-|t|/2}$ for $|t|>1$.

We observe next that the relation \eqref{eq:ident} remains valid for
any bounded operator $S$ of class $C^{\beta}(A)$. Indeed using
\[
 [S, \e^{\i t A}]= \e^{\i t \ca_{\ell}} (\e^{\i t \ca}-1)S, 
\]
and (\ref{e1.3}),
 we have $\|[S,\rme^{\i tA}]\|\leq C\rme^{|t|/8} |t|^{\beta}$ and it is
easy to construct a sequence of operators $S_{n}\in C^{2}(A)$
satisfying a similar estimate uniformly in $n$ and $\|S_{n}-S\|\to0$
as $n\to\infty$. We apply \eqref{eq:ident} to each $S_{n}$ and then
pass to the limit.

Replacing $A$ by $\veps A$ in \eqref{eq:ident} and using 
\[
\|[S,t^{-1}\rme^{\i t\veps A}]\| \leq C
\rme^{\varepsilon\gamma|t|}\veps^{\beta}|t|^{\beta-1}, 
\] we complete the proof of the proposition.
\qed

We set now
\beq\label{eq:ef}
\begin{array}{rl}
E(\tau):=& E_{1}(\tau)= \frac{\rme^{\i\tau}-1}{\i\tau}
=\int_{0}^{1}\rme^{\i\tau t}dt, \\[2mm]
 F(\tau):= &E(\tau)-1=\frac{\rme^{\i\tau}-1-i\tau}{\i\tau}.
\end{array}
\eeq
From Lemma \ref{zlub} we know that $E, F\in \cN$. Moreover  $\wtilde F(\tau)= \tau F'(\tau)= \rme^{\i \tau} -E(\tau)$
so $\wtilde{F}\in\cn$.

\begin{proposition}\label{pr:estime}
Let  $S\in C^{\alpha}(A)$ for $\frac{3}{2}<\alpha<2$  and  set $S'=[S,iA]$.
Then for any number $s$ such that $1/2< s <\alpha-1$ and
any function $f$ such that $f'(\tau)=\jap{\tau}^{-2s}$ we have
\begin{equation}\label{eq:nice}
[S,\i f(\veps  A)] = \jap{\veps A}^{-s}
\left( \veps S'+O(\veps^{\alpha})\right) \jap{\veps A}^{-s}.
\end{equation}
\end{proposition}
In the usual Hilbert space setup when $A$ is a selfadjoint operator 
and $S$ is of class $C^{2}(A)$, this proposition  was
proved in \cite[Prop. 2.4]{G1} using a general commutator expansion
due to Gol\'enia and Jecko \cite{GJ}.

\proof
Since $s>1/2$ the function $f$ is bounded. 
We assume, without loss of generality,  $6\gamma<1$ and $0<\veps\leq1$.
To simplify notations
we set $g(\tau)=\jap{\tau}^{-s}$ and $f_{\veps}=f(\veps A)$,
$g_{\veps}=g(\veps A)$. Assume that we have proved that:
\begin{equation}\label{e1.8}
[S,\i f_{\veps} ] = g_{\veps}^{2} \veps  S'+
g_{\veps}O(\veps^{\alpha})g_{\veps}.
\end{equation}
 If $\beta=\alpha-1$ then $S'\in C^{\beta}(A)$ hence
from Prop. \ref{pr:est1} we get
$[g_{\veps},S']=g_{\veps}O(\veps^{\beta})g_{\veps}$. By using  
$g_{\veps}^{2} S'=g_{\veps} S'g_{\veps}+g_{\veps}[g_{\veps},S']$
we then obtain \eqref{eq:nice}. Thus 
it remains to prove (\ref{e1.8}).

As before, we first include $\veps$ in $A$, so we take $\veps=1$,
and then discuss the dependence on $\veps$. Obviously:
\begin{align}\label{eq:new1}
f(\ca_r)-f(\ca_\ell)
&=\int \rme^{\i t\ca_\ell} \frac{1}{\i t}\big( \rme^{\i t\ca}-1\big)
\what{f'}(t) dt
=\int \rme^{\i t\ca_\ell} \ca E(t\ca) \what{f'}(t) dt \\
&= \Big(  f'(\ca_{\ell})
+\int \rme^{\i t\ca_\ell} F(t\ca) \what{f'}(t) dt \Big) \ca.
\label{eq:new3}
\end{align}
Thus if $S$ is a bounded operator of class $C^{1}(A)$
we get the first order commutator expansion with remainder
\begin{equation}\label{eq:1.5}
[S,\i f(A)]=f'(A)S' + \cR(S') \quad\text{with}\quad
\cR= \int \rme^{\i t\ca_{\ell}} F(t\ca)\what{f'}(t) dt.
\end{equation}
We have $\|\rme^{\i t\ca_{\ell}}\|\leq M\rme^{\gamma|t|}$ and 
$\|F(t\ca)\|\leq C\rme^{2\gamma|t|}$ by Lemma
\ref{lm:normal}. On the other hand, $\what{f'}$ decays like
$\rme^{-|t|/2}$, so there is no convergence problem at infinity and
the integral defining $\cR(S')$ is norm convergent. Then
\begin{equation}\label{eq:1.5eps}
[S,\i f(\veps A)]=\veps f'(\veps A)S' + 
\veps \cR^{\veps}(S') \quad\text{with}\quad
\cR^{\veps}= 
\int \rme^{\i\veps t\ca_{\ell}} F(\veps t\ca)\what{f'}(t) dt
\end{equation}
and (\ref{e1.8}) follows if we prove that (recall that $\beta= \alpha-1>\12$):
\begin{equation}\label{e1.9}
\|\langle\veps A\rangle^{s}
\cR^{\veps}(T)\langle\veps A\rangle^{s}\|\leq C \varepsilon^\beta, \
T\in C^{\beta}(A), \ \12 < s<\beta, 
\end{equation} 
We shall in fact prove a stronger estimate, namely
\begin{equation}\label{e1.10}
\|(1-\i\veps A)\cR^{\veps}(T)\jap{\veps A}^{s}\|\leq C\veps^{\beta}.
\end{equation}
We set $\psi(t):= \what{f'}(t)=G_{2s}(t)$ and recall from Lemma
\ref{1.1} (4) that since $ 2s >1$: 
\begin{equation}\label{e1.11}
\e^{ c|t|}\psi', \e^{c|t|}\delta^{(k)}\psi\in L^{1}(\rr), \ 0\leq c<\12, \ k\in \nn.
\end{equation}
Using $\ca_\ell=\ca_r-\ca$ we get:
\begin{align*}
(1-\i\veps\ca_\ell)\cR^{\veps} & = \cR^{\veps}-
\int\left(\frac{d}{dt}
\rme^{\i\veps t\ca_{\ell}}\right) F(\veps t\ca)\psi(t)dt \\
& =
\int \rme^{\i\veps t\ca_{\ell}}\left( 
F(\veps t\ca)\big(t\psi(t))+\wtilde{\psi}(t)\big) +
\wtilde{F}(\veps t\ca)\psi(t) 
\right)\frac{dt}{t}  \\
& =
\int \rme^{\i\veps t\ca_r}\big( F_1(\veps t\ca)\psi_1(t) 
+F_2(\veps t\ca)\psi(t)\big)
\frac{dt}{t}
\end{align*}

where $F_1(\tau)=\rme^{-\i \tau}F(\tau), F_2(\tau)=\rme^{-\i
  \tau}\wtilde{F}(\tau)$, and
$\psi_1(t)=\big(t\psi(t)+\wtilde{\psi}(t)\big)$. By taking into
account the explicit expressions given in (\ref{eq:ef}) for
$F,\wtilde{F}$ we obtain $F_1(\tau)=F(-\tau)+ (1-\rme^{-\i   \tau})$ and
$F_2(\tau)=-F(-\tau)$. In order to justify the integration by parts
argument we have used the estimates on $\psi$ recalled in
(\ref{e1.11}).

Thus we see that 
$(1-i\veps\ca_\ell)\cR^{\veps}$ is a linear combination of terms
of the form 
$\cj_{\veps}=\int \rme^{\i \veps t\ca_r} K(\veps t\ca)\zeta(t)\frac{dt}{t}$
with $K(\tau)$ equal to one of the functions $F(-\tau)$ or $1-\rme^{-\i   \tau}$
and $\zeta(t)$ either $\psi(t)$, or $t\psi(t)$, or $\delta\psi(t)$.

In all three cases the function $\zeta$ verifies
$|\zeta(t)|+|\delta\zeta(t)|\leq C_{\mu} \rme^{-\mu|t|}$ for any
$\mu<1/2$, and $K(\tau)$ satisfies the conditions in Lemma
\ref{lm:ave}.  We now apply Lemma \ref{lm:ave} where $m>0$ may be
taken as small as we wish. This proves (\ref{e1.10}) and completes
the proof of the proposition. \qed

\section{Boundary values of resolvents}\label{s:bvr} \init

In this section we prove the main result of this paper, described in
Thm. \ref{th:bvr}. We show that if $H$ is a selfadjoint operator on
a Krein space $\cK$, satisfying a positive commutator estimate in
the Krein sense on some interval, then weighted resolvent estimates
near the real axis (analogous to the well-known Hilbert space case)
hold on this interval.

\subsection{Putnam argument and beyond}\label{ss:put}

To get a better perspective on the positive commutator methods we
make some preliminary comments in the context of a theorem due to
Putnam, see \cite{P1} or \cite[Thm. 2.2.4]{P2}.  \emph{In this
  subsection we assume that $\ch$ is a Hilbert space and $H$ is a
  selfadjoint operator on it.} We denote $\one_J(H)$ the spectral
projections of $H$ and set $R(z)=(H-z)^{-1}$.

Putnam discovered that if one may construct a (bounded) selfadjoint
operator $B$ such that $[H,\i B]\geq0$ (in form sense) then $H$ has
a rich absolutely continuous spectrum. We recall here his argument
\cite[p.\ 20]{P2}. This is the proof of the implication
$\eqref{eq:hypo}\Rightarrow\eqref{eq:put11}$ below and is very
simple but gives only an estimate on the imaginary part of the
resolvent $\Im R(z)$ for $z=\lambda+\i\mu, \mu\downarrow0$. Next we
explain how to modify it such as to control the whole resolvent
$R(z)$.

\begin{proposition} \label{pr:put} Let $B=B^{*}$ and $C$ be bounded
  operators and let us consider the following assertions:
\begin{align}
& CC^{*} \leq [H,\i B] \quad \text{ as quadratic forms on } \Dom H,
\label{eq:hypo}
\\ 
& C^{*}\one_{J}(H)C \leq \|B\|  |J| \quad \text{ for any Borel set } J,
\label{eq:put11} \\
& C^{*}\big(\Im R(z)\big)C \leq \pi \|B\| \quad \text{ for all } z 
\text{ with } \Im z>0 \label{eq:put1},
\end{align}
where $|J|$ is the Lebesgue measure of $J$.  Then we have
(\ref{eq:hypo}) $\Rightarrow$ (\ref{eq:put11})$\Leftrightarrow$
(\ref{eq:put1}).
\end{proposition} 

\proof If $J$ is an interval with midpoint $\lambda$ then
\[
\one_{J}(H)CC^{*}\one_{J}(H)\leq \one_{J}(H)[H-\lambda,\i B]\one_{J}(H)=2\Re\left(\one_{J}(H)(H-\lambda)\i B\one_{J}(H)\right),
\]
hence for any  $u\in\ch$ we have
\[
\begin{array}{rl}
\|C^{*}\one_{J}(H)u\|^{2}&\leq 2 \Re\braket{(H-\lambda)\one_{J}(H)u}{\i B\one_{J}(H))u}\\[2mm]
& \leq |J|\|\one_{J}(H)u\|\|B\one_{J}(H))u\|  \leq |J|\|B\|\|\one_{J}(H)u\|^{2}.
\end{array}
\]
This is equivalent to 
\[
\one_{J}(H)CC^{*}\one_{J}(H)\leq \|B\| |J| \one_{J}(H) \leq \|B\| |J|,
\] 
hence $\|C^{*}\one_{J}(H)\|^{2}\leq \|B\| |J|$. Obviously, if
\eqref{eq:put11} holds for intervals then it holds for any Borel
set. Note also that \eqref{eq:put11} can be stated as
$\|\one_{J}(H)C\|\leq \|B\|^{1/2} |J|^{1/2}$.

Now we prove $\eqref{eq:put1}\Leftrightarrow\eqref{eq:put11}$.  If
$E_{u}$ is the measure $E_{u}(J)=\braket{u}{\one_{J}(H)u}$ then
\begin{equation*}
\frac{1}{\pi}\Im \braket{u}{R(\lambda+\i\mu)u}=\frac{1}{\pi}
\int \frac{\mu}{(x-\lambda)^{2}+\mu^{2}} d E_{u}(x).
\end{equation*}
Now clearly $\Im\braket{u}{R(z)u}\leq \pi M$ holds for all $z$ with
$\Im z> 0$ if and only if $E_{u}$ is an absolutely continuous
measure with derivative $E'_{u}(\lambda)\leq M$ for a.e.  $\lambda$.
\qed

\begin{remark}{\rm The relation \eqref{eq:put1} says that the
    imaginary part of the holomorphic function $C^{*}R(z)C$ in $\Im
    z>0$ is bounded, and this is equivalent to the boundedness of
    the boundary value $C^{*}\big(\Im
    R(\lambda+i0)\big)C$. Unfortunately, from the boundedness of the
    imaginary part of a function holomorphic in the upper half-plane
    it is not possible to deduce the boundedness of the real part,
    hence of the function, because the Hilbert transform is not
    bounded in $L^{\infty}(\rr)$. However, if $C^{*}\big(\Im
    R(\lambda+i0)\big)C$ is a H\"older continuous function of
    $\lambda$ on a real open set $J$, then $C^{*}R(z)C$ extends to a
    H\"older continuous function on the union of the upper
    half-plane and $J$.  }\end{remark}

We now modify Putnam's argument such as to estimate $C^{*}R(z)C$ and
not only the imaginary part. This is related to the \emph{energy
  estimate} as presented in \cite{G1}.

\begin{proposition}\label{pr:putd}
  Let $B=B^{*}$ and $C,D$ be bounded operators with $BC=CD$ and
\begin{equation}\label{eq:hyp}
CC^{*} \leq [H,\i B] \quad \text{ as quadratic forms on } \Dom H.
\end{equation}
Then we have
\begin{equation}\label{eq:put12}
  \|C^{*}R(z)C\| \leq 2 (\|B\|+\|D\|) \quad \text{if } \Im z\neq0.
\end{equation}
A bounded operator $D$ such that $BC=CD$ exists if and only if $B$
leaves the range of $C$ invariant.
\end{proposition}
\proof Let $\Im z>0$ and $b=-\|B\|$ (if $\Im z<0$ let
$b=\|B\|$). Denote $R=R(z)$ and $L=C^*RC$. Then
\begin{align*}
L^*L &= C^*R^*CC^*RC \leq C^*R^*[H,\i B] RC 
= C^*R^*[H-z,\i (B+b)] RC \\
& = C^*R^*(H-z)\i (B+b) RC - C^*R^*\i (B+b) (H-z) RC \\
& = C^*\i (B+b) RC + C^*R^*(\bar{z}-z)\i (B+b) RC
- C^*R^*\i (B+b) C \\
& =2\Im\big( C^* R^* (B+b) C\big) + C^*R^*(2\Im z)(B+b) RC \\
& =2\Im\big( C^* R^*C (D+b)\big) + C^*R^*(2\Im z)(B+b) RC.
\end{align*}
Since $(2\Im z)(B+b)\leq0$ we get with $\alpha=\|L\|/\|D+b\|$:
$$
L^*L\leq 2\Im\big(L (D+b)\big)\leq \alpha L^*L +\alpha^{-1} (D+b)^2
\leq\alpha \|L\|^2 +\alpha^{-1} \|D+b\|^2
=2\|L\|\|D+b\|
$$
which is better than \eqref{eq:put12}. For the last assertion note
that by the closed graph theorem we may take $D=C^{-1}_{0}BC$ with
$C_{0}=C|(\Ker{C})^\perp$, cf.  \cite[Thm. 1]{Do}.  \qed

Prop. \ref{pr:putd} and ideas from \cite{G1} give the
following extension of Mourre's theorem \cite{Mo}.

\begin{theoreme}\label{th:mourre}
  Let $A$ be a selfadjoint operator on the Hilbert space $\ch$ such
  that $H$ is of class $C^\alpha(A)$ for some $\alpha>3/2$ and let
  $I$ be a real bounded open interval such that 
  \[
  E(I)[H,\i
  A]E(I)\geq a E(I)
  \] for some number $a>0$. Then for each compact
  interval $J\subset I$ and each $s>1/2$ there is a number $C$ such
  that
  \begin{equation}\label{eq:bvrh}
  \|\jap{A}^{-s} R(z)\jap{A}^{-s}\|\leq C \quad\text{if \ } \Re z\in J
  \text{ and \ } \Im z\neq0. 
  \end{equation} 
  If some $\phi\in\coinf(\R)$ with $\phi(\lambda)=\lambda$ near $I$ 
  is fixed, then $C$ depends only on $a$ and on an upper bound for
  the $C^{\alpha}(A)$ norm of $\phi(H)$.   
\end{theoreme}

We sketch only the main idea of the proof to explain the r\^ole of 
Prop. \ref{pr:putd}; details are given in a more general context 
in Subsect.  \ref{ss:bvestimate}. Note that it suffices to prove
$\sup_{z\nin\rr}\|\jap{A}^{-s} R(z)\xi(H)^{2}\jap{A}^{-s}\|\leq C$
if $\xi\in \coinf(I)$ real. Clearly one may replace here $A$ by
$\veps A$ with $\veps>0$. Let $f$ be a function with
$f'(\tau)=\jap{\tau}^{-2s}$. Then \eqref{eq:hyp} is
satisfied by $B=\frac{2}{a\veps}\xi(H)f(\veps A)\xi(H)$
and $C=\xi(H)\jap{\veps A}^{-s}$ if $\veps$ is small and $1/2<s<1$.

\begin{remark}\label{re:surprise}{\rm 
	In \cite{Mo} it is assumed that $\alpha=2$ and $\e^{\i
  	tA}\Dom H=\Dom H$ for all $t$. The extension from $C^2(A)$ to
    $C^\alpha(A)$ with $\alpha>3/2$ is not really significant in
    applications ($\alpha>1$ is the natural condition and such an
    improvement would be practically relevant). We included, however, this
    generalization because it is rather surprising that the method of
    \cite{G1} allows one to pass from the class $C^2(A)$ to
    the class $C^\alpha(A)$ with $\alpha>3/2$ without any change in
    the strategy of the proof. Indeed, the case $\alpha>1$ as
    treated in \cite{ABG} requires a rather substantial modification
    of the ``method of differential inequalities'' of Mourre, while here 
    the restriction $\alpha>3/2$ comes only from the proof of
    \eqref{eq:nice}.  
}\end{remark}

\subsection{Positive commutators in Krein spaces}

We now extend the techniques and results of Subsect. \ref{ss:put} to
the Krein space setting. We begin with a Putnam type assertion.

\begin{proposition}\label{pr:putnam-krein}
  Let $H$ be a selfadjoint operator with $\rho(H)\neq\emptyset$ on the
  $K-$space $\ch$.  Let $\Pi$ be a positive projection which commutes
  with $H$ and let $B,C,D$ be bounded operators such that
  \[
\begin{array}{rl}
  (1)&  B= B^{*}, \ C=\Pi C,\\[2mm]
(2)& BC= CD, \\[2mm]   
(3)&     CC^{*} \leq\Pi[H,\i B]\Pi \hbox{ as quadratic forms on }\Dom H.
\end{array}
\] 
Then the operator $L(z)=C^{*}R(z)C$ satisfies
\[ 
\braket{L(z)u}{L(z)u}\leq c (\|B\|+\|D\|)\|L(z)u\|\|u\|
\quad\text{for }  \ u \in\ch, \ z\in \rho(H), 
\]  where $c$ depends only on $H$ and 
  $\Pi$.
\end{proposition}

\proof Set $R=R(z)$, $L= L(z)$ and assume that $\Im z\geq 0$ (the
proof is similar $\Im z\leq 0$).  Note that if $z\in\rho(H)$ then $\overline{z}\in \rho(H)$
 and $R^{*}=(H-\bar{z})^{-1}$. For $b\in\R$ we
have:
\begin{align*}
R^{*}[H,\i B]R & = R^{*}[H-z,\i (B+b)]R = 
\i (B+b)R - R^{*} \i(B+b) +(2\Im z) R^{*}(B+b)R \\
& = 2{\Im}\big(R^{*}(B+b)\big) + (2{\Im} z) R^{*}(B+b)R.
\end{align*}
Since $(B+b)C=C(D+b)$ we get
\begin{equation}\label{e4.2}
C^{*}R^{*}[H,\i B]RC = 
2{\Im}\big(C^{*}R^{*}C(D+b)\big) + (2{\Im} z) C^{*}R^{*}(B+b)RC.
\end{equation}
Since $C=\Pi C$ and $\Pi$ commutes with $H$ we have
\[
C^{*}R^{*}(B+b)RC=C^{*}R^{*}\Pi (B+b)\Pi RC.
\]
Using (\ref{e2.1}) we may choose $b=-\|B\|_{\Pi}$ such that \(
(2{\Im} z) C^{*}R^{*}(B+b)RC\leq 0, \) hence from (\ref{e4.2}) we
get:
\[
C^{*}R^{*}[H,\i B]RC \leq 2{\Im}\big(L^{*}(D+b)\big).
\] 
Now observe that $C^{*}R^{*}[H,\i B]RC=C^{*}R^{*}\Pi[H,\i B]\Pi
RC$ hence from hypothesis (3), we get
$$
L^*L=C^{*}R^{*}CC^{*}RC \leq 2{\rm Im}\big(L^*(D+b)\big).
$$
This yields for $u\in\ch$, with a constant $m$ depending only on
$\ch$:
$$
\braket{Lu}{Lu}\leq 2{\rm Im} \braket{Lu}{(D+b)u}
\leq m\|Lu\|\|(D+b)u\|\leq m\|Lu\|(\|D\| +\|B\|_{\Pi})\|u\|,
$$
using that $b= - \|B\|_{\Pi}$.  Since $\|B\|_{\Pi}\leq d \| B\|$,
for some constant $d$ depending only on $ \Pi$, this gives the
required estimate for $c= \max(m, md)$. \qed

\begin{remark}\label{rm:D}{\rm If $\ch$ is a Krein space then there
    is a bounded operator $D$ such that hypothesis (2) in
    Prop. \ref{pr:putnam-krein} is satisfied if and only if $B$
    leaves the range of $C$ invariant, cf.\ \cite[Thm. 1]{Do}.
    Indeed, since $\ch$ is Hilbertizable, we may choose a closed
    subspace $\ck$ in $\ch$ such that $\ch=\Ker C\oplus\ck$; then
    take $D=C^{-1}_{0}BC$ where $C_{0}=C|\ck$.  }\end{remark}

\begin{corollary}\label{co:putnam-krein}
  Let $\ch$ be a Krein space and $\Pi$ a positive projection which
  commutes with $H$. Assume that $B,C$ are bounded operators with
  $B=B^{*},C=\Pi C$, and such that $B$ leaves invariant the range of
  $C$.  If the inequality $\Pi[H,iB]\Pi\geq CC^{*}$ holds in
  quadratic form sense on $\Dom H$ and if we set $L(z)=C^{*}R(z)C$
  then $\braket{L(z)u}{L(z)u}\leq c\|L(z)u\|\|u\|\,\forall u\in\ch$,
  where the number $c$ depends only on $\Pi,B,C$.
\end{corollary}

\subsection{Boundary value estimates}\label{ss:bvestimate}

We refer to Definition \ref{df:bv} for the open real set $\beta(H)$
on which $H$ admits a smooth functional calculus. For example, if
$H$ is a definitizable operator on a Krein space then by Proposition
\ref{pr:reso} we have $\beta(H)=\R$.

The following theorem is the main result of our work. 
\begin{theoreme}\label{th:bvr}
  Let $\ch$ be a Krein space and $A$ the generator of a $C_0$-group
on $\ch$ such that the Krein structure is of class
  $C^1(A)$. Let $H$ be a selfadjoint operator on $\ch$ and $\Pi$ a
  positive projection which commutes with $H$ such that the
  following conditions are satisfied:
\begin{compactenum}

\item $H$ is of class $C^\alpha(A)$ for some $\alpha>3/2$, in
  particular $H'=[H,\i A]$ is well defined;

\item there is $\varphi\in\coinf(\beta(H))$ real with $\varphi(\lambda)=1$
  on a neighborhood of a compact interval $J$ such that $\varphi(H)\Pi=\varphi(H)$ and:
  \beq\label{mourre}
   \varphi(H)(\Re H')\varphi(H)\geq a
  \varphi(H)^2, \ a>0.
  \eeq

\end{compactenum}
Then if $s>1/2$ and $\varepsilon>0$ is small enough, we have
\begin{equation}\label{eq:bvr}
{\textstyle\sup_{J\pm \i ]0, \nu]}}\|\jap{\varepsilon A}^{-s} R(z)\jap{\varepsilon
  A}^{-s}\|<\infty, \hbox{ for some }\nu>0.
\end{equation}
\end{theoreme}

Even though our framework is much more general than the familiar
Hilbertian one, we will adopt the usual terminology and call an
estimate like (\ref{mourre}) a {\em Mourre estimate}.

\begin{remark}\label{remi}
 {\rm In applications one often assumes that $H$ admits a Borel functional calculus on an interval $I\supset J$ and that  $\Pi= \one_{I}(H)$. If $\one_{I}(H)\leq 0$ then  the assumption  (\ref{mourre})  should be replaced by 
 \[
\varphi(H)(\Re H')\varphi(H)\leq a\varphi^{2}(H), \ a>0.
\]
Multiplying  the Krein structure by $-1$ one is then reduced to the situation of the theorem. }
\end{remark}
\proof Let $I$ be  a neighborhood of $J$ on which
$\varphi(\lambda)=1$.  We notice that it suffices to show
$$
\sup_{z\nin\rr}\|\jap{\varepsilon A}^{-s}
R(z)\xi(H)^2\jap{\varepsilon A}^{-s}\|<\infty
$$
for each real $\xi\in\coinf(I)$. Indeed, if $\Re z\in J$ and we
choose $\xi$ such that $0\leq\xi\leq1$ and $\xi(\lambda)=1$ when
$\lambda$ is at distance less then $\nu$ of $J$, then
$R(z)=R(z)\xi(H)^2+R(z)(1-\xi(H)^2)$ and
$\|R(z)(1-\xi(H)^2)\|\leq\nu^{-k}$ for some finite number $k$.

Clearly we may assume $s<\beta=\alpha-1<1$. We shall use the
notations introduced in the proof of Prop. \ref{pr:estime}:
$g(\tau)=\jap{\tau}^{-s}$, $f$ is a function such that $f'=g^2$, and
$g_\varepsilon=g(\varepsilon A), f_\varepsilon=f(\varepsilon A)$.
Note that $f_\varepsilon$ is a bounded operator by Prop.
\ref{pr:bdd}.  For Greek letters $\xi, \eta$, etc, we often adopt
the abbreviations $\eta\equiv\eta(H), \xi=\xi(H)$, etc. 

 If
$X_\varepsilon,Y_\varepsilon$ are bounded operators defined for
small $\veps$ we write $X_\varepsilon\sim Y_\veps$ if
$X_\varepsilon-Y_\varepsilon=g_\varepsilon
O(\varepsilon^\beta)g_\varepsilon^*$ and $X_\varepsilon\prec
Y_\varepsilon$  if
$X_\varepsilon-Y_\varepsilon\leq g_\varepsilon
O(\varepsilon^\beta)g_\varepsilon^*$. For example, Prop.
\ref{pr:est2} gives $g_\varepsilon\sim g_\varepsilon^*$ and from
Prop. \ref{pr:est1} we obtain $\xi g_\varepsilon \sim
g_\varepsilon\xi$ if $\xi\in\coinf(\beta(S))$.

Fix $\phi\in C_0^\infty(\R)$ real such that $\phi(\lambda)=1$ on a
neighborhood of the support of $\varphi$ and set $S=\phi(H)$. Then
$S$ is a bounded symmetric operator of class $C^\alpha(A)$ and we
have $\eta S'\eta =\eta H'\eta$ for all $\eta\in C_0^\infty(I)$.
From Prop. \ref{pr:estime} we get $[S,\i\varepsilon^{-1}f_{\veps}] \sim g_{\veps} S' g_\varepsilon^*$,
 hence if we denote
$F_\varepsilon=\varepsilon^{-1}\Re f_{\veps}$ we obtain:
\[
[S,\i
F_\varepsilon] \sim g_{\veps} (\Re S')g_\varepsilon^*.
\]
 Then if
$\eta\in\coinf(I)$ we get: 
\[
\begin{array}{rl}
[S,\i\eta F_\varepsilon \eta] &\sim \eta g_{\veps} (\Re
S')g_\varepsilon^*\eta \sim g_{\veps} \eta (\Re S')\eta
g_\varepsilon^* \\[2mm]
&= g_{\veps} \eta (\Re H')\eta g_\varepsilon^* \succ
a g_{\veps} \eta^2 g_\varepsilon^* \sim a \eta g_{\veps}
g_\varepsilon^* \eta.
\end{array}
\]If $\eta$ is chosen such that $\xi\eta=\xi$ then we get finally 
\[
[S,\i\xi F_\varepsilon \xi] \geq \frac{a}{2} \xi g_{\veps}
g_\varepsilon^* \xi
\]
 for $\varepsilon$ small enough.

In Prop. \ref{pr:putnam-krein} we take $B=\xi F_\varepsilon
\xi$ and $C=\xi g_{\veps}$. Observe that
$\xi\Pi=\xi\varphi\Pi=\xi\varphi=\xi$ hence, by taking adjoints,
$\Pi\xi=\xi\Pi=\xi$. To find $D$ we note that $BC=CD$ means $\xi
F_\varepsilon\xi^2 g_\varepsilon= \xi g_\varepsilon D$ hence follows
from $F_\varepsilon\xi^2 g_\varepsilon= g_\varepsilon D$ so it
suffices to take $D=g_\varepsilon^{-1}F_\varepsilon\xi^2
g_\varepsilon$. This is a bounded operator because $\xi$ is of class
$C^1(A)$ and $0<s<1$, so $F_\varepsilon\xi^2$ leaves invariant the
range of $g_\varepsilon$. Now we apply Prop.
\ref{pr:putnam-krein} and obtain
\[
\begin{array}{rl}
\braket{L_\varepsilon u}{L_\varepsilon u}\leq& K
(\|B_\varepsilon \|+\|D_\varepsilon \|)\|L_\varepsilon u\|\|u\|\\[2mm]
\leq& \delta \|L_\varepsilon u\|^2+
(4\delta)^{-1} (\|B_\varepsilon \|+\|D_\varepsilon \|)^2 \|u\|^2, \  u\in \cH,
\end{array}
\]
for some  $\delta>0$, where we have
indicated the dependence in $\varepsilon $ for clarity, in
particular $L_\varepsilon=g_\varepsilon^*\xi^2 R g_\varepsilon$. We
write this as 
\[
 \braket{L_\varepsilon u}{L_\varepsilon u} \leq
\delta \|L_\varepsilon u\|^2+ c \|u\|^2,
\] where
$c=c(\delta,\varepsilon)$.  With the notation $\eta_\perp=1-\eta$ we
have $\xi\eta_\perp=0$ hence
$$
\eta_\perp L_\varepsilon  
=\eta_\perp g_\varepsilon^*\xi^2 R g_\varepsilon
=[g_\varepsilon^*,\eta]\xi^2 R g_\varepsilon
=g_\varepsilon^* O(\varepsilon)g_\varepsilon^* \xi^2 R g_\varepsilon
=O(\varepsilon)L_\varepsilon. 
$$
Thus we have $\eta L_\varepsilon = L_\varepsilon -\eta^\perp
L_\varepsilon = L_\varepsilon+O(\varepsilon) L_\varepsilon$.  Since
the projection $\Pi$ is positive, there is a constant $N$ such that
$N^{-1}\|v\|^2\leq\braket{v}{v}$ for $v\in\Pi\ch$. Thus from
$\eta=\Pi\eta$ we get:
\begin{align*}
N^{-1}\|\eta L_\varepsilon u\|^2 & \leq
\braket{\eta L_\varepsilon u}{\eta L_\varepsilon u}
= \braket{L_\varepsilon u+O(\varepsilon) L_\varepsilon
  u}{L_\varepsilon u+O(\varepsilon) L_\varepsilon u}  \\
& \leq \braket{L_\varepsilon u}{L_\varepsilon u} 
+ O(\varepsilon) \|L_\varepsilon u\|^2
\leq (\delta+ O(\varepsilon)) \|L_\varepsilon u\|^2+ 
c(\delta,\varepsilon) \|u\|^2.
\end{align*}
But $L_\varepsilon = \eta L_\varepsilon
+O(\varepsilon)L_\varepsilon$ hence
$(1-O(\varepsilon))\|L_\varepsilon u\|\leq \|\eta L_\varepsilon u\|
$. Inserting this above we get for $\varepsilon$ small enough the
estimate 
\[
\|L_\varepsilon u\|^2 \leq 2N(\delta+ O(\varepsilon))
\|L_\varepsilon u\|^2+ 2Nc(\delta,\varepsilon) \|u\|^2.
\]
 Finally,
taking both $\delta$ and $\veps$ small we obtain $\|L_\varepsilon
u\| \leq C \|u\|$ for some constant $C$. Thus
$\|g_\varepsilon^*\xi^2 R g_\varepsilon u\| \leq C\|u\|$ and
\eqref{eq:soc} gives $\|g_\varepsilon\xi^2 R g_\varepsilon u\| \leq
C \|u\|$.  \qed

\begin{remark}\label{re:bdd}{\rm We were forced to ask $\ch$ to be a
    Krein space, and not an arbitrary $K-$space, only because of
    hilbertizability assumption in Prop.
    \ref{pr:bdd}. }\end{remark}

\subsection{Virial theorem}\label{ss:virial}

In order to check the positive commutator estimate (\ref{mourre}),
one needs to extend to $K$-spaces some facts related to the {\em
  virial theorem}. We do this in this subsection.  Let $H$ be a
selfadjoint operator in a $K-$space with a not empty resolvent
set. \emph{In all this subsection we fix an open real set $I$ on
  which $H$ admits a $C^0$-functional calculus.}

Then, as shown in Thm. \ref{th:Borel}, the calculus extends to a
bounded Borel functional calculus on $I$, so $\varphi(H)$ is well
defined if $\vphi$ is a bounded Borel function on $I$.

\begin{lemma}\label{lm:evin}

If $\lambda\in I$  then
$\one_{\{\lambda\}}(H)$ is the orthogonal  projection onto $\Ker(H-\lambda)$.
\end{lemma}

\proof $\one_{\{\lambda\}}(H)$ is a projection because
$\one_{\{\lambda\}}^2=\one_{\{\lambda\}}$. Recall that $r_{z}$ for  $z\in \rho(H)$ is the function $r_{z}(x)= (x-z)^{-1}$. Then
$r_z(H)=R(z)$ and clearly $\Ker(H-\lambda)$ is exactly the set of
vectors $u\in\ch$ such that $r_z(H)u=r_z(\lambda)u$. Since the Borel
functional calculus is multiplicative we have
$$
r_z(H)\one_{\{\lambda\}}(H)=\one_{\{\lambda\}}(H)r_z(H)=(\one_{\{\lambda\}} r_z)(H)=(\one_{\{\lambda\}}
r_z(\lambda))(H)= r_z(\lambda)\one_{\{\lambda\}}(H)  .
$$
Thus $\one_{\{\lambda\}}(H)\ch\subset\Ker(H-\lambda)$. Reciprocally, if
$u\in\Ker(H-\lambda)$ then $r_z(H)u=r_z(\lambda)u$ hence
$\varphi(H)u=\varphi(\lambda)u$ for any rational function with poles
only in the resolvent set of $H$. From \eqref{eq:sfc} for example,
we then get $\varphi(H)u=\varphi(\lambda)u$ for any
$\varphi\in\coinf(I)$, and finally by taking limits we get it for
any bounded Borel function on $I$. In particular $\one_{\{\lambda\}}(H)u=u$. \qed

Now let $A$ be the generator of a $C^0$-group such that $H$ is of
class $C^1(A)$. If we interpret $H'=[H,iA]$ as a sesquilinear form on
$\Dom H$, then we have the following \emph{virial theorem}.

\begin{lemma}\label{lm:virial}
For any $\lambda\in I$ we have
$\one_{\{\lambda\}}(H)H'\one_{\{\lambda\}}(H)=0$. 
\end{lemma}

\proof Let $z\in\rho(H)$ and $R=(z-H)^{-1}$. Then
$R'\equiv[R,\i A]=RH'R$ and for any bounded Borel $\varphi$ with
support in $I$ we get $\varphi(H)H'\varphi(H)=
\varphi_z(H)R'\varphi_z(H)$ with
$\varphi_z(x)=\varphi(x)(z-x)$. Thus we have:
\[
\one_{\{\lambda\}}(H)H'\one_{\{\lambda\}}(H)=
(z-\lambda)^2\one_{\{\lambda\}}(H)R'\one_{\{\lambda\}}(H)=
(z-\lambda)^2\lim_{\tau\to0}\one_{\{\lambda\}}(H)[R,A_\tau]\one_{\{\lambda\}}(H)
\] 
where $A_\tau=(\rme^{\i\tau A}-1)/\tau$. Since
\[
\begin{array}{rl}
&\one_{\{\lambda\}}(H)[R,A_\tau]\one_{\{\lambda\}}(H) =
\one_{\{\lambda\}}(H)RA_\tau\one_{\{\lambda\}}(H)
-\one_{\{\lambda\}}(H)A_\tau R\one_{\{\lambda\}}(H)\\[2mm]
 =&\one_{\{\lambda\}}(H)(z-\lambda)A_\tau\one_{\{\lambda\}}(H)
-\one_{\{\lambda\}}(H)A_\tau (z-\lambda)\one_{\{\lambda\}}(H)=0,
\end{array}
\]we get the required result.
\qed

\begin{corollary}\label{co:virial}Let $H$ be a selfadjoint operator
  on the Krein space $\cH$ and let $I\subset \beta(H)$. 
  Assume that for some $J\subset I$ we have $\one_J(H)\geq 0$ and
  that there is a number $a>0$ and a compact operator $K$ such that
  \[
  \one_J(H) H' \one_J(H) \geq a\one_J(H)+K.
  \] Then the point
  spectrum of $H$ in $J$ is finite and consists of eigenvalues of
  finite multiplicity.
   Moreover, if $\lambda\in J$ is not an
  eigenvalue of $H$ and $b<a$ then there is a compact neighborhood
  $I$ of $\lambda$ in $J$ such that
\[\one_I(H) H' \one_I(H) \geq
  b\one_I(H).
  \]
\end{corollary}
\proof The range of $\one_J(H)$ is a Hilbert space (for the induced
Krein structure) stable under $H$, so the usual proof (see
e.g. \cite{Mo}) applies. \qed

We shall need one more technical fact for applications in Section
\ref{kgsect}. We write $S\simeq T$ if $S,T$ are operators and $S-T$
is compact.  Recall that $ C^\alpha(A)\subset C^1_\rmu(A)$ for
$\alpha>1$.

\begin{lemma}\label{lm:cmp}
  Assume $H\in C^1_\rmu(A)$. Let $H_0$ be a second operator (not
  necessarily selfadjoint) of class $C^1_\rmu(A)$ such that
  $(H-z)^{-1}\simeq(H_0-z)^{-1}$ for some
  $z\in\rho(H)\cap\rho(H_0)$. If $H_0$ admits a smooth functional
  calculus on $J$ then for any $\varphi\in C^\infty_{0}(J)$ we have
  $\varphi(H)H'\varphi(H)\simeq\varphi(H_0)H'_0\varphi(H_0)$.
\end{lemma}
\proof
Let $R=(z-H)^{-1}, R_0=(z-H_0)^{-1}$, and $\varphi_z$ as above. Then
\[
\varphi(H)H'\varphi(H)-\varphi(H_0)H'_0\varphi(H_0)=
\varphi_z(H)R'\varphi_z(H)-\varphi_z(H_0)R'_0\varphi_z(H_0).
\]
The operator $R'-R'_0$ is compact as norm limit of compact
operators, using that $H, H_{0}\in C^{1}_{\rm u}(A)$, and $\varphi_z(H)-\varphi_z(H_0)$ is compact by a
standard argument.
\qed

\section{Klein-Gordon operators}\init\label{kgsect}

In this section we discuss various Krein spaces and operators on
them associated to the following {\em abstract Klein-Gordon
  equation}:
\begin{equation}
\label{kigi}
\p_{t}^{2}\phi(t)- 2\i k\p_{t}\phi(t)+ h\phi(t)=0,
\end{equation}
where $\phi: \rr\to \cH$, $\cH$ is a Hilbert space and $h$, $k$ are
selfadjoint, resp. symmetric operators on $\cH$.
Our main references in connection with the
  spectral theory of the Klein-Gordon equation in the Krein space
  framework are the papers \cite{LNT1,LNT2}, where one may also find
  a thorough discussion of the previous works on this subject. But
  our approach to these questions is rather different and, of
  course, our main purpose is to study the boundary behavior of
  resolvents of operators associated to these equations.

We first introduce an abstract setting which
  allows one to treat in a unified way the {\em charge} and {\em
    energy} versions of the Klein-Gordon operators. More precisely,
  the energy space $\re$ (with the norm topology) is equipped with a
  hermitian form (the charge) which is not a Krein structure but
  allows one to embed $\re$ into its adjoint space $\re^*$, so that
  $\re\subset\re^*$ densely and continuously, see \eqref{eq:rk} and
  \eqref{eq:kphsp}.  Then we define the \emph{maximal} Klein-Gordon
  operator $K_{\text{max}}$ as a closed unbounded operator in the
  space $\re^*$, this is the operator \eqref{eq:kg0} considered as
  operator in $\re^*$ with domain $\re$. In Subsect.\ \ref{ss:ukg}
  we show that the Klein-Gordon operators usually considered in the
  literature are restrictions $L$ of $K_{\text{max}}$ to spaces
  $\rl$ such that $\re\subset\rl\subset\re^*$ continuously and
  densely. For example the energy space is $\re$ (and is a priori
  not a Krein space) and the usual energy Klein-Gordon operator is
  isomorphic to the operator $K_{\text{min}}$ induced by
  $K_{\text{max}}$ in $\re$.  The other realizations obviously
  verify $K_{\text{min}}\subset L \subset K_{\text{max}}$ in
  operator sense. For example, the charge space $\rh_\theta$ is an
  intermediate space between $\re$ and $\re^*$ (but, in general, not
  an interpolation space) and the corresponding charge Klein-Gordon
  operator $K_\theta$ is just the restriction of $K_{\text{max}}$
  to $\rh_\theta$. We study these operators in some detail, in fact
  in this framework it is easy to show that they have the same
  spectrum and it is also quite straightforward to describe their
  domains. The natural Krein space structure on the charge spaces
  does not play any role here.
 
We emphasize that these results are valid under
  quite general conditions: we assume (A1), which is clearly needed
  to give a meaning to the Klein-Gordon operators, and (A2), without
  which the resolvents of these operators would be empty. Moreover,
  by Lemma \ref{lm:idiot} the condition (A2) is automatically
  satisfied if $h$ is bounded from below, which is a quite weak
  assumption (the Stark effect model treated in \cite{V2} seems to
  be the only physically interesting case with $h$ not bounded from
  below).

In Subsect.\ \ref{ss:free} we then study the
  functional calculus of the various {\em free } Klein-Gordon
  operators, which corresponds to the case $k=0$ in (\ref{kigi}).
  Finally, in Subsect.\ \ref{ss:cops} we introduce some abstract
  conditions under which a Mourre estimate can be shown for the
  charge Klein-Gordon operator.  This section contains only our
  simplest application of Theorem \ref{th:bvr} and is somewhat
  complementary to our paper \cite{GGH1}, where resolvent estimates
  for energy Klein-Gordon operators are obtained, although the
  method to obtain a Mourre estimate is quite different.

\subsection{Notations}\label{notato}

We need some new notations and terminology.

{\em Linear operators}

We write $f:X\tilde\to Y$ if $X,Y$ are sets and $f:X\to Y$ is
bijective.  If $X,Y,Z$ are Banach spaces with $X\subset Y\subset Z$
continuously and densely then to each continuous operator $S:X\to Z$
we associate a densely defined operator $\what S$ acting in $Y$,
namely the restriction of $S$ to the domain $\Dom \what
S=S^{-1}(Y)$. We say that $\what S$ is the operator induced by $S$
in $Y$ and use the same notation for $S$ and $\what S$ unless this
abuse of notations leads to confusions.

{\em Scale of Sobolev  spaces}

Let  $\cH$  be a Hilbert space with norm $\|\cdot\|$ and scalar
product $(\cdot| \cdot)$. We identify $\cH$ with its adjoint
space $\cH^{*}=\cH$ via the Riesz isomorphism.   Let $h$ be a selfadjoint operator on $\cH$. 

We can associate to it the 
{\em non-homogeneous Sobolev spaces}
\[
 \langle h\rangle^{-s}\cH:= \Dom |h|^{s}, \   \langle h\rangle^{s}\cH:= ( \langle h\rangle^{-s}\cH)^{*}, \ s\geq 0.
\]
The spaces $ \langle h\rangle^{-s}\cH$ are equipped with the graph norm $\| \langle h\rangle^{s}u\|$.
We will use the notation 
\[
(u| v), \ u\in \langle h\rangle^{-s}\cH, \ v\in \langle h\rangle^{s}\cH, \ s\geq 0,
\]
 to denote the duality bracket between $\jh^{-s}\cH$ and $\jh^{s}\cH$.
\def\jh{\langle h\rangle}

\subsection{Quadratic pencils}\label{s:pencil}

We fix a Hilbert space $\ch$ with $\ch^*=\ch$ and consider two
operators $h,k$ such that:
\[
\leqno{\mbox{(A1)}} \hspace{1mm}
\left\{ 
\hspace{-4mm}
\begin{array}{ll}
& h\hbox{ is selfadjoint on }\cH,\\[1mm]
&k\in B(\jh^{-\12}\cH, \cH)\hbox{ is symmetric.}
\end{array}
\right.
\]

The unique continuous extension of $k: \cH\to \jh^{\12}\cH$ will still be denoted by $k$.

We set also:
 \begin{itemize}
\item[]$h_0:=h+k^2:\jh^{-\12}\cH\to\jh^{\12}\cH$,
 \item[] $p(z)= h+z(2k-z)=h_0-(k-z)^2 : \jh^{-\12}\cH\to\jh^{\12}\cH$ for $z\in\C$.
\end{itemize}
The map $z\mapsto p(z)$ is called a {\em quadratic pencil}.

Note that formally   $\phi(t)= \e^{\i z t}\phi$ solves  the Klein-Gordon equation (\ref{kigi}) iff $p(z)\phi=0$.

Obviously $p(z)$ is also a well defined operator in $B(\jh^{-1}\cH,\ch)$
and $B(\ch,\jh\cH)$. Moreover, \emph{the domain in $\ch$ of the 
operator $p(z):\jh^{-\12}\cH\to\jh^{\12}\cH$ is precisely $\jh^{-1}\cH$, i.e.
$\jh^{-1}\cH=p(z)^{-1}\ch$}. Indeed, for $u\in\jh^{-\12}\cH$ we have
$p(z)u=hu+z(2k-z)u$ and the last term belongs to $\ch$, hence
$p(z)u\in\ch$ if and only if $hu\in\ch$.

Clearly $p(z)^{*}=p(\bar{z})$ in $B(\jh^{-\12}\cH,\jh^{\12}\cH)$. We shall
prove below that this relation also holds for the operators in $\ch$
induced by $p(z)$ and $p(\bar{z})$.

\begin{lemma}\label{lm:kg} Assume (A1). Then 
  the operator induced by $p(z)$ in $\ch$ is a closed operator
  and its Hilbert space adjoint is the operator induced by
  $p(\bar{z})$ in $\ch$. In other terms, the relation
  $p(z)^*=p(\bar{z})$ also holds in the sense of closed
  operators in $\ch$. The following six conditions are equivalent: \\[1mm]

\begin{tabular}{ll}
(1) $p(z):\jh^{-1}\cH\tilde\to\ch$; &(2) $p(\bar{z}):\jh^{-1}\cH\tilde\to\ch$; \\
(3) $p(z):\ch\tilde\to\jh\cH$;& (4) $p(\bar{z}):\ch\tilde\to\jh\cH$;\\
(5) $p(z):\jh^{-\12}\cH\to\jh^{\12}\cH$; &(6) $p(\bar{z}):\jh^{-\12}\cH\to\jh^{\12}\cH$.
\end{tabular}\\[1mm]
In particular, the set 
\begin{equation}\label{eq:rhohk}
\rho(h,k):=\{z\in\C \mid p(z):\jh^{-\12}\cH\tilde\to\jh^{\12}\cH\}
=\{z\in\C \mid p(z):\jh^{-1}\cH\tilde\to\ch\}
\end{equation}
is invariant under conjugation.
\end{lemma}
\proof If we set $\ell=\zeta(2k-\zeta)\in B(\h^{-\12}\cH,\cH)$ then
$\ell:\h^{-\12}\cH\to\cH$ and its adjoint in $\cH$ satisfies $\ell^{*}\supset
\bar{\zeta}(2k-\bar{\zeta})\in B(\h^{-\12}\cH,\cH)$.  In particular, $\ell$
and $\ell^*$ are $h$-bounded with relative bound zero, hence there is a
real number $n$ such that $\|\ell(h+\i n)^{-1}\|<1$ and
$\|\ell^*(h-\i n)^{-1}\|<1$. From $h+\ell+ \i n=\big(1+\ell(h+\i n)^{-1}\big)
(h+\i n)$ it follows that $h+\ell+\i n:\cH^1\tilde\to\cH$ from which we get
that $(h+\ell+\i n)^*$ is a bijection from its domain onto $\cH$, see e.g. \cite[Thms. 4.17, 5.12]{Weid}.

Clearly $(h+\ell+\i n)^*\supset h+\ell^*-\i n$,  and
an argument similar to that already used implies
$h+\ell^*-\i n:\h^{-1}\cH\tilde\to\cH$. Thus $(h+\ell)^*=h+\ell^*$ which means
$p(\zeta)^*=p(\bar{\zeta})$. 

Now the equivalence $p(\zeta):\h^{-1}\cH\tilde\to\cH \Leftrightarrow
p(\bar{\zeta}):\h^{-1}\cH\tilde\to\cH$ is immediate (see again \cite{Weid}).  If these relations hold, then
$p(\zeta):\cH\tilde\to\h^{1}\cH$ because this operator is the adjoint
of $p(\bar{\zeta}):\h^{-1}\cH\tilde\to\cH$, and then by interpolation we
obtain $p(\zeta):\h^{-\12}\cH\tilde\to\h^{\12}\cH$ hence
$p(\bar{\zeta}):\h^{-\12}\cH\tilde\to\h^{\12}\cH$. Reciprocally, if
$p(\zeta):\h^{-\12}\cH\tilde\to\h^{\12}\cH$ then $p(\zeta):\h^{-1}\cH\tilde\to\cH$
because the domain of the operator in $\cH$ associated to
$p(\zeta):\h^{-\12}\cH\to\h^{\12}\cH$ is $\h^{-1}\cH$. \qed

In the sequel we will assume
\[
\leqno{\mbox{(A2)}} \hspace{1mm}\quad
\rho(h,k)\neq\emptyset.
\]
Let us state an easy lemma  which allows to check (A2).
\begin{lemma}\label{lm:idiot}
  If (A1) holds and $h$ is bounded below, then there is $c_{0}>0$
  such that
 \[
\{ z\ : |{\rm Im} z| > |{\rm Re}z| + c_{0}\}\subset \rho(h,k).
\]
\end{lemma}

\proof Consider $p(z)$ as a linear operator on $\cH$ with domain
$\langle h\rangle^{-1}\cH$.  Let $c$ be such that $h+ c^{2}\geq 1$
and $\delta= \| k(h+c^{2})^{-\12}\|$.  For $ z= a+\i b$, $\alpha>0$:
\[
\begin{array}{rl}
{\rm Re} p( z)=& h + b^{2}-a^{2} +2ka\\[2mm]
\geq& h+ c^{2} + b^{2}- a^{2}- c^{2} - \alpha a^{2} - \alpha^{-1}k^{2}\\[2mm]
\geq&(1- \alpha^{-1}\delta^{2})(h+ c^{2})+ b^{2}-a^{2}-c^{2}- \alpha a^{2}\\[2mm]
\end{array}
\]
For $\alpha= \delta^{2}$ this yields
\[
{\rm Re}p( z)\geq b^{2}- (1+\delta^{2})a^{2}- c^{2} \geq c_{1}>0,
\]
if $|b|>|a|+ c_{0}$ for $c_{0}>0$. If we set $p:= p(z)$ then for all
$u\in\jap{h}^{-1}\cH$ we shall have \( c_1 \| u\|^{2}\leq \Re
(u|pu)\leq\|u\| \|pu\| \) hence \( c_1 \| u\|\leq \|pu\| \) and
similarly \( c_1 \| u\|\leq \|p^*u\| \). Since $p$ is closed this
implies $p:\langle h\rangle^{-1}\cH \tilde\to \cH$. \qed

\begin{remark}{\rm We now make some comments
      on the connection between our assumptions and those of Langer,
      Najman, and Tretter \cite{LNT1,LNT2}. The main differences
      with respect to \cite{LNT1,LNT2} come from the fact that our
      conditions are stated in terms of $h$ instead of $h_0$.  They
      assume that $h_{0}$ is a given strictly positive self-adjoint
      operator (in particular, the wave equation is not considered)
      and that $k$ is a symmetric operator whose domain contains
      $\jap{h_{0}}^{-1/2}\ch$. Other conditions are needed to get
      deeper facts, e.g.\ in \cite{LNT1} it is required that
      $1-h_{0}^{-1/2}k^{2}h_{0}^{-1/2}$ be a boundedly invertible
      operator on $\ch$ to get a convenient definition of $h$.  In
      our setting, $h$ is given (with no assumptions on its
      spectrum) and we require $\jap{h}^{-1/2}\ch\subset D(k)$. If
      $h_0$ and $h$ are bounded from below then from Lemma
      \ref{lm:idiot} it follows that
      $\jap{h}^{-1/2}\ch=\jap{h_{0}}^{-1/2}\ch$ and we do not need
      any other assumption for the rest of our results. However, the
      operator $h$ in \cite{LNT1,LNT2} could be unbounded from below
      and then there is no straightforward relation between their
      results and ours.  }\end{remark}

\subsection{Spaces}\label{ss:kgs}
The following two spaces play a fundamental role in what follows:
\begin{equation}\label{eq:rk}
\rk:=\jh^{-\12}\cH\oplus\ch \quad\text{and}\quad  \rk^*:=\ch\oplus\jh^{\12}\cH.
\end{equation}
One often calls $\rk$ the \emph{energy space}.
Observe that $\rk\subset\rk^*$. As decided in Subsect. \ref{ss:not},
the space $\rk^*$ is identified with the adjoint space of $\rk$ with
the help of the sesquilinear form: 
\begin{equation}\label{eq:kphsp}
\braket{u}{v}=
(u_{0}| v_{1})+ (u_{1}| v_{0}),
\quad \text{for } 
u=(\begin{smallmatrix}u_0\\u_1\end{smallmatrix})\in\rk , \ 
v=(\begin{smallmatrix}v_0\\v_1\end{smallmatrix})\in\rk^*,
 \end{equation}
 usually called the {\em charge}.

 We identify $\rk^{**}=\rk$ as in the Hilbert space case by setting
 $\braket{v}{u}=\overline{\braket{u}{v}}$. This allows us to speak
 about symmetric or positive operators $S:\rk\to\rk^*$.

 Observe that we have dense and continuous embeddings $\rk\subset
 \ch\oplus\ch \subset \rk^*$ and the identification of $\rk^*$ with
 the adjoint of $\rk$ is determined by the Krein structure of
 $\ch\oplus\ch$ exactly as in the case of Friedrichs couples in the
 category of Hilbert spaces.  Note however that $\ch\oplus\ch$ is
 not an interpolation space between $\rk$ and $\rk^*$ if $\ch$ is
 infinite dimensional (see below). In any case, by complex
 interpolation we get for any $0\leq\sigma\leq1$:
\begin{equation}\label{eq:inter}
[\rk,\rk^*]_\sigma = \jh^{(\sigma-1)/2}\ch\oplus\jh^{\sigma/2}\ch,
\end{equation}
so we cannot obtain $\ch\oplus\ch$ in this way. 

We define the \emph{charge space of order
  $\theta$} for $0\leq\theta\leq1/2$ by
\begin{equation}\label{eq:theta}
\rh_{\theta}=\jh^{-\theta}\ch\oplus\jh^{\theta}\ch .
\end{equation}
Then $\rk\subset\rh_{\theta}\subset\rk^{*}$ strictly and two
such spaces are not comparable (if $\ch$ is infinite dimensional,
which is implicitly assumed in all this work). Observe that the
middle space defined by complex interpolation
\begin{equation}\label{eq:middle}
[\rk,\rk^{*}]_{1/2}=\jh^{-1/4}\ch\oplus\jh^{1/4}\ch 
\end{equation}
equals $\rh_{1/4}$ and we shall see
that it plays a remarkable  role in the theory. If $\theta\neq1/4$
then $\rh_{\theta}$ is not an interpolation space between $\rk$ and $\rk^*$:
in  Remark \ref{re:notbdd}
we give examples of bounded operators on $\rk^*$ which leave
$\rk$ invariant but not $\rh_{\theta}$ if $\theta\neq1/4$.

Since $(\jh^{-\theta}\ch)^{*}=\jh^{\theta}\ch$, the spaces
$(\rh_{\theta}, \braket{\cdot}{\cdot})$ are examples of Krein spaces
as in Subsect. \ref{ss:phspace}.

Below, when we speak of
selfadjointness of operators in $\rh_{\theta}$, we refer to this Krein
structure.

Since $\rk\subset \rk^{*}$,  the sesquilinear form $\braket{\cdot}{\cdot}$ restricts to a hermitian form on $\rk$. Note however that
$(\rk, \braket{\cdot}{\cdot})$ is {\em not} a Krein space, since $\braket{\cdot}{\cdot}$ is not non-degenerate on $\rk$.

\subsection{Operators }\label{ss:kgo}
It is easy to extend the relations from Subsect. \ref{ss:phspace}
 to the present setting. For example, since we think of elements of
 $\rk$ as column matrices, we may represent operators $\rk\to\rk^*$
 as matrices of operators:
\[
S=\mat{a}{b}{c}{d}\text{ with }
\left\{
\hspace{-1mm}
\begin{array}{l} a\in
  B(\jh^{-\12}\cH,\ch),\ b\in B(\ch),\\[2mm] c\in
  B(\jh^{-\12}\cH,\jh^{\12}\cH), \ d\in B(\ch,\jh^{\12}\cH).
\end{array} 
\right.
 \]
A computation gives $S^*=(\begin{smallmatrix} d^* & b^{*} \\ c^{*} &
  a^{*} \end{smallmatrix})$ hence $S$ is symmetric if and only if
\begin{equation}\label{eq:ksyms}
S=\mat{a}{b}{c}{a^{*}}\text{ with }
\left\{
\hspace{-1mm}
\begin{array}{l}
a\in B(\jh^{-\12}\cH,\ch),\ b=b^{*}\in B(\ch),\\[2mm] c=c^{*}\in
B(\jh^{-\12}\cH,\jh^{\12}\cH), \  d\in B(\ch,\jh^{\12}\cH). 
\end{array} 
\right.
 \end{equation}
Lemma \ref{lm:kpositive} also has a natural version in the present  
context.

We may view any $S\in B(\rk,\rk^{*})$ as operator on $\rk^*$ with
domain $\rk$, hence its resolvent set and spectrum are well
defined. More precisely, the \emph{resolvent set} $\rho(S)$ of $S$
is the set of $z\in\C$ such that $S-z:\rk\to\rk^*$ is bijective and
the \emph{spectrum} of $S$ is $\sigma(S)=\C\setminus\rho(S)$.

\subsection{Klein-Gordon operators}\label{ss:ukg}

The \emph{Klein-Gordon operator} is the continuous map
$\hat K:\rk\to\rk^{*}$ defined by
\begin{equation}\label{eq:kg0}
  \hat K= \mat{k}{1}{h_{0}}{k} \quad\text{hence}\quad
\hat K\lin{u_{0}}{u_{1}}=\lin{ku_{0}+u_{1}}{h_{0}u_{0}+ku_{1}}.
\end{equation}

Formally we see that if $\phi(t)$ is a solution of (\ref{kigi}) and
we set 
\beq\label{cauchy1} f(t)=
\lin{\phi(t)}{\i^{-1}\p_{t}\phi(t)-k\phi(t)}, \eeq 
then $f(t)=
\e^{\i t\hat{K}}f(0)$, hence $\hat{K}$ (or more precisely some of
its restrictions) is the generator of the group associated to
(\ref{kigi}) for the parametrization (\ref{cauchy1}) of Cauchy
data. The choice (\ref{cauchy1}) is natural when one wants to
emphasize the {\em symplectic aspect} of the Klein-Gordon equation
(\ref{kigi}).

From \eqref{eq:ksyms} it follows that $\hat{K}$ is a symmetric
operator and that for all $u\in\rk$:
\begin{equation}\label{eq:symf}
\braket{u}{\hat{K}u}=(u_0 |h_0u_0)+\|u_1\|^2+2\Re(ku_0|u_1)
=(u_0 |hu_0)+\|ku_0+ u_1\|^2. 
\end{equation}

Note that we may, and we shall, think of $\hat K$ as closed densely
defined operator in $\rk^*$. There is no a priori given Krein
structure on $\rk^*$ but various charge and energy Klein-Gordon
operators will be obtained as operators induced by $\hat K$ in Krein
spaces continuously embedded in $\rk^{*}$. 

\begin{proposition}\label{pr:srk} Assume (A1). Then 
 $\rho(\hat K)=\rho(h,k)$ and if $z\in\rho(\hat K)$ we have: 
\begin{equation}\label{eq:kinverse}
  (\hat K-z)^{-1}=:R_{\hat K}(z)=
  \mat{p(z)^{-1}(z-k)}{p(z)^{-1}}{1+(z-k)p(z)^{-1}(z-k)}{(z-k)p(z)^{-1}}.  
\end{equation} 
\end{proposition}
\proof We shall prove that $\hat K-z:\rk\tilde\to\rk^* \Leftrightarrow
p(z):\jh^{-\12}\cH\tilde\to\jh^{\12}\cH$ and if these conditions are
satisfied then we shall justify the formally obvious relation
\eqref{eq:kinverse}.  Assume first
$p(z):\jh^{-\12}\cH\tilde\to\jh^{\12}\cH$. Set $q =p(z)^{-1}$, $\ell=k-z$,
and let $G$ be the right hand of \eqref{eq:kinverse}, so that
\begin{equation}\label{eq:kinverse0}
G=\mat{-q\ell}{q }{1+\ell q\ell }{-\ell q} \quad\text{and}\quad
G\lin{a}{b}=\lin{-q(\ell a-b)}{a+\ell q (\ell a -b)}.
\end{equation}
This clearly defines a continuous operator $\rk^*\to\rk$ and a
simple computation gives $(\hat K-z)G=1$ on $\rk^*$ and
$G(\hat K-z)=1$ on $\rk$.  So $G$ is the inverse of
$\hat K-z:\rk\to\rk^*$.  

Reciprocally, assume that $\hat K-z:\rk\tilde\to\rk^{*}$.  If
$u_0\in\jh^{-\12}\cH$ and $u_1=-\ell u_0$ then $u_{1}\in\ch$ and
$h_{0}u_0+\ell u_1=(h_{0}-\ell^{2})u_0=p(z)u_0$ hence
$(\hat K-z) (\begin{smallmatrix} u_0 \\ u_1 \end{smallmatrix})
=(\begin{smallmatrix}0 \\ p(z)u_0 \end{smallmatrix})$. 
Thus if $p(z)u_0=0$ then
$(\hat K-z)(\begin{smallmatrix} u_0 \\ u_1 \end{smallmatrix})=0$ and 
so $u_0=0$. Hence $p(z):\jh^{-\12}\cH\to\jh^{\12}\cH$ is injective. Now
let $v_{1}\in\jh^{\12}\cH$. Since $(\hat K-z)\rk=\rk^{*}$ and 
$(\begin{smallmatrix}0\\v_{1}\end{smallmatrix})\in\rk^{*}$, there
are $u_0\in\jh^{-\12}\cH$ and $u_1\in\ch$ such that
$(\hat K-z)(\begin{smallmatrix} u_0 \\ u_1 \end{smallmatrix})=
(\begin{smallmatrix} 0 \\ v_1 \end{smallmatrix})$, or $\ell u_0+u_1=0$ and
$h_{0}u_0+\ell u_1=v_{1}$, hence $p(z)u_0=v_{1}$. This proves that
$p(z)\jh^{-\12}\cH=\jh^{\12}\cH$ and so $p(z):\jh^{-\12}\cH\tilde\to\jh^{\12}\cH$.
\qed

We now realize the Klein-Gordon operator as a closed densely defined
operator in other Banach spaces.

\begin{proposition}\label{pr:lrl}
  Let $\rl$ be a Banach space such that $\rk\subset\rl\subset\rk^*$
  continuously and densely. \emph{The operator $L$ induced by $\hat K$ in
    $\rl$} is the restriction of $\hat K$ to $\Dom L=\{u\in\rk \mid
  \hat Ku\in\rl\}$ considered as operator in $\rl$. This is a closed
  densely defined operator such that $\rho(L)\supset\rho(h,k)$ and
  $R_L(z):=(L-z)^{-1}=R_{\hat K}(z)|\rl$ for any $z\in\rho(h,k)$, in
  particular $\Dom L=R_{\hat K}(z)\rl$ for any such $z$.
\end{proposition}
\proof If $u\in\rk\subset\rl$ and $z\in\rho(h,k)$ then
${\hat K}u\in\rl$ if and only if $({\hat K}-z)u\in\rl$ hence if and only if
$u\in({\hat K}-z)^{-1}\rl=R_{\hat K}(z)\rl$.  

Since $R_{\hat K}(z)$ is a continuous surjection and
$\rl$ is dense in $\ck^*$, the space $\Dom L$ is dense in $\rk$, which
is dense in $\rl$, hence $\Dom L$ is also dense in $\rl$. By the
closed graph theorem, the restriction of $R_{\hat K}(z)$ to $\rl$ is a
continuous operator in $\rl$, so $L$ is a closed  densely defined
operator in $\rl$.
\qed

Let us now discuss several natural operators obtained from
Prop. \ref{pr:lrl} for various choices of $\rl$.

The largest possible choice of $\rl$ is $\rl = \rk^{*}$. In this
case the operator $L$ equals $\hat K$.  When we want to stress that
we look at ${\hat K}$ as closed densely defined operator in $\rk^*$
we denote it by $K_{\text{max}}$.

We have ${\hat K}^*={\hat K}$ if we consider ${\hat K}$ as an
operator $\rk\to\rk^*$ but as we shall see below $K_{\mathrm{min}}=
K_{\mathrm{max}}^*$ is a quite different object.

The smallest possible choice of $\rl$ is $\rl=\rk$. We shall denote
$K_{\text{min}}$ the operator induced by ${\hat K}$ in $\rk$.  Note  that
\[
K_{\text{min}}\subset L \subset K_{\text{max}},
\]  for any realization $L$ of the Klein-Gordon operator.

In the next proposition we describe explicitly the domain of
$K_{\mathrm{min}}$, its resolvent set, and we compute its
adjoint. Recall that we identified the adjoint space of $\rk$ with
$\rk^*$  with the help of the sesquilinear form \eqref{eq:kphsp}. In
particular, if $S$ is a closed densely defined operator in $\rk$
then the domain of $S^*$ is the set of $v\in\rk^*$ such that the map
$u\mapsto\braket{Su}{v}$ is continuous for the $\rk$-topology and
then $S^*v$ is the unique $w\in\rk^*$ such that
$u\mapsto\braket{Su}{v}=\braket{u}{w}$ for all $u\in \Dom S$. 

\begin{proposition}\label{pr:energy} Assume (A1), (A2). 
Let $K_{\mathrm{min}}$ be the operator induced by
${\hat K}$ in $\rk$. Then $K_{\mathrm{min}}^*=K_{\mathrm{max}}$,
  $\rho(K_{{\mathrm{min}}})=\rho(h,k)$ and 
\begin{equation}\label{eq:kge}
  \Dom K_{\mathrm{min}}=\{ (\begin{smallmatrix}u_0\\u_1\end{smallmatrix}) \mid 
  u_0\in\jh^{-1}\cH,\ u_1\in\ch,\  ku_0+u_1\in\jh^{-\12}\cH\}
\end{equation}
\end{proposition}
\proof We denote by $\rd$ the right hand of \eqref{eq:kge} and first
prove $\Dom K_{\mathrm{min}}=\rd$. 

We have $u\in \Dom K_{\mathrm{min}}$ if and only if $u\in\rk$ and
${\hat K}u\in\rk$, i.e. $ku_0+u_1\in\jh^{-\12}\cH$ and
$h_0u_0+ku_1\in\ch$. These conditions are satisfied if $u\in\rd$
because $h_0u_0+ku_1=hu_0+k(ku_0+u_1)$ and
$hu_0\in\ch,k(ku_0+u_1)\in\ch$. Thus $\rd\subset \Dom
K_{\mathrm{min}}$. Reciprocally, if $u\in \Dom K_{\mathrm{min}}$
then $hu_0=h_0u_0-k^2u_0=(h_0u_0+ku_1)-k(ku_0+u_1)$ belongs to
$\ch$, hence $u_0\in\ch^1$. This proves that $\Dom
K_{\mathrm{min}}\subset\rd$ hence \eqref{eq:kge} is true.
 
Next we prove $K_{\mathrm{min}}^*=K_{\mathrm{max}}$.  For any $u\in
\Dom K_{\mathrm{min}}$ and $v\in\rk^*$ we have 
\[
\braket{Ku}{v}=(ku_0+u_1|v_1)+(h_0u_0+ku_1|v_0).
\]
If $v\in\rk=\Dom K_{\mathrm{max}}$ then it is clear that the right
hand side is continuous for the $\rk$-topology and the right hand
side above is just $\braket{u}{K_{\mathrm{max}}v}$. Therefore $K_{\mathrm{max}}\subset K_{\mathrm{min}}^{*}$.

Reciprocally, we would like to
show that 
\begin{equation}
\label{suto}
|\braket{K_{\mathrm{min}}u}{v}|\leq C\|u\|_\rk, \ \forall \ u\in\Dom K_{\mathrm{min}} 
\end{equation}
implies $v\in\rk$.
 Fix $z\in\rho(h,k)$ and let
$R=R_{\hat K}(z)$. Then $(K_{\mathrm{min}} -z)^{-1}=R_{\mid \rk}$ by Prop.
\ref{pr:lrl}. Note that (\ref{suto}) is equivalent
to \[
|\braket{({\hat K}-z)u}{v}|\leq C'\|u\|_\rk \ , \forall \ u\in
\Dom K_{\mathrm{min}},
\]
 for some constant $C'$ and this is
equivalent to 
\[
|\braket{w}{v}|=|\braket{({\hat K}-z)Rw}{v}|\leq
C'\|Rw\|_\rk, \ \forall \ w\in\rk.
\] But $R:\rk^*\to\rk$ is continuous,
so we obtain $|\braket{w}{v}|\leq C''\|w\|_{\rk^*}$ for all $w\in\rk$.
Since $\rk$ is dense in $\rk^*$ we see that $\braket{\cdot}{v}$
extends to a continuous form on $\rk^*$, hence $v\in\rk$. 

Finally, we have
$\rho(K_{\mathrm{min}})=\rho(K_{\mathrm{max}})^*=\rho(h,k)^*=\rho(h,k)$.  
\qed

\begin{proposition}\label{pr:c0group}
  If (A1) holds and if $h$ is bounded from below then
  $K_{\mathrm{min}}$ and $K_{\mathrm{max}}$ are generators of
  $C_0$-groups.
\end{proposition}
\proof Since $K_{\mathrm{min}}^*=K_{\mathrm{max}}$ it suffices to
consider the case of $K_{\mathrm{min}}$. The rest of the proof is a
variation on the proof of \cite[Thm.\ 3.2]{Kako}. First we show that
it suffices to assume $h\geq1$. Indeed, if $c$ is a number such that
$h+c\geq1$ and if we replace everywhere $h$ by $h+c$ then $h_0$ gets
replaced by $h_0+c$ and we have
$$
\hat K= \mat{k}{1}{h_{0}+c}{k} -\mat{0}{0}{c}{0}.
$$
Since the last term is a bounded operator, it suffices to show that
the first term on the right hand side is a generator of $C_0$-group.
So from now on we may assume $h\geq1$. Then $0\in\rho(h,k)$ and due
to \eqref{eq:kinverse} we have
$$
K_{\mathrm{min}}^{-1}=
\mat{-h^{-1}k}{h^{-1}}{1+kh^{-1}k}{-kh^{-1}}.  
$$
We know that this is a bounded operator on $\rk$. On the other hand,
it is easy to check that the ``energy'' hermitian form
$\braket{u}{\hat{K}u}=(u_0 |hu_0)+\|ku_0+ u_1\|^2$ introduced in
\eqref{eq:symf} is an admissible scalar product on $\rk$, i.e. $\rk$
equipped with this form is a Hilbert space.  Since
$\braket{u}{\hat{K}K_{\mathrm{min}}^{-1}u}=\braket{u}{u}\in\R$, the
operator $K_{\mathrm{min}}^{-1}$ is symmetric, hence
$K_{\mathrm{min}}$ is a selfadjoint operator on this Hilbert
space. \qed

Another case of interest is $\rl=\rh_{\theta}$, $0\leq \theta<\12$,
which we now discuss.

\begin{proposition}\label{pr:kgc}  Assume (A1), (A2).
  Let $K_{\theta}$ be the operator induced by ${\hat K}$ in the
  space $\rh_{\theta}$ defined in \eqref{eq:theta}. Then
\begin{equation}\label{eq:kgc1}
  \Dom K_{\theta}=\{ (\begin{smallmatrix}u_0\\u_1\end{smallmatrix}) \mid 
  u_0\in\jh^{-\12}\cH,\ u_1\in\ch,\  ku_0+u_1\in\jh^{-\theta}\ch,\ 
  h_{0}u_0+ku_1\in\jh^{\theta}\ch \}. 
\end{equation}
Moreover $K_{\theta}$ is selfadjoint on the Krein space
$(\rh_{\theta}, \braket{\cdot}{\cdot})$ and
$\rho(K_{\theta})=\rho(h,k)$.
\end{proposition}
\proof If $v_0\in\jh^{-\theta}\cH$, $v_1\in\jh^{\theta}\ch$ and
$(\begin{smallmatrix}u_0\\u_1\end{smallmatrix}):=
R_{\hat K}(z)(\begin{smallmatrix}v_0\\v_1\end{smallmatrix})$ then, with the
notations of the proof of Prop. \ref{pr:srk}, we have $\ell
u_0+u_1=v_{0}$ and $h_0u_0+\ell u_1=v_{1}$ hence
$(\begin{smallmatrix}u_0\\u_1\end{smallmatrix})$ belongs to the set
$\rd$ defined by the right hand side of \eqref{eq:kgc1}. Thus
$R_{\hat K}(z)\rh\subset\rd$. Reciprocally, if $u_0,u_1$ are as in
\eqref{eq:kgc1} then
$(\begin{smallmatrix}v_{0}\\v_{1}\end{smallmatrix}):=
({\hat K}-z)(\begin{smallmatrix}u_{0}\\u_{1}\end{smallmatrix})$ belongs to
$\rh_{\theta}$ and $R_{\hat K}(z)(\begin{smallmatrix}v_{0}\\v_{1}\end{smallmatrix})=
(\begin{smallmatrix}u_{0}\\u_{1}\end{smallmatrix})$, thus
$\rd\subset R_{\hat K}(z)\rd$. This proves \eqref{eq:kgc1}. 

To prove the selfadjointness of $K_{\theta}$ it suffices to show
$R_{K_{\theta}}(z)^*=R_{K_{\theta}}(\bar{z})$ for some $z\in\rho(h,k)$, which is not
empty, by (A2). But this is obvious, see the line
before \eqref{eq:ksyms}. 

Since by Prop. \ref{pr:lrl} we know that $\rho(h,k)\subset \rho(K_{\theta})$, 
it remains to prove that $\rho(K_{\theta})\subset\rho(h,k)$.  Assume that
$K_{\theta}-z:\Dom K_{\theta}\tilde\to\rh_{\theta}$ and argue as in the proof of Prop.
\ref{pr:srk}. We first show that $p(z):\jh^{-\12}\cH\to\jh^{\12}\cH$ is
injective. If $u_0\in\jh^{-\12}\cH$ and $p(z)u_0=0$ set $u_1=-\ell
u_0$.  Then $u_{1}\in\ch$ and \[
h_{0}u_0+\ell
u_1=(h_{0}-\ell^{2})u_0=p(z)u_0=0,
\]
 hence 
$({\hat K}-z) (\begin{smallmatrix} u_0 \\ u_1 \end{smallmatrix})=0$.  
Also:
\[
\begin{array}{rl}
ku_0+u_1=&lu_0+u_1+zu_0=zu_0\in\jh^{-\12}\cH\subset\jh^{-\theta}\cH,\\[2mm]
h_0u_0+ku_1=&h_0u_0+lu_1+zu_1=zu_1\in\ch\subset\jh^{-\theta}\ch .
\end{array}
\]Thus $(\begin{smallmatrix} u_0 \\ u_1 \end{smallmatrix})\in \Dom K_{\theta}$
and $(K_{\theta}-z) (\begin{smallmatrix} u_0 \\ u_1 \end{smallmatrix})=0$, so
$u_0=0$. This proves the injectivity of
$p(z):\jh^{-\12}\cH\to\jh^{\12}\cH$.
 In particular, $p(z):\jh^{-1}\cH\to\ch$
is injective.  

According to Lemma \ref{lm:kg}, it remains to prove
that this map is also surjective. Let $v_1\in\ch$. Since
$(K_{\theta}-z)\Dom K_{\theta}=\rh_{\theta}$ and 
$(\begin{smallmatrix}0\\v_1\end{smallmatrix})\in\rh_{\theta}$, 
there is $u\in \Dom K_{\theta}$ 
such that 
$(K_{\theta}-z)u
=(\begin{smallmatrix}0\\v_1\end{smallmatrix})$, hence
$lu_0+u_1=0$ and $h_{0}u_0+lu_1=v_1$, thus $p(z)u_0=v_1$. But
$p(z)=h-z^{2}+2z k$ hence $hu_0=v_1+z^{2}u_0-2z ku_0\in\ch$ so
$u_0\in\jh^{-1}\cH$. Thus $p(z)\jh^{-1}\cH=\ch$.
\qed

\begin{remark}\label{re:1/4}{\rm As explained before, we have
    $K_{\mathrm{min}}\subset K_{\theta}\subset K_{\mathrm{max}}$ for
    any $0\leq \theta\leq \12$ and the spectrum of all these
    operators coincide. But for $\theta=1/4$ we have more: from
    \eqref{eq:middle} it follows that in this case the operator
    $K_{1/4}$ is obtained by interpolation of order $1/2$ between
    $K_{\mathrm{min}}$ and $K_{\mathrm{max}}=K_{\mathrm{min}}^*$ (in
    resolvent sense). In particular, these operators should have
    similar spectral properties and functional calculus, fact which
    will be confirmed by later developments.  }
\end{remark}

As an example, from Proposition \ref{pr:c0group}
  we get the following extension of \cite[Thm. 6.5]{LNT2}:

\begin{corollary}\label{cr:c0group}
  If (A1) holds and $h$ is bounded from below then the operator
  $K_{1/4}$ generates a $C_0$-group.
\end{corollary}

\subsection{Charge and energy operators }\label{ss:ceops}

The selfadjoint operator $K_{\theta}$ in the Krein space
$\rh_{\theta}$ will be called \emph{charge Klein-Gordon} operator,
although this terminology is often reserved to the case
$\theta=1/4$.

If  $\phi(t)$ is a solution of (\ref{kigi}) and we set instead of (\ref{cauchy1}):
\begin{equation}
\label{cauchy2}
f(t)= \lin{\phi(t)}{\i^{-1}\p_{t}\phi(t)},
\end{equation}
then formally $f(t)= \e^{\i t \hat{H}}f(0)$ for 
\[
\hat{H}= \mat{0}{1}{h}{2k}.
\]
The choice  (\ref{cauchy2}) of Cauchy data is the standard one in the PDE literature and is convenient when one wants to emphasize the
{\em energy conservation} of the Klein-Gordon equation (\ref{kigi}).
 
We now show that the 
operator $K_{\mathrm{min}}$ is isomorphic to the usual 
\emph{energy Klein-Gordon} operator $H$,  which is the  realization of $\hat{H}$ on $\rk$, so we could say that 
$K_{\mathrm{min}}$ is the \emph{energy Klein-Gordon operator in the charge
representation}. 

Note first that if $a:\jh^{-\12}\cH\to\ch$ is a continuous symmetric map then  
the operator $\Phi(a)=(\begin{smallmatrix}{1}&{0}\\{a}&{1}\end{smallmatrix})$
is a well defined continuous map $\rk^{*}\to\rk^{*}$ which leaves $\rk$
invariant. 
 Thus $\Phi(a)$ is an isomorphism 
$\rk^{*}\to\rk^{*}$ with $\Phi(-a)$ as inverse, which clearly implies
that $\Phi(a):\rk\to\rk$ is also an isomorphism. Observe that $\Phi(a)$ is
symmetric when considered as operator $\rk\to\rk^{*}$. 

Set $\Phi=\Phi(k)$. Then
\[
\hat{H}:=\mat{0}{1}{h}{2k}:\rk\to \rk^{*}
\]
is a continuous (not symmetric) operator and 
$\Phi \hat K=\hat{H}\Phi$.  

The usual {\em energy Klein-Gordon operator} $H$ is the closed operator in $\rk$
induced by $\hat H$.
Clearly 
\[
\Dom H= \jh^{-1}\cH\oplus\jh^{-\12}\cH\hbox{ and }\Phi K_{\mathrm{min}}\Phi^{-1}=H,
\]
 where $\Phi$ is considered as an
automorphism of $\rk$. Thus we immediately get $\rho(H)=\rho(h,k)$
and, more generally, $K_{\mathrm{min}}$ and $H$ have the same spectral properties.

We mention that the preceding relation
  $\rho(H)=\rho(h,k)$ is the analog in our context of the assertion
  $\rho(A)=\rho(L)$ in  \cite[Lemma 5.1]{LNT1}.

Assume now that $0\in\rho(h,k)$. According to Lemma \ref{lm:kg} this
is equivalent to $h:\jh^{-\12}\cH\tilde\to\jh^{\12}\cH$ hence
$(\begin{smallmatrix}{0}&{1}\\{h}&{0}\end{smallmatrix}):
\rk\tilde\to\rk^{*}$.  Then $\rk$, equipped with the form
\begin{equation}\label{eq:enre}
\braket{u}{v}_{\re}=
\braket{u}{(\begin{smallmatrix}{0}&{1}\\{h}&{0}\end{smallmatrix})v}=
(u_{0}|hv_{0})+(u_{1}|v_{1})
\end{equation}
is a Krein space.  It is easy to check that $H$ is selfadjoint on
$(\re, \braket{\cdot}{\cdot}_{\rk})$. Indeed, we have
$0\in\rho(H)=\rho(h,k)$ and
$H^{-1}=(\begin{smallmatrix}{-2h^{-1}k}&{h^{-1}}\\{1}&{0}\end{smallmatrix})$
is a bounded symmetric operator because
$\braket{u}{H^{-1}u}_{\re}=2\Re(u_{0}|u_{1})-2(u_{0}|u_{0})$. 
This selfadjointness result should be compared with \cite[Thm. 4.3]{LNT1}.

This is the usual energy Klein-Gordon setting. We now express it in
the charge representation, i.e. in terms of the operator
$K_{\mathrm{\min}}$. Since $\Phi^{-1}:\re\to\rk$ is an isomorphism
which intertwines $E$ and $K_{\mathrm{min}}$ we see that the
\emph{energy Krein structure} on $\rk$ is given by \eqref{eq:symf}
and that $K_{\mathrm{min}}$ is selfadjoint for it.

\subsection{Free operators}\label{ss:free}
We now discuss the  {\em free operator}
\[
\hat{K}_{0} :=\mat{0}{1}{h_{0}}{0}:\rk\to\rk^*,
\] obtained for $k=0$. In this case $h_{0}= h$ and we will formulate
the various results below in terms of $h_{0}$.
Our purpose is to give some details on the
  functional calculus of the various free Klein-Gordon operators.
  We included this topic for completeness but also because the
  explicit formulas are important in Subsect.\
  \ref{ss:cops}. Moreover, they allow one to understand the
  optimality of the estimates in Theorem \ref{th:main}.

Denote by $L_{0}$ any of the operators $K_{0,\mathrm{min}}$ and
$K_{0, \theta}$ induced by $\hat{K}_{0}$ in $\rk$ and $\rh_\theta$
respectively.  Note that the operator $K_{0, \mathrm{max}}$ has the
same properties as $K_{0,\mathrm{min}}$ because
$K_{0,\mathrm{max}}=(K_{0,\mathrm{min}})^*$. 
\begin{lemma}\label{astula}
  Set $\sigma_{\pm}(h_{0}):= \sigma(h_{0})\cap \rr^{\pm}$ and
  $R_{h_{0}}(z):= (h_{0}-z)^{-1}$. Then: \beq\label{astalu}
  \sigma(L_{0})=\big(\sigma_+(h_{0})^{1/2}\big) \cup
  \big(-\sigma_+(h_{0})^{1/2}\big) \cup
  \big(\i|\sigma_-(h_{0})|^{1/2}\big) \cup
  \big(-\i|\sigma_-(h_{0})|^{1/2}\big), \eeq
\begin{equation}\label{eq:oinverse}
\begin{array}{rl}
  R_{L_{0}}(z)=&
  \mat{z R_{h_{0}}(z^2)}{R_{h_{0}}(z^2)}{1+z^2
    R_{h_{0}}(z^2)}{z R_{h_{0}}(z^2)}\\[3mm]
    =&  \mat{z
    R_{h_{0}}(z^2)}{R_{h_{0}}(z^2)}{{h_{0}}R_{h_{0}}(z^2)}{z
    R_{h_{0}}(z^2)} = 
  (L_{0}+z)R_{h_{0}}(z^2).
  \end{array}
\end{equation}
\end{lemma}
\proof By Props. \ref{pr:energy} and \ref{pr:kgc} we have
$\sigma(L_{0})=\{z\in \cc \ : \ z^{2}\in \sigma(h_{0})\}$, which implies
(\ref{astalu}). Then (\ref{eq:oinverse}) follows from
\[
(L_{0}-z)(L_{0}+z)=L_{0}^2-z^2= {h_{0}}-z^2,
\]
where  ${h_{0}}$ is identified with the diagonal matrix having ${h_{0}}$ 
on the diagonal.\qed

\begin{remark}\label{re:bad}{\rm Note that the resolvent of the
    operator $K_{0,0}$ has a rather unusual behavior: if $h_{0}$ is
    positive and unbounded and if we equip $\rh_0=\ch\oplus\ch$ with
    the Hilbert direct sum norm, then \eqref{eq:oinverse} implies
    $\|R_{K_{0,0}}(z)\|\geq \|h_{0}R_{h_{0}}(z^2)\|\geq1 \ \forall
    z$.  }\end{remark}

We now compute $\varphi(L_{0})$ for entire functions $\vphi$ by
using the relations
$$
L_{0}^{2n}=\mat{{h_{0}}^n}{0}{0}{{h_{0}}^n} \quad\text{and}\quad
L_{0}^{2n+1}=\mat{0}{{h_{0}}^{n}}{{h_{0}}^{n+1}}{0}, \ n\in \nn.
$$
If $\varphi(z)=\sum_{n\geq0}a_nz^n$ and if we define 
\begin{align}
  \varphi_\rmc (z) &=\frac{1}{2}\big(\varphi(\sqrt z)+ \varphi(-\sqrt
  z) \big)=
  {\textstyle\sum_{n\geq0}}a_{2n}z^{n}, \\
  \varphi_\rms (z) & =\frac{1}{2\sqrt z}\big(\varphi(\sqrt z)-
  \varphi(-\sqrt z) \big)= {\textstyle\sum_{n\geq0}}a_{2n+1}z^{n}
\end{align}
then by working with the set of entire vectors of the selfadjoint
operator ${h_{0}}$ in $\ch$ we obtain
\begin{equation}\label{eq:fct}
\varphi(L_{0})=\mat{\varphi_\rmc ({h_{0}})}{\varphi_\rms
  ({h_{0}})}{{h_{0}}\varphi_\rms ({h_{0}})}{\varphi_\rmc ({h_{0}})} .
\end{equation}

For example, if ${h_{0}}=\veps^2$ for some operator $\veps$, not
necessarily selfadjoint, then
\begin{equation}
\rme^{\i tL_{0}}=
\left(
\begin{array}{cc}
\cos(t\veps) &  \i\veps^{-1}\sin(t\veps) \\
\i\veps\sin(t\veps)  &   \cos(t\veps)
\end{array}
\right).
\end{equation}

Let us now assume $h_{0}=\epsilon^{2}$ for $\epsilon\geq 0$. 
Then
$\sigma(L_{0})=\sigma(\veps)\cup-\sigma(\veps)$ and \eqref{eq:fct}
becomes
\begin{equation}\label{eq:funct}
\varphi(L_{0})=\mat{\frac{\varphi(\veps)+\varphi(-\veps)}{2}}
{\frac{\varphi(\veps)-\varphi(-\veps)}{2\veps}}
{\veps\frac{\varphi(\veps)-\varphi(-\veps)}{2}}
{\frac{\varphi(\veps)+\varphi(-\veps)}{2}}
=\mat{\varphi_{+}(\veps)}{\varphi_{-}(\veps)/\veps}
{\varphi_{-}(\veps)\veps}{\varphi_{+}(\veps)}
\end{equation}
where \[
\vphi_{\pm}(x)=(\vphi(x)\pm\vphi(-x))/2,
\] are the even and odd
parts of the function $\vphi$. The value of
$(\vphi(x)-\vphi(-x))/2x$ at $x=0$ is $\vphi'(0)$ by definition.

We now discuss bounds for the Borel functional calculus of $L_{0}$.

The bounds in the case of $K_{0,\mathrm{min}}$ and $K_{0,
  \mathrm{max}}$ are of a different nature than those for $K_{0,
  \theta}$ (unless $\theta=1/4$).  We introduce the following spaces
$\Lambda$, $\Lambda_{\theta}$ of bounded Borel functions. Recall
that $\varphi_{\pm}$ denote the even/odd parts of $\varphi$.
\begin{definition}
  We denote by $\Lambda$, resp. $\Lambda_{\theta}$, the spaces of
  Borel functions $\vphi:\R\to\C$ such that:
\begin{equation}\label{eq:rest}
\|\vphi\|_{\Lambda}:=
\sup_{x\in\R}|\vphi(x)|
+\sup_{x\geq 0}|\varphi_{-}(x)/x|<\infty,
\end{equation} 
resp.
\begin{equation}\label{eq:ktst}
\|\vphi\|_{\Lambda_{\theta}}:=
\|\vphi\|_{\Lambda}+\sup_{x\geq 0}|\varphi_{-}(x)/x|
+\sup_{x\in\R}|\varphi_-(x)|\jap{x}^{|4\theta-1|}<\infty. 
\end{equation}
 \end{definition}
 Note that $\Lambda_{1/4}= \Lambda$.
 
\begin{lemma}\label{lm:efctcalc}
 Assume $h_{0}=\epsilon^{2}$ for some $\epsilon\geq 0$.   
 Then there is a unique
  linear map   $\Lambda\ni\vphi\mapsto\vphi(K_{0,\mathrm{min}})\in B(\rk)$
   such that
  $\vphi(K_{0,\mathrm{min}})=(K_{0,\mathrm{min}}-z)^{-1}$ if
  $\vphi(x)=(x-z)^{-1}$ with $z\nin\R$ and such that the
  following continuity property is satisfied: 
  
  if $\vphi_{n}$ is a bounded sequence in $\Lambda$ with
  $\vphi_{n}(x)\to\vphi(x)$ for each real $x$, then
  $\vphi_{n}(K_{0,\mathrm{min}})\to\vphi(K_{0,\mathrm{min}})$
  weakly.
  
  The map $\Lambda\ni \vphi\mapsto\vphi(K_{0,\mathrm{min}})\in B(\rk)$ is an
  algebra morphism and \eqref{eq:funct} holds. Moreover:  \begin{equation}
\label{irtu}
\|\vphi(K_{0,\mathrm{min}})\|_{B(\rk)}\leq C\|\vphi\|_{\Lambda}, \ C\geq 0.
\end{equation}
\end{lemma}

\begin{lemma}\label{lm:kfctcalc}
  Assume $h_{0}=\epsilon^{2}$ for some $\epsilon\geq 0$.  Then there
  is a unique linear map
  $\Lambda_{\theta}\ni\vphi\mapsto\vphi(K_{0,\theta})\in
  B(\rh_{\theta})$ such that
  $\vphi(K_{0,\theta})=(K_{0,\theta}-z)^{-1}$ if
  $\vphi(x)=(x-z)^{-1}$ with $z\nin\R$ and such that the following
  continuity property is satisfied:
  
  if $\vphi_{n}$ is a
  bounded sequence in $\Lambda_{\theta}$ with $\vphi_{n}(x)\to\vphi(x)$ for
  each real $x$, then
  $\vphi_{n}(K_{0,\theta})\to\vphi(K_{0,\theta})$
  weakly. 
  
  The map $\Lambda_{\theta}\ni\vphi\mapsto\vphi(K_{0,\theta})\in
  B(\rh_{\theta})$ is an algebra morphism and \eqref{eq:funct}
  holds. Moreover:
  \begin{equation}
\label{irti}
\|\vphi(K_{0,\theta})\|_{B(\rh_{\theta})}\leq C\|\vphi\|_{\Lambda_{\theta}}, \ C\geq 0.
\end{equation}
\end{lemma}

{\it Proof of Lemmas \ref{lm:efctcalc}, \ref{lm:kfctcalc}.}  For
later use we note the following easy facts:
 \begin{equation}
\label{zahia-1}
\sup_{x\in \rr}|\varphi(x)|\sim \sup_{x\geq 0}|\varphi_{+}(x)|+ \sup_{x\geq 0}|\varphi_{-}(x)|,
\end{equation}
 \begin{equation}
\label{zahia0}
\begin{array}{rl}
&\sup_{x\geq 0}|\langle x\rangle\varphi_{-}(x)/x| + \sup_{x\geq 0}|x\varphi_{-}(x)/\langle x\rangle|\\[2mm]
\sim& \sup_{x\geq 0}| \varphi_{-}(x)| + \sup_{x\geq 0}|\varphi_{-}(x)/x|,
\end{array}
\end{equation}
\begin{equation}
\label{zahia1}
\begin{array}{rl}
&\sup_{x\geq 0}|\langle x\rangle^{4\theta}\varphi_{-}(x)/x|+\sup_{x\geq 0}|x\varphi_{-}(x)/\langle x\rangle^{4\theta}|\\[2mm]
\sim& \sup_{x\geq 0}|\varphi_{-}(x)|+\sup_{x\geq 0}|\varphi_{-}(x)\varphi_{-}(x)/x|+\sup_{x\geq 0}|\langle x\rangle^{|4\theta-1|}\varphi_{-}(x)|.
\end{array}
\end{equation}
Let us first prove Lemma \ref{lm:efctcalc}.
We consider on $\rk$  the admissible norm 
defined by $\|u\|_{\rk}^{2}= \|\jap{\veps}u_{0}\|^{2}
+\|u_{1}\|^{2}$. The diagonal matrix with
coefficients $\jap{\veps}$ and $1$ is an isometric bijection
$\rk\to\rh_{0}=\ch\oplus\ch$. It follows from (\ref{eq:funct}) that if $\varphi$ is an entire function, bounded on $\rr$,   the norm of the operator
$\vphi(K_{0,\mathrm{min}})$ in $\rk$ is equal to the norm in
$\rh_{0}$ of the operator
\[
\begin{array}{rl}
&\mat{\jap{\veps}}{0}{0}{1}
\mat{\varphi_{+}(\veps)}{\varphi_{-}(\veps)/\veps}
{\varphi_{-}(\veps)\veps}{\varphi_{+}(\veps)}
\mat{\jap{\veps}^{-1}}{0}{0}{1}\\[3mm]
=&\mat{\varphi_{+}(\veps)}
{\jap{\veps}\varphi_{-}(\veps)/\veps}
{\varphi_{-}(\veps)\veps/\jap{\veps}}
{\varphi_{+}(\veps)},
\end{array}
\]
with a convention as stated above for $\vphi_{-}(0)/0$.  Hence there
is a number $c>0$ such that
\begin{equation}\label{eq:reste}
c\|\vphi(K_{0,\mathrm{min}})\|_{\rk}\leq \sup_{x\geq0}|\vphi_{+}(x)| 
+ \sup_{x\geq0}|\jap{x}\vphi_{-}(x)/x|
+ \sup_{x\geq0}|x\vphi_{-}(x)/\jap{x}|.
\end{equation}
Applying (\ref{zahia-1}), (\ref{zahia0}) we obtain (\ref{irtu}).  We extend the functional calculus from entire functions in $\Lambda$ to Borel functions in $\Lambda$ in the standard way.

To prove Lemma \ref{lm:kfctcalc} we argue similarly, introducing the compatible norm 
$\|u\|_{\rh_{\theta}}^{2}=
\|\jap{\veps}^{2\theta}u_{0}\|^{2}
+\|\jap{\veps}^{-2\theta}u_{1}\|^{2}$ on $\rh_{\theta}$. 
The diagonal matrix with coefficients
$\jap{\veps}^{2\theta}$ and $\jap{\veps}^{-2\theta}$ is an isometric
bijection $\rh_{\theta}\to\rh_{0}$. Hence the norm of $\varphi(K_{0, \theta})$ in $\rh$
is equal to the norm in $\rh_{0}$ of the operator
\[
\begin{array}{rl}
&\mat{\jap{\veps}^{2\theta}}{0}{0}{\jap{\veps}^{-2\theta}}
\mat{\varphi_{+}(\veps)}{\varphi_{-}(\veps)/\veps}
{\varphi_{-}(\veps)\veps}{\varphi_{+}(\veps)}
\mat{\jap{\veps}^{-2\theta}}{0}{0}{\jap{\veps}^{2\theta}}\\[3mm]
=&\mat{\varphi_{+}(\veps)}
{\jap{\veps}^{4\theta}\varphi_{-}(\veps)/\veps}
{\varphi_{-}(\veps)\veps/\jap{\veps}^{4\theta}}
{\varphi_{+}(\veps)}.
\end{array}
\]
Thus there is a number $c>0$ such that
\begin{equation}\label{eq:ktste}
c\|\vphi(K_{0, \theta})\|_{\rh}\leq \sup_{x\geq0}|\vphi_{+}(x)| 
+ \sup_{x\geq0}|\vphi_{-}(x)/x|\jap{x}^{4\theta}
+ \sup_{x\geq0}|x\vphi_{-}(x)|/\jap{x}^{4\theta}.
\end{equation}
Using (\ref{zahia-1}), (\ref{zahia1}) we obtain (\ref{irti}). \qed

\begin{remark}\label{re:notbdd}{\rm If $\varepsilon$ is not bounded
    we see that the lack of regularity at infinity of the function
    $\vphi(x)=\rme^{\i tx}$ makes $\rme^{\i tK_{0, \theta}}$
    unbounded if $t\neq0$ and $\theta\neq1/4$. This fact also allows
    us to show that the spaces $\rh_{\theta}$ with $\theta\neq1/4$
    are not interpolation spaces between $\rk$ and $\rk^*$. Indeed,
    if $t\neq0$ then $\rme^{\i tK_{\mathrm{max}}}$ is bounded in
    $\rk^*$, leaves $\rk$ invariant and induces there the bounded
    operator $\rme^{\i tK_{\mathrm{min}}}$. It induces in $\rh$ the
    densely defined operator $\rme^{\i tK_{0, \theta}}$ which is
    unbounded if $\theta\neq1/4$.  }\end{remark}

\begin{remark}\label{re:chipm}{\rm
One may clearly give sense to the right hand side of \eqref{eq:funct}
as a closed densely defined operator for a large class of functions
$\vphi$ and so to give a meaning to $\vphi(L_{0})$ as (unbounded) operator.
For example, if $\varepsilon>0$ then
\begin{equation}
\one_{\R^{\pm}}(L_{0})=\frac{1}{2}\mat{1}{\pm\veps^{-1}}{\pm\veps}{1}=:\Pi_{\pm}
\end{equation}
and these are the spectral projections of $L_{0}$ corresponding to the
half lines $\R^{\pm}$.  By the preceding lemmas or by a simple
direct argument the operators $\one_{\R^{\pm}}(K_{\min}^{0})$ are
bounded operators on $\rk$ if and only if $\inf\veps >0$ while the
$\one_{\R^{\pm}}(K_{0,\theta})$ are bounded operators on
$\rh_\theta$ if and only if $\inf\veps>0$ and $\theta=1/4$. In any
case, the $\Pi_{\pm}$ are projections
(i.e. $\Pi_{\pm}^{2}=\Pi_{\pm}$) such that
$\Pi_{+}\Pi_{-}=\Pi_{-}\Pi_{+}=0$ and $\Pi_{+}+\Pi_{-}=1$ at least
on dense domains.  It is easy to check that
$\one_{\R^{+}}(K_{0,\theta})\geq0$ and $\one_{\R^{-}}(K_{0,\theta})\leq0$ (by
Lemma \ref{lm:kpositive} in the bounded case and a direct argument
in general). The case of $\Pi_{+}=\one_{\R^{+}}(K_{0,\theta})$ for
$\theta\neq1/4$ (e.g. let $\theta=0$ and $\inf\veps>0$) is
particularly interesting: this is a positive selfadjoint operator
on $\rh$ which is an (unbounded) orthogonal projection whose
resolvent set is empty.  Indeed, for any $z\neq0,1$ the operator
$z(\Pi_{+}-z)^{-1}=(1-z)^{-1}\Pi_{+}-1$ is not bounded.
}\end{remark}

It is easy to compute the boundary values of the resolvent and the
``spectral measure'' of $L_{0}$. From \eqref{eq:oinverse} we see that if
$\lambda>0$ then, in the sense of distributions,
\begin{equation}\label{eq:+}
R_{L_{0}}(\lambda+ \i0)=
\left( \begin{array}{cc}
\lambda R_{h_{0}}(\lambda^{2}+\i0) & R_{h_{0}}(\lambda^{2}+\i 0)\\
{h_{0}}R_{h_{0}}(\lambda^{2}+\i 0) & \lambda R_{h_{0}}(\lambda^{2}+\i 0)
\end{array}\right),
\end{equation}
while if $\lambda<0$ then
\begin{equation}\label{eq:-}
R_{L_{0}}(\lambda+ \i 0)=
\left( \begin{array}{cc}
\lambda R_{h_{0}}(\lambda^{2}-\i 0) & R_{h_{0}}(\lambda^{2}-\i 0)\\
h_{0}R_{h_{0}}(\lambda^{2}-\i 0) & \lambda R_{h_{0}}(\lambda^{2}-\i 0)
\end{array}\right).
\end{equation}
Recall that, if $S$ is a selfadjoint (in the usual sense) operator with 
resolvent $R_{S}$ and spectral measure $E_{S}$ then
\[
E_{S}'(\lambda)=\frac{1}{2\pi \i }\big( R_{S}(\lambda+\i 0)-R_{S}(\lambda-\i 0)\big)
\]
by which we mean
$\vphi(S)=\int \vphi(\lambda) d E_{S}(\lambda)
=\int \vphi(\lambda) E_{S}'(\lambda)d\lambda$
where the second equality holds in the sense of distributions for smooth 
$\vphi$. If $S>0$ (i.e.\ $S\geq0$ and is injective) then we get:
\[
\begin{array}{rl}
&\int \vphi(\lambda) E_{S}'(\lambda^{2})d\lambda
=\int \frac{1}{2\lambda^{1/2}}\vphi(\lambda^{1/2}) 
E_{S}'(\lambda)d\lambda\\[2mm]
=&\frac{1}{2S^{1/2}}\vphi(S^{1/2})
=\frac{1}{2S^{1/2}}\int\vphi(\lambda)E_{S^{1/2}}'(\lambda),
\end{array}
\]  
which can be written 
\[
E_{S}'(\lambda^{2})=\frac{1}{2S^{1/2}}E_{S^{1/2}}'(\lambda)
=\frac{1}{2\lambda}E_{S^{1/2}}'(\lambda).
\]
By using this in \eqref{eq:+} and \eqref{eq:-} we get for $\lambda>0$:
\begin{equation}\label{eq:+sp}
E_{L_{0}}'(\lambda)=
\left( \begin{array}{cc}
\lambda E_{h_{0}}'(\lambda^{2}) & E_{h_{0}}'(\lambda^{2})\\
h_{0}E_{h_{0}}'(\lambda^{2}) & \lambda E_{h_{0}}'(\lambda^{2})
\end{array}\right)
=\frac{1}{2}\left( \begin{array}{cc}
E_\veps'(\lambda) & \veps^{-1}E_\veps'(\lambda)\\
\veps E_\veps'(\lambda) & E_\veps'(\lambda)
\end{array}\right)
\end{equation}
and
\begin{equation}\label{eq:-sp}
E_{L_{0}}'(-\lambda)= -
\left( \begin{array}{cc}
-\lambda E_{h_{0}}'(\lambda^{2}) & E_{h_{0}}'(\lambda^{2})\\
h_{0} E_{h_{0}}'(\lambda^{2}) & -\lambda E_{h_{0}}'(\lambda^{2})
\end{array}\right)
=\frac{1}{2}\left( \begin{array}{cc}
E_\veps'(\lambda) & -\veps^{-1}E_\veps'(\lambda)\\
-\veps E_\veps'(\lambda) & E_\veps'(\lambda)
\end{array}\right).
\end{equation}

\subsection{Conjugate operators for $K_\theta$}\label{ss:cops}

We now construct conjugate operators for the free and total
Hamiltonian.  The treatment is cleaner for the charge Klein-Gordon
operators $K_{0, \theta}$, $K_{\theta}$ because they are
selfadjoint for the same Krein structure so we concentrate on this
case.

Several types of conjugate operators can be considered in this
context, here we shall work only with those of scalar type. To be
precise, operators of the form $S=s\oplus s$, i.e. diagonal matrices
$S=(\begin{smallmatrix}s&0\\0&s\end{smallmatrix})$, will be called
\emph{scalar operators}.  We use the same notation for an operator
$s$ in $\jap{h}^{\theta}\ch$ which leaves $\jap{h}^{-\theta}\ch$
invariant and the diagonal operator $S=s\oplus s$ in $\rh_\theta$.

We introduce the assumptions (the first one appears already in
\cite{J1}, see also\cite{LNT2}):
\[
\leqno{\mbox{(E)}} \hspace{1mm}
\left\{ 
\hspace{-4mm}
\begin{array}{ll}
& \veps \text{ is a positive 
  selfadjoint operator on } \ch, \\[1mm]
&k:\Dom\varepsilon\to\cH  \text{ is compact and symmetric as
  operator in } \ch.
\end{array}
\right.
\]
\[
\leqno{\mbox{(M)}} \hspace{1mm}
\left\{ 
\hspace{-4mm}
\begin{array}{ll}
& a \text{ is a selfadjoint operator on } 
\ch \text{ such that } 
\rme^{\i ta} \Dom\varepsilon\subset\Dom\varepsilon
\text{ for all } t\in\R, \\[1mm]
& \varepsilon \text{ and } k 
\text{ considered as operators } \Dom\varepsilon\to\cH
\text{ are of class } C^{1}_{\rmu}(a).
\end{array}
\right.
\]

If (E) holds the quadratic form $\varepsilon^2-k^{2}$ on $D(\veps)$ is
closed and bounded from below.  If $h$ is the associated
selfadjoint operator, $h$ is bounded below and its spectrum is
discrete below $\inf\varepsilon^2$. As before, we set
$h_0=\varepsilon^2$ and we have 
$\jap{h}^{-1/2}\ch=\jap{h_0}^{-1/2}\ch=\Dom\veps$.  This implies
$\jh^{s}\cH= \langle\epsilon\rangle^{2s}\cH$ for $|s|\leq 1/2$.

In particular (A1), (A2) of Sect.\ \ref{s:pencil} are satisfied, 
by  Lemma \ref{lm:idiot}.

If (M) holds  $\rme^{\i ta}$ induces a
$C_{0}$-group in $\Dom\veps$ hence in all $\jap{h}^{\sigma}\ch$  with $|\sigma|\leq \12$. This gives a meaning to the regularity
condition on $\veps$ and $k$. As before we use notations like
$\veps':=[\veps,\i a]$, etc.

Our purpose is to study the selfadjoint operators 
\begin{equation}\label{eq:HH}
K_{0,\theta}=\mat{0}{1}{\veps^{2}}{0} 
\quad\text{and}\quad
K_{\theta}=\mat{k}{1}{\veps^{2}}{k}
\end{equation}
acting in the Krein space $\rh_{\theta}$. The conjugate operator
will be 
\[
A:=\mat{a}{0}{0}{a}=a\oplus a .
\]
Clearly $A$ is the generator of the $C_{0}$-group of scalar
operators $\rme^{\i tA}=\rme^{\i ta}\oplus\rme^{\i ta}$ on
$\rh_{\theta}$.  More generally:

\begin{lemma}\label{lm:kreinreg}
  Let $A=a\oplus a$. Then $\rme^{\i tA}=\rme^{\i ta}\oplus\rme^{\i
    ta}$ is a $C_{0}$-group on $\rk^*$ which leaves invariant the
  spaces $\rk$ and $\rh$ and induces $C_0$-groups on them.  The
  Krein structure of $\rh_\theta$ is of class $C^1(A)$.
\end{lemma}

In fact $\rme^{\i tA}$ is unitary on $\rh_\theta$, i.e.  we have
$\braket{\rme^{\i tA}u}{\rme^{\i tA}v}=\braket{u}{v}$ for all
$u,v\in\rh_\theta$.

The resolvent of $K_\theta$ is the restriction of the resolvent
$R_{\hat{K}}(z):\rk^*\to\rk$ explicitly described in
\eqref{eq:kinverse} and it is easier to work with $R_{\hat{K}}(z)$.
Here and below $z$ is a fixed point in
$\rho(h,k)\cap\rho(h_0,0)$. Note that $\hat{K}-\hat{K}_{0}=
(\begin{smallmatrix}k&0\\0&k\end{smallmatrix}):\rk\to\rk^{*}$ is
compact hence $R_{\hat{K}}(z)-R_{\hat{K}_0}(z):\rk^*\to\rk$ is a
compact operator too. In particular 
$R_{K_\theta}(z)-R_{K_{0,\theta}}(z)$ \emph{is a
  compact operator on $\rh_\theta$}.

\begin{lemma}
  $K_\theta$ and $K_{0,\theta}$ are of class $C^{1}_{\rmu}(A)$.
\end{lemma}
\proof 
It suffices to prove the stronger property that the map 
\[
\rr\ni t\mapsto \e^{\i t A}R_{\hat{K}}(z)\e^{-\i tA}\in B(\rk^{*}, \rk)
\]
is norm differentiable.
 If we set $K(t)=\rme^{\i tA} K \rme^{-\i
  tA}$, this is clearly equivalent to the norm differentiability of
$t\mapsto K(t)\in B(\rk,\rk^{*})$. But this is obvious because if
$h_{t}=\rme^{-\i ta}h\rme^{\i ta}$ and $k_{t}$ is defined similarly,
then we have
$K(t)=(\begin{smallmatrix}k_t&1\\h_t&k_t\end{smallmatrix})$ and
$h_{t},k_{t}$ clearly are norm differentiable when considered as
$B(\jh^{-\12}\cH,\jh^{\12}\cH)$ valued functions.  \qed

We saw before that $K_{0,\theta}\geq0$ and
$\sigma(K_{0,\theta})=\sigma(\veps)\cup\sigma(-\veps)$.  Our first
purpose is to construct $a$ such that $A$ be conjugate to
$K_{0,\theta}$ on some subsets of its spectrum. Our choice of $A$
does not seem convenient because
\begin{equation}\label{eq:ha}
[K_{0,\theta},\i A]=\mat{0}{0}{{[\veps^{2},\i a]}}{0},
\end{equation}
but the restriction to positive or negative energies of this
commutator satisfies the Mourre estimate. It is here that positivity
properties of functions of $K_{0,\theta}$ with respect to the Krein
structure of $\rh_\theta$ will play a role.

\begin{lemma}\label{lm:hpos}
  Let $\vphi\in\Lambda_{\theta}$ with $\vphi\geq0$.  If
  $\vphi(\lambda)=0$ for $\lambda\leq0$ then
  $\vphi(K_{0,\theta})\geq0$.  If $\vphi(\lambda)=0$ for
  $\lambda\geq0$ then $\vphi(K_{0,\theta})\leq0$.
\end{lemma}
\proof In the first case we obtain from \eqref{eq:funct}
\begin{equation}\label{eq:posfct>0}
\varphi(K_{0,\theta})=\frac{1}{2}
\mat{\varphi(\veps)}{\varphi(\veps)/\veps}
{\veps\varphi(\veps)}{\varphi(\veps)} 
\end{equation}
while in the second case we get
\begin{equation}\label{eq:posfct<0}
\varphi(K_{0,\theta})=\frac{1}{2}
\mat{\varphi(-\veps)}{-\varphi(-\veps)/\veps}
{-\veps\varphi(-\veps)}{\varphi(-\veps)} 
\end{equation}
and Lemma \ref{lm:kpositive} gives the stated results. \qed

\begin{remark}\label{re:scalariz}{\rm By using
    the ``spectral projections'' $\Pi_{\pm}=\one_{\R^\pm}(H_{0})$
    associated to the intervals $\R^\pm$ discussed in Remark
    \ref{re:chipm} we see that the operator $H_{0}$ is ``scalar'' on
    each of the regions $\lambda>0$ and $\lambda<0$ in the following
    sense: if $\vphi$ is a bounded function with compact support in
    one of the regions $\lambda>0$ or $\lambda<0$ then
\begin{equation}\label{eq:scalar}
H_{0}\Pi_{\pm}=\pm \veps \Pi_{\pm} \quad\text{and}\quad
\vphi(H_{0})=\vphi(H_{0})\Pi_{\pm}=\vphi(\pm\veps)\Pi_{\pm}
\end{equation}
This is a simple computation based on \eqref{eq:posfct>0} and
\eqref{eq:posfct<0}. Note however that the second equality above is
also a direct consequence of the first one, i.e. the explicit
relations \eqref{eq:posfct>0} and \eqref{eq:posfct<0} are not really
needed.  }\end{remark}

\begin{remark}\label{re:tcare}{\rm If $\inf\veps>0$ and $\theta=1/4$
    then $\Pi_{\pm}$ are bounded orthogonal projections on
    $\rh_{1/4}$ with $\Pi_{+}\Pi_{-}=\Pi_{-}\Pi_{+}=0$,
    $\Pi_{+}+\Pi_{-}=\one$, and $\pm\Pi_{\pm}\geq0$. Then
    $\rh_{\pm}=\pm\Pi_{\pm}\rh_{1/4}$ are Hilbert spaces (the minus
    sign means that we change the sign of the scalar product), we
    have $\rh_{1/4}=\rh_{+}\oplus \rh_{-}$ topologically, and the operator
    $K_{0,1/4}$ leaves $\rh_{\pm}$ invariant and induces there
    selfadjoint operators in the usual sense. But the operators
    $\rme^{\i tA}$ do note leave invariant this direct sum if the
    commutator $[K_{0,1/4},\i A]$ is not trivial.  }\end{remark}

\begin{lemma}\label{lm:commut}
Let $\vphi,\psi\in C_{0}^{\infty}(]0,\infty[)$ with $\vphi\psi=\vphi$. 
Then \[
\vphi(K_{0,\theta})=\vphi(K_{0,\theta})\psi(\veps)=
\psi(\veps)\vphi(K_{0,\theta}),
\] and 
\begin{equation}\label{eq:commut}
\vphi(K_{0,\theta})[K_{0,\theta},\i A]\vphi(K_{0,\theta})  =  
\vphi(K_{0,\theta})\psi(\veps)\veps'\psi(\veps)\vphi(K_{0,\theta}).
\end{equation}
\end{lemma}
\proof
Clearly
$$
\vphi(K_{0,\theta})
=\vphi(K_{0,\theta})\psi(K_{0,\theta})\Pi_{+}
=\vphi(K_{0,\theta})\Pi_{+}\psi(\veps) 
=\vphi(K_{0,\theta})\psi(\veps).
$$ Then the left hand side above is
\[
\begin{array}{rl}
&\vphi(K_{0,\theta})K_{0,\theta} \i A\vphi(K_{0,\theta}) - \vphi(K_{0,\theta})\i
AK_{0,\theta}\vphi(K_{0,\theta}) \\[2mm]
=&\vphi(K_{0,\theta})\psi(\veps)\veps \i a\psi(\veps)\vphi(K_{0,\theta})-
\vphi(K_{0,\theta})\psi(\veps)\i a\veps\psi(\veps)\vphi(K_{0,\theta}), 
\end{array}
\]
which is equal to $\vphi(K_{0,\theta})\psi(\veps)[\veps,\i a]
\psi(\veps)\vphi(K_{0,\theta})$.  \qed

\begin{lemma}\label{lm:mest}
Assume that
$\one_{U}(\veps)\veps'\one_{U}(\veps)=\phi(\varepsilon)\one_{U}(\veps)$
for some open set $U\subset \rr^{+}$  and some 
$\phi\in C_{0}(]0,\infty[)$. Then 
\[
\vphi(K_{\theta})K_{\theta}'\vphi(K_{\theta})\simeq\vphi(K_{\theta})\phi(K_{\theta})\vphi(K_{\theta}), \ \forall \vphi\in C^{\infty}_{0}(U).
\]\end{lemma}
\proof
Due to Lemma \ref{lm:cmp} we have   
$\vphi(K_{\theta})K_{\theta}'\vphi(K_{\theta})\simeq\vphi(K_{0,\theta})K_{0,\theta}'\vphi(K_{0,\theta})$. 
Let $\psi\in C^{\infty}_{0}(U)$ such that $\vphi\psi=\vphi$. 
Then Lemma \ref{lm:commut} implies
\[  
\begin{array}{rl}
&\vphi(K_{\theta})K_{\theta}'\vphi(K_{\theta})\simeq\vphi(K_{0,\theta})\psi(\veps)\veps'\psi(\veps)\vphi(K_{0,\theta})\\[2mm]
=&\vphi(K_{0,\theta})\psi(\veps)\phi(\veps)\psi(\veps)\vphi(K_{0,\theta})\\[2mm]
=&\vphi(K_{0,\theta})\phi(K_{0,\theta})\vphi(K_{0,\theta})\\[2mm]
\simeq&\vphi(K_{\theta})\phi(K_{\theta})\vphi(K_{\theta}). \ \Box
\end{array}
\]

In the next proposition, we prove a Mourre estimate for
$K_{\theta}$, assuming that $K_{\theta}$ is definitizable.

\begin{proposition}\label{pr:mest}
Assume that (E), (M) are satisfied and that $K_{\theta}$ is definitizable on 
$\cK_{\theta}$. Let $J\subset ]0, +\infty[$ be a compact interval with 
$\one_{J}(K_{\theta})\geq 0$.  Assume finally that 
\beq\label{turlututu}
\one_{U}(\veps)\veps'\one_{U}(\veps)=\phi(\varepsilon)\one_{U}(\veps),
\eeq
with $U\subset]0,\infty[$ open and some $\phi\in C_{0}(]0,\infty[)$,  $\phi(x)>0$ on $J$. 
Then:
\ben
\item  $J$ contains at most a finite number 
of eigenvalues of $K_{\theta}$,
\item if $\lambda\in J$ is not an eigenvalue of $K_{\theta}$
then there is a number $c>0$ and a neighborhood $I$ of $\lambda$ in $J$ 
such that 
\[
\one_{I}(K_{\theta})\Re(K_{\theta}')\one_{I}(K_{\theta})\geq c \one_{I}(K_{\theta}).
\] 
\een
\end{proposition}
\proof
If $\vphi\in C^{\infty}_{0}(U)$ then from Lemma \ref{lm:mest} we get
\[
\begin{array}{rl}
&\vphi(K_{\theta})\Re(K_{\theta}')\vphi(K_{\theta})=
\Re (\vphi(K_{\theta})K_{\theta}'\vphi(K_{\theta}))\\[2mm]
\simeq& \Re (\vphi(K_{\theta})\phi(K_{\theta})\vphi(K_{\theta}))=
\vphi(K_{\theta})\phi(K_{\theta})\vphi(K_{\theta}).
\end{array}
\] 
By taking $\vphi$ equal to $1$ on $J$ we get
\[
\one_{I}(K_{\theta})\Re(K_{\theta}')\one_{I}(K_{\theta})\simeq \phi(K_{\theta})\one_{J}K_{\theta})\geq (\inf_{J}\phi) \one_{J}(K_{\theta}). 
\]
Then we apply the virial theorem proved in Corollary \ref{co:virial}. \qed

\subsection{Definitizability of charge Klein-Gordon operators}

In Prop. \ref{pr:mest} we assumed that $K_{\theta}$ was
definitizable. We state here a rather standard result in this
direction, see \cite{J1}, \cite{LNT2}. Note that the condition
$0\not\in \sigma(\varepsilon)$ below can be interpreted as (strict)
{\em positivity of the mass}.
\begin{proposition}
  Assume (A1), (A2) of Sect.\ \ref{s:pencil} and $0\not\in
  \sigma(\varepsilon)$. Then $K_{1/4}$ is definitizable on
  $(\rh_{1/4}, \braket{\cdot}{\cdot})$. Moreover the critical points
  of $K_{1/4}$ are eigenvalues.
\end{proposition}
\proof The result follows directly from \cite{J1}, provided we check
the hypotheses there.  Let us denote for simplicity $\rh_{1/4}$,
$K_{0, 1/4}$ and $K_{1/4}$ simply by $\rh$, $K_{0}$ and $K$.  Since
$0\not\in \sigma(\varepsilon)$, we can equip $\rh$ with the
Hilbertian scalar product
\[
(u|v)_{\rh}:= (u_{0}| \varepsilon^{\12}v_{0})+ (u_{1}| \varepsilon^{-\12}v_{1}),
\]
which induces the same topology on $\rh$.  $K_{0}$ is selfadjoint
for $(\cdot|\cdot)_{\rh}$, hence has no singular critical points
(see \cite{J1} for this notion). Moreover since $|K_{0}|=
\left(\begin{smallmatrix} \varepsilon&0\\0&\varepsilon
\end{smallmatrix}\right)
$ the spaces $\cH_{\pm1}$ in \cite[Sect. 1.2]{J1} are equal to
$\langle K_{0}\rangle^{\mp\12}\cH$. In particular we have
\begin{equation}
\label{jono}
\cH_{1}= \rk, \ \cH_{-1}= \rk^{*}.
\end{equation}
We have $K= K_{0}+V$, for $V= \left(\begin{smallmatrix}
 k&0\\0&k
\end{smallmatrix}\right)$. By (\ref{jono}) we see that $V:
\cH_{1}\to \cH_{-1}$ is compact iff $k:\langle
\varepsilon\rangle^{-1}\cH\to \cH$ is compact, which holds by
(E2). Therefore we can apply \cite[Thm. 3]{J1} to obtain the
proposition. \qed

\subsection{Examples}

We now give some concrete examples. Let us consider the {\em charged
  Klein-Gordon equation} on Minkowski space:
\[
(\p_{t}-\i v(x))^{2}\phi(t,x)-\Delta_{x}\phi(t,x)+ m^{2}\phi(t,x)=0,
\hbox{ in }\rr^{1+d}. 
\]
It is an example of (\ref{kigi}) for $\cH= L^{2}(\rr^{d}, dx)$,
$k=v(x)$ a real electric potential, and $h= -\Delta_{x}+ m^{2}-
v^{2}(x)$, $m>0$ is the mass of the Klein-Gordon field.  Concerning
the electric potential we assume
\begin{equation}
\label{taratata}
v\varepsilon^{-1} \text{ is compact on }L^{2}(\rr^{d}),
\end{equation}
Let us consider the charge Klein-Gordon operator $K= K_{1/4}$.

We have $h_{0}= -\Delta_{x}+ m^{2}$, $\veps=
(-\Delta_{x}+m^{2})^{\12}$ hence (E) is satisfied and
$\varepsilon^{-1}\cH$ equals the Sobolev space $H^{1}(\rr^{d})$.

As conjugate operator we take
\[
a= \12(f(|p|)p\cdot x + x\cdot pf(|p|)), 
\hbox{ with }f\in C^{\infty}_{0}(0, \infty), \ p= \i^{-1}\nabla_{x}.
\]
Clearly (M) is satisfied. Moreover $[\varepsilon, \i a]=
f(|p|)p^{2}\varepsilon^{-1}$. This implies that condition
(\ref{turlututu}) in Prop. \ref{pr:mest} is satisfied for all
$U\subset \rr\backslash\{0\}$.

The operator $\veps$ is clearly of class $C^{\infty}(a)$. If we
assume that
\begin{equation}
\label{tirlititi}
\langle x\rangle^{\alpha} v \varepsilon^{-1}\hbox{ is bounded on }L^{2}(\rr^{d}),
\end{equation} 
then $k$ is of class $C^{\alpha}_{\mathrm{u}}(a)$. Therefore for
$\alpha\geq 1$ condition (M2) is satisfied.  Moreover we easily see
that $K$ is of class $C^{\alpha}(A)$. Therefore if (\ref{tirlititi})
holds for some $\alpha>3/2$ we can apply Thm. \ref{th:bvr}.  Note
that one may add in the standard way a long-range component $v_{\rm
  l}(x)$ to $v(x)$, by imposing a decay condition on
$\p_{x}^{\alpha}v_{\rm }(x)$ for $|\alpha|\leq 2$.

Note that the operator $A$, hence the weights $\langle
A\rangle^{-s}$ are scalar operators. Again by standard arguments,
one obtains finally the following resolvent estimate on $K$, for $I$
a compact interval disjoint from eigenvalues of $K_{1/4}$:
\[
\sup_{z\in I\pm \i ]0, \nu]}\| \langle
x\rangle^{-s}(K-z)^{-1}\langle
x\rangle^{-s}\|_{B(\cK_{1/4})}<\infty, \ \forall s>\12. 
\]
Note that these estimates are also obtained in \cite{GGH1}, by a
different method.

\end{document}